\documentclass[twocolumn]{aastex63}
\usepackage{graphicx}
\usepackage{comment}

\def\griz{\hbox{$griz$}}		

\def\grizy{\hbox{$grizy$}}

\newcommand{\OII}{[\ion{O}{2}]}
\newcommand{\OIII}{[\ion{O}{3}]}
\begin{document}

\title{The Hobby-Eberly Telescope Dark Energy Experiment (HETDEX) \\ Survey Design, Reductions, and Detections
\footnote{Based on observations obtained with the Hobby-Eberly Telescope, which is a joint project of the University of Texas at Austin, the Pennsylvania State University, Ludwig-Maximilians-Universit\"at M\"unchen, and Georg-August-Universit\"at G\"ottingen.}
}

\author[0000-0002-8433-8185]{Karl Gebhardt}
\affiliation{Department of Astronomy, The University of Texas at Austin, 2515 Speedway Boulevard, Austin, TX 78712, USA}
\email{gebhardt@utexas.edu}


\author[0000-0002-2307-0146]{Erin Mentuch Cooper}
\affiliation{Department of Astronomy, The University of Texas at Austin, 2515 Speedway Boulevard, Austin, TX 78712, USA}
\affiliation{McDonald Observatory, The University of Texas at Austin, Austin, TX 78712}

\author[0000-0002-1328-0211]{Robin Ciardullo}
\affiliation{Department of Astronomy \& Astrophysics, The Pennsylvania State University, University Park, PA 16802, USA}
\affiliation{Institute for Gravitation and the Cosmos, The Pennsylvania State University, University Park, PA 16802, USA}

\author{Viviana Acquaviva}
\affiliation{Physics Department, CUNY NYC College of Technology, Brooklyn, NY 11201, USA}
\affiliation{Center for Computational Astrophysics, Flatiron Institute, New York, NY 10010, USA}

\author{Ralf Bender}
\affiliation{Max-Planck Institut f\"ur extraterrestrische Physik, Giessenbachstrasse 1, 85748 Garching, Germany}
\affiliation{University Observatory, Fakult\"at f\"ur Physik, Ludwig-Maximilians University Munich, Scheiner Strasse 1, 81679 Munich, Germany}

\author[0000-0003-4381-5245]{William P. Bowman}
\affiliation{Department of Astronomy \& Astrophysics, The Pennsylvania
State University, University Park, PA 16802, USA}
\affiliation{Institute for Gravitation and the Cosmos, The Pennsylvania State University, University Park, PA 16802, USA}

\author{Barbara G. Castanheira}
\affiliation{Department of Physics, Baylor University, One Bear Place 97316, Waco, TX 76798, USA}

\author{Gavin Dalton}
\affiliation{University of Oxford, Denys Wilkinson Building, Keble Road, Oxford, OX1 3RH, UK}

\author{Dustin Davis}
\affiliation{Department of Astronomy, The University of Texas at Austin, 2515 Speedway Boulevard, Austin, TX 78712, USA}

\author[0000-0001-6982-4081]{Roelof S. de Jong}
\affiliation{Leibniz-Institut f\"ur Astrophysik Potsdam (AIP), An der Sternwarte 16, 14482 Potsdam, Germany}

\author{D. L. DePoy}
\affiliation{Department of Physics and Astronomy, Texas A\&M University, College Station, TX, 77843-4242 USA}
\affiliation{George P.\ and Cynthia Woods Mitchell Institute for Fundamental Physics and Astronomy, Texas A\&M University, College Station, TX, 77843-4242 USA}

\author{Yaswant Devarakonda}
\affiliation{Department of Physics and Astronomy, Texas A\&M University, College Station, TX, 77843-4242 USA}

\author{Sun Dongsheng}
\affiliation{Institute for Cosmic Ray Research, The University of Tokyo, 5-1-5 Kashiwanoha, Kashiwa, Chiba 277-8582, Japan}

\author{Niv Drory}
\affiliation{McDonald Observatory, The University of Texas at Austin, Austin, TX 78712}

\author{Maximilian Fabricius}
\affiliation{Max-Planck Institut f\"ur extraterrestrische Physik, Giessenbachstrasse 1, 85748 Garching, Germany}
\affiliation{University Observatory, Fakult\"at f\"ur Physik, Ludwig-Maximilians University Munich, Scheiner Strasse 1, 81679 Munich, Germany}

\author[0000-0003-2575-0652]{Daniel J. Farrow}
\affiliation{Max-Planck Institut f\"ur extraterrestrische Physik, Giessenbachstrasse 1, 85748 Garching, Germany}
\affiliation{University Observatory, Fakult\"at f\"ur Physik, Ludwig-Maximilians University Munich, Scheiner Strasse 1, 81679 Munich, Germany}

\author{John Feldmeier}
\affiliation{Department of Physics, Astronomy, Geology \& Environmental Sciences, Youngstown State University Youngstown, OH 44555}

\author{Steven L. Finkelstein}
\affiliation{Department of Astronomy, The University of Texas at Austin, 2515 Speedway Boulevard, Austin, TX 78712, USA}

\author{Cynthia S. Froning}
\affiliation{McDonald Observatory, The University of Texas at Austin, Austin, TX 78712}

\author{Eric Gawiser}
\affiliation{Physics and Astronomy Department, Rutgers, The State University, Piscataway, NJ 08854-8019}

\author{Caryl Gronwall}
\affiliation{Department of Astronomy \& Astrophysics, The Pennsylvania
State University, University Park, PA 16802, USA}
\affiliation{Institute for Gravitation and the Cosmos, The Pennsylvania State University, University Park, PA 16802, USA}

\author{Laura Herold}
\affiliation{Max-Planck-Institut f\"ur Astrophysik, Karl-Schwarzschild-Str. 1, 85748 Garching, Germany}

\author{Gary J. Hill}
\affiliation{Department of Astronomy, The University of Texas at Austin, 2515 Speedway Boulevard, Austin, TX 78712, USA}
\affiliation{McDonald Observatory, The University of Texas at Austin, Austin, TX 78712}

\author{Ulrich Hopp}
\affiliation{University Observatory, Fakult\"at f\"ur Physik, Ludwig-Maximilians University Munich, Scheiner Strasse 1, 81679 Munich, Germany}
\affiliation{Max-Planck Institut f\"ur extraterrestrische Physik, Giessenbachstrasse 1, 85748 Garching, Germany}

\author{Lindsay R. House}
\affiliation{Department of Astronomy, The University of Texas at Austin, 2515 Speedway Boulevard, Austin, TX 78712, USA}

\author{Steven Janowiecki}
\affiliation{McDonald Observatory, The University of Texas at Austin, Austin, TX 78712}

\author{Matthew Jarvis}
\affiliation{University of Oxford, Denys Wilkinson Building, Keble Road, Oxford, OX1 3RH, UK}

\author[0000-0002-8434-979X]{Donghui Jeong}
\affiliation{Department of Astronomy \& Astrophysics, The Pennsylvania State University, University Park, PA 16802}
\affiliation{Institute for Gravitation and the Cosmos, The Pennsylvania State University, University Park, PA 16802}

\author{Shardha Jogee}
\affiliation{Department of Astronomy, The University of Texas at Austin, 2515 Speedway Boulevard, Austin, TX 78712, USA}

\author{Ryota Kakuma}
\affiliation{Institute for Cosmic Ray Research, The University of Tokyo, 5-1-5 Kashiwanoha, Kashiwa, Chiba 277-8582, Japan}

\author{Andreas Kelz}
\affiliation{Leibniz-Institut f\"ur Astrophysik Potsdam (AIP), An der Sternwarte 16, 14482 Potsdam, Germany}

\author{W. Kollatschny}
\affiliation{Institut f\"ur Astrophysik, Universit\"at G\"ottingen, Friedrich-Hund Platz 1, D-37077 G\"ottingen, Germany}

\author[0000-0002-0136-2404]{Eiichiro Komatsu}
\affiliation{Max-Planck-Institut f\"ur Astrophysik, Karl-Schwarzschild-Str. 1, 85748 Garching, Germany}
\affiliation{Kavli Institute for the Physics and Mathematics of the Universe (Kavli IPMU, WPI), the University of Tokyo, Kashiwa, Chiba 277-8583, Japan}

\author{Mirko Krumpe}
\affiliation{Leibniz-Institut f\"ur Astrophysik Potsdam (AIP), An der Sternwarte 16, 14482 Potsdam, Germany}

\author{Martin Landriau}
\affiliation{Lawrence Berkeley National Laboratory, 1 Cyclotron Road, Berkeley, CA 94720, USA}

\author{Chenxu Liu}
\affiliation{Department of Astronomy, The University of Texas at Austin, 2515 Speedway Boulevard, Austin, TX 78712, USA}

\author{Maja Lujan Niemeyer}
\affiliation{Max-Planck-Institut f\"ur Astrophysik, Karl-Schwarzschild-Str. 1, 85748 Garching, Germany}

\author{Phillip MacQueen}
\affiliation{McDonald Observatory, The University of Texas at Austin, Austin, TX 78712}

\author{Jennifer Marshall}
\affiliation{Department of Physics and Astronomy, Texas A\&M University, College Station, TX, 77843-4242 USA}
\affiliation{George P.\ and Cynthia Woods Mitchell Institute for Fundamental Physics and Astronomy, Texas A\&M University, College Station, TX, 77843-4242 USA}

\author{Ken Mawatari}
\affiliation{Institute for Cosmic Ray Research, The University of Tokyo, 5-1-5 Kashiwanoha, Kashiwa, Chiba 277-8582, Japan}

\author{Emily M. McLinden}
\affiliation{Department of Astronomy, The University of Texas at Austin, 2515 Speedway Boulevard, Austin, TX 78712, USA}

\author{Shiro Mukae}
\affiliation{Institute for Cosmic Ray Research, The University of Tokyo, 5-1-5 Kashiwanoha, Kashiwa, Chiba 277-8582, Japan}

\author{Gautam Nagaraj}
\affiliation{Department of Astronomy \& Astrophysics, The Pennsylvania
State University, University Park, PA 16802, USA}
\affiliation{Institute for Gravitation and the Cosmos, The Pennsylvania State University, University Park, PA 16802, USA}

\author{Yoshiaki Ono}
\affiliation{Institute for Cosmic Ray Research, The University of Tokyo, 5-1-5 Kashiwanoha, Kashiwa, Chiba 277-8582, Japan}

\author{Masami Ouchi}
\affiliation{National Astronomical Observatory of Japan, 2-21-1 Osawa, Mitaka, Tokyo 181-8588, Japan}
\affiliation{Institute for Cosmic Ray Research, The University of Tokyo, 5-1-5 Kashiwanoha, Kashiwa, Chiba 277-8582, Japan}
\affiliation{Kavli Institute for the Physics and Mathematics of the Universe (Kavli IPMU, WPI), the University of Tokyo, Kashiwa, Chiba 277-8583, Japan}

\author{Casey Papovich}
\affiliation{Department of Physics and Astronomy, Texas A\&M University, College Station, TX, 77843-4242 USA}
\affiliation{George P.\ and Cynthia Woods Mitchell Institute for Fundamental Physics and Astronomy, Texas A\&M University, College Station, TX, 77843-4242 USA}

\author{Nao Sakai}
\affiliation{Institute for Cosmic Ray Research, The University of Tokyo, 5-1-5 Kashiwanoha, Kashiwa, Chiba 277-8582, Japan}

\author{Shun Saito}
\affiliation{Institute for Multi-messenger Astrophysics and Cosmology, Department of Physics, Missouri University of Science and Technology, 1315 N Pine St, Rolla, MO 65409}
\affiliation{Kavli Institute for the Physics and Mathematics of the Universe (Kavli IPMU, WPI), the University of Tokyo, Kashiwa, Chiba 277-8583, Japan}

\author{Donald P. Schneider}
\affiliation{Department of Astronomy \& Astrophysics, The Pennsylvania State University, University Park, PA 16802, USA}
\affiliation{Institute for Gravitation and the Cosmos, The Pennsylvania State University, University Park, PA 16802, USA}

\author{Andreas Schulze}
\affiliation{National Astronomical Observatory of Japan, 2-21-1 Osawa, Mitaka, Tokyo 181-8588, Japan}
\affiliation{Kavli Institute for the Physics and Mathematics of the Universe (Kavli IPMU, WPI), the University of Tokyo, Kashiwa, Chiba 277-8583, Japan}
 
\author{Khavvia Shanmugasundararaj}
\affiliation{Department of Astronomy, The University of Texas at Austin, 2515 Speedway Boulevard, Austin, TX 78712, USA}

\author{Matthew Shetrone}
\affiliation{UC Observatories UC Santa Cruz 1156 High Street Santa Cruz, CA 95064}

\author{Chris Sneden}
\affiliation{Department of Astronomy, The University of Texas at Austin, 2515 Speedway Boulevard, Austin, TX 78712, USA}

\author{Jan Snigula}
\affiliation{Max-Planck Institut f\"ur extraterrestrische Physik, Giessenbachstrasse 1, 85748 Garching, Germany}
\affiliation{University Observatory, Fakult\"at f\"ur Physik, Ludwig-Maximilians University Munich, Scheiner Strasse 1, 81679 Munich, Germany}

\author{Matthias Steinmetz}
\affiliation{Leibniz-Institut f\"ur Astrophysik Potsdam (AIP), An der Sternwarte 16, 14482 Potsdam, Germany}

\author{Benjamin P. Thomas}
\affiliation{Department of Astronomy, The University of Texas at Austin, 2515 Speedway Boulevard, Austin, TX 78712, USA}

\author{Brianna Thomas}
\affiliation{Department of Astronomy, University of Washington, Seattle, 3910 15th Ave NE, Room C319, Seattle WA 98195-0002}

\author{Sarah Tuttle}
\affiliation{Department of Astronomy, University of Washington, Seattle, 3910 15th Ave NE, Room C319, Seattle WA 98195-0002}

\author{Tanya Urrutia}
\affiliation{Leibniz-Institut f\"ur Astrophysik Potsdam (AIP), An der Sternwarte 16, 14482 Potsdam, Germany}

\author{Lutz Wisotzki}
\affiliation{Leibniz-Institut f\"ur Astrophysik Potsdam (AIP), An der Sternwarte 16, 14482 Potsdam, Germany}

\author{Isak Wold}
\affiliation{Astrophysics Division, NASA Goddard Space Flight Center, Greenbelt, MD 20771, USA}

\author{Gregory Zeimann}
\affiliation{Hobby Eberly Telescope, University of Texas, Austin, Austin, TX, 78712}

\author{Yechi Zhang}
\affiliation{Institute for Cosmic Ray Research, The University of Tokyo, 5-1-5 Kashiwanoha, Kashiwa, Chiba 277-8582, Japan}


\begin{abstract}

We describe the survey design, calibration, commissioning, and emission-line detection algorithms for the Hobby-Eberly Telescope Dark Energy Experiment (HETDEX)\null. The goal of HETDEX is to measure the redshifts of over a million Ly$\alpha$ emitting galaxies between $1.88<z<3.52$, in a 540~deg$^2$ area encompassing a co-moving volume of 10.9~Gpc$^3$. No pre-selection of targets is involved; instead the HETDEX measurements are accomplished via a spectroscopic survey using a suite of wide-field integral field units distributed over the focal plane of the telescope. This survey measures the Hubble expansion parameter and angular diameter distance, with a final expected accuracy of better than 1\%. We detail the project's observational strategy, reduction pipeline, source detection, and catalog generation, and present initial results for science verification in the COSMOS, Extended Groth Strip, and GOODS-N fields. We demonstrate that our data reach the required specifications in throughput, astrometric accuracy, flux limit, and object detection, with the end products being a catalog of emission-line sources, their object classifications, and flux-calibrated spectra. 
\end{abstract}



\section{Overview} 
\label{sec:intro}

Data from supernova distances \citep{riess+98, perlmutter+99, riess+21} combined with the results from the Cosmic Microwave Background radiation \citep{Bennett+13, komatsu+14, Planck2020} show that the universe is undergoing accelerated expansion compared to what is expected for a universe with only radiation (photons and massless neutrinos) and matter (baryons, dark matter, and massive neutrinos). The additional acceleration accounts for about 70\% of the current expansion and thus the universe's mass-energy content.  The source of this acceleration is called ``dark energy'', a designation that reflects the gross ignorance of the scientific community.  Yet dark energy has profound implications for the formation of galaxies and the evolution of the universe.  The theoretical community have proposed a number of equally-compelling ideas, such as vacuum energy (a cosmological constant), a new scalar field (evolving dark energy or quintessence), and a modification to gravity \citep[see the review by][]{Turner+08}.  Any one of these proposals requires a fundamental change in our understanding of the laws of physics, and the general consensus is that the resolution of this problem will be nothing less than revolutionary.

Despite significant differences in magnitude from the theoretical prediction, the most promising candidate for dark energy remains the vacuum energy, or Einstein's ``cosmological constant", as estimates from supernovae observations and the cosmic microwave background appear consistent with this interpretation \citep[e.g.,][]{betoule+14, scolnic+18, Planck2020, adachi+20, aiola+20, dutcher+21}. Studies based on measurements of the large-scale clustering of galaxies provide similar conclusions \citep[e.g.,][]{alam2021, des2021a}. There are a number of uncertainties concerning the vacuum energy interpretation, with the most serious being the difference in the Hubble Constant ($H_0$) using different measurement techniques. Specifically, there are lingering differences between local measurements of $H_0$ and the value inferred from the power spectrum of the early universe and the assumption of a $\Lambda$CDM cosmology \citep{freedman+19, riess+21, wong+20, Planck2020, divalentino21, divalentino+21, hikage+19, joudaki+20, heymans+21}. While these differences are small, it the era of precision cosmology, they may signify new physics. To address the problem, a large number of programs have been designed to measure the effect of dark energy, with most focused at late times when dark energy is expected to dominate \citep[see Figure~2 of][]{vargas-magana+19}. Te best way to constrain the history of dark energy is to measure the expansion rate of the universe over as wide a time baseline as possible. A cosmological tracer that can be used for this purpose is Ly$\alpha$ emitting galaxies (LAEs).  LAEs have been detected over a large redshift range, and their redshifted 1215.67\,\AA\ line can easily be detected using low-resolution spectroscopy or narrow-band imaging \citep[e.g.,][]{cowie+98, gronwall+07, ouchi+08, Adams2011}.

\begin{figure*}[ht!]
\includegraphics[width=\textwidth]{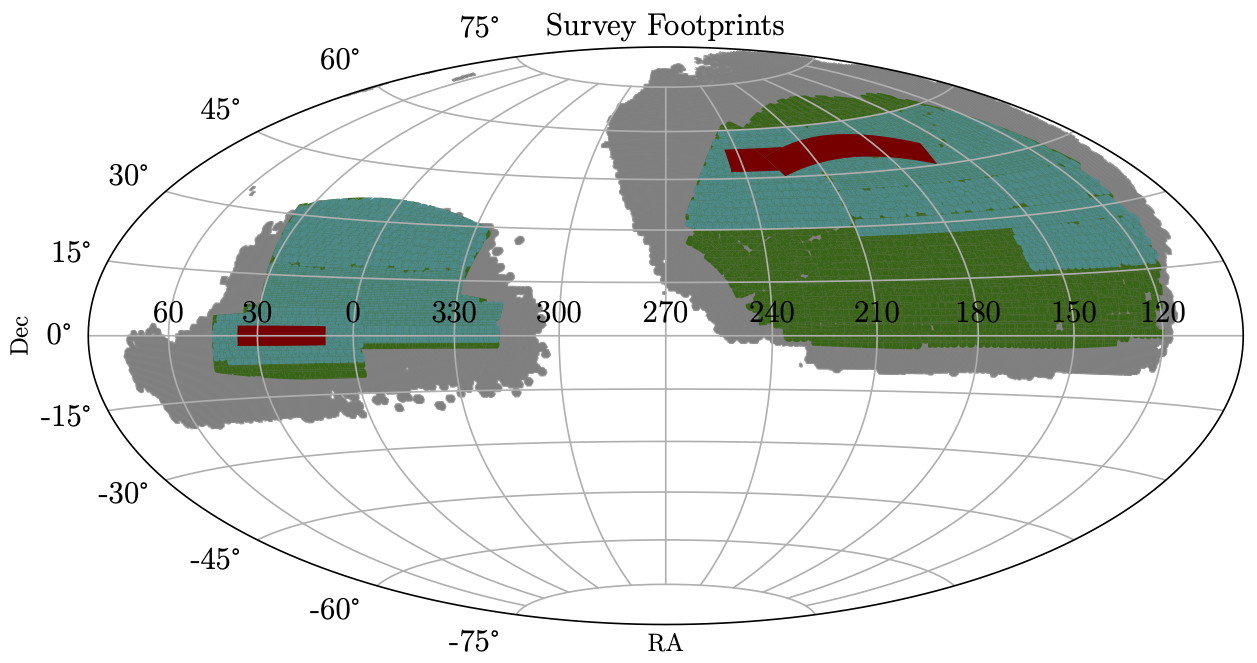}
\caption{The HETDEX field compared to overlapping large-area surveys.  The red regions display the 540~deg$^2$ baseline fields of HETDEX\null. The Green, Cyan and Gray areas show, respectively, the BOSS \citep{BOSS}, eBOSS \citep{eBOSS}, and DESI \citep{DESI} footprints.}
\label{fig:fields}
\end{figure*}

The Hobby-Eberly Telescope Dark Energy Experiment (HETDEX) is a spectroscopic survey aimed at measuring the Hubble parameter, $H(z)$, and the angular diameter distance, $D_A(z)$, in order to determine potential evolution of the dark energy density. Without compelling theoretical guidance or preference for one specific cosmological model compared to any other, we focus HETDEX on being able to provide a direct measure of the dark energy density for a cosmological constant model. Our target specification is a 3$\sigma$ detection of the dark energy density at $z=2.4$ assuming a cosmological constant. This accuracy translates into a measurement precision of 0.9\% on $H(z=2.4)$ and 0.8\% on $D_A(z=2.4)$ between redshifts $1.88 < z < 3.52$.  All instrumental, observational, and calibration requirements follow from this top-level specification.

To put HETDEX in context with other published, on-going and planned experiments, the cleanest comparison is to use the expected uncertainties on the distance estimates. Currently, the uncertainties on $H$ and $D_A$ at different redshifts range from 1.8--3\% \citep[see the summary in][]{des2021a}. The Baryon Oscillation Spectroscopic Survey (BOSS) and its extension, eBOSS, measure $D_A$ with accuracy at 1.8\% at $z=0.5$ \citep{bautista2020, gilmarin2020}, 2.0\% at $z=0.7$ \citep{demattia2021}, and $\sim 3\%$ at $z=2.3$ \citep{dumas2020}, and the Dark Energy Survey (DES) gives an uncertainty of 2.7\% at $z=0.84$ \citep{des2021a}. Various on-going missions, such as the Dark Energy Spectroscopic Instrument \citep[DESI;][]{DESI}, the Prime Focus Spectrograph \citep[PFS;][]{PFS}, and \textit{Euclid} \citep{EUCLID} are expected to achieve a precision of $\sim 0.5\%$ at $z \sim 1$. Eventually, the \textit{Nancy Grace Roman Telescope} \citep{RST} will produce an uncertainty below 0.5\% at $z \sim 1$. The most comparable experiment to HETDEX is DESI, which plans to push out to $z > 2$ with a precision similar to that of HETDEX, i.e., below 1\%. The techniques used by DESI and HETDEX are different and well complement each other. The design of HETDEX is to provide a measure of the distance scales at $z>2$ that is comparable to the most accurate low-redshift measurements. By not relying on theoretical models to design the survey, we aim to provide the most accurate observational comparison to low-redshift programs.

A survey the size of HETDEX will provide significant science beyond just a measure of $H(z)$ and $D_A(z)$, including measurements of additional cosmological parameters, constraints on galactic and AGN evolution, and information on the halo populations of the Milky Way.  For this paper, we do not discuss the full scientific benefit from HETDEX; instead we focus on how the our goals for $H(z)$ and $D_A(z)$ set the requirements for all calibrations and analysis. The redshift range and the accuracies expected on the expansion rates are designed to provide a unique and significant measure of the evolution of dark energy.

HETDEX will use a set of 74 integral-field unit (IFU) fiber arrays, which are currently installed at the focal surface of the 10-m class Hobby-Eberly Telescope (HET) (see \citealt{hill2016}, \citealt{indahl2016}, and Hill et al.\ 2021, submitted). For the data presented in this paper, the IFUs were installed over several years, and the population of the focal plane ranges from 20 to 71 active IFUs.  Each IFU contains a bundle of 448 $1\farcs 5$ diameter fibers as described in \cite{kelz2014}. The IFUs feed two low-resolution Visible Integral-field Replicable Unit Spectrographs (VIRUS) covering the wavelength range between 3500\,\AA\ and 5500\,\AA\null. The full set of 74 IFUs will contain 33,152 fibers, distributed over the central $18\arcmin$ of the telescope's  $22\arcmin$ diameter field of view.  When used with a standard 3-point dither pattern, the instrument produces a focal surface filling factor of about 1 in 4.6.

The methodology of HETDEX is straightforward. At each location in the sky, three 6-minute exposures are taken in a triangular dithering pattern to fill in the gaps between the fibers. Twilight sky frames produce both the wavelength calibration and the fiber-to-fiber normalization, while field stars with known magnitudes and positions primarily from the Sloan Digital Sky Survey \citep[SDSS;][]{york+00, SDSS-7} determine the overall flux calibration and astrometric solution.  Each emission-line source falling within an IFU is found via custom detection software, and the objects' counterparts are identified on complementary images acquired from the Blanco, Mayall, Subaru, and Hubble Space Telescopes.  Using the Bayesian analysis of \cite{leung+17}, \cite{ouchi+20}, and \cite{farrow+21} with updates in Davis et al.\ 2021 (in preparation), each emission-line source is then classified as either a $1.88 < z < 3.52$ Ly$\alpha$ emitter, a $z < 0.5$ \OII\ emitting galaxy, or less often, some other type of object.

\begin{figure*}
\includegraphics[width=\textwidth]{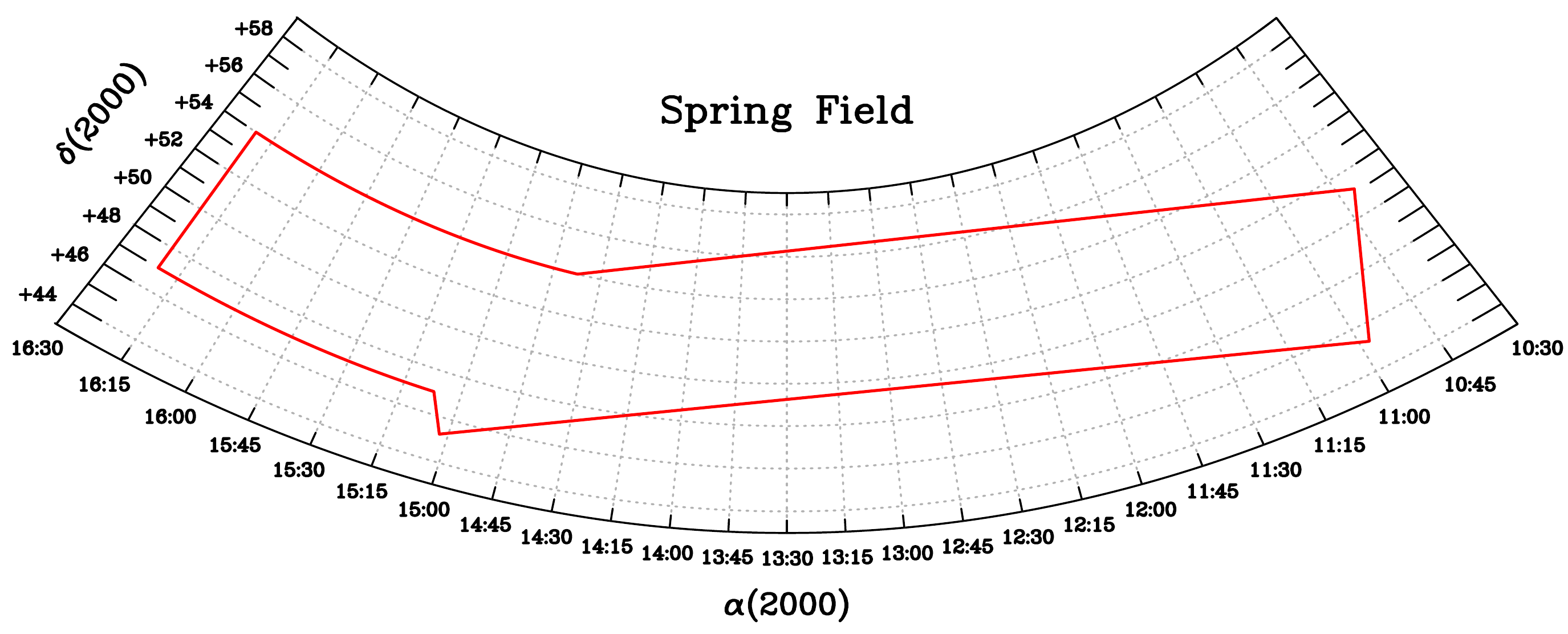}
\includegraphics[width=\textwidth]{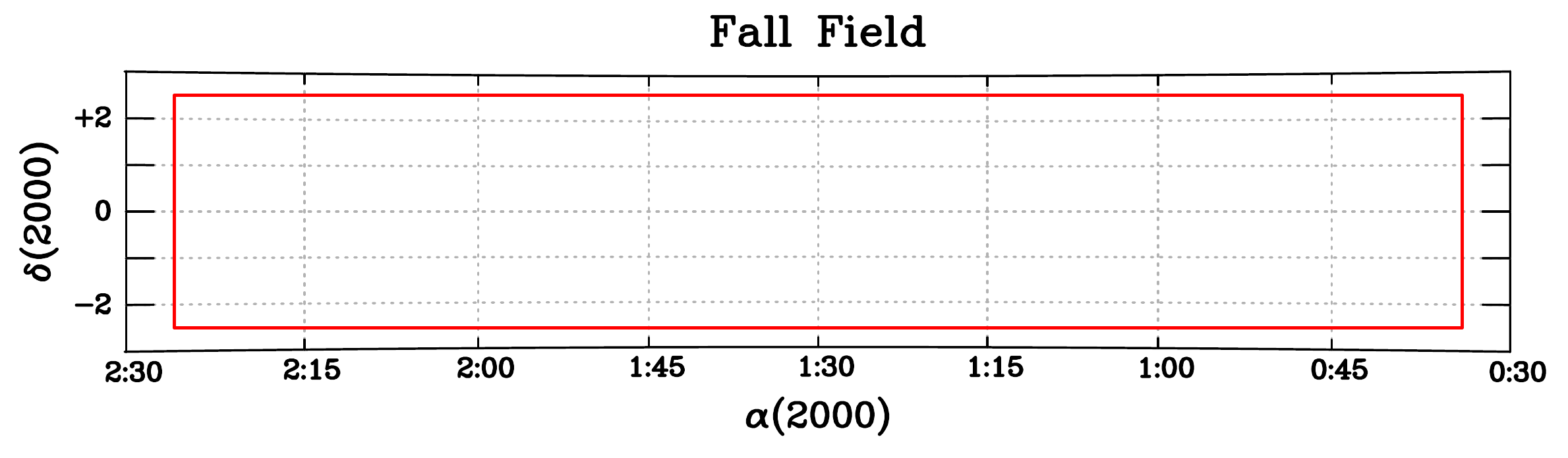}
\caption{The outline of the HETDEX spring and fall fields. The spring field extends over $\sim 390$~deg$^2$ of sky; the fall field, $\sim 150$~deg$^2$.  Our baseline design contains 4000 pointings in the spring field (covering 15\% of the total area) and 2000 pointings in the fall field (covering 21\% of the area).  Thus our nominal survey will have 90~deg$^2$ of spectra when considering the 1/4.6 fill in the focal plane. The lower fill factor in the spring field provides contingency for the survey.}
\label{fig:footprint}
\end{figure*}

The HETDEX survey covers two distinct regions of the sky extending 540~deg$^2$. The high-declination region, referred to as the ``spring'' field, covers 390 deg$^2$, while the equatorial ``fall'' field, extends over 150~deg$^2$.  Figure~\ref{fig:fields} and \ref{fig:footprint} show the field locations. We expect that around 460,000 IFU observations will be taken within these boundaries, with the exact placement of the fields dependent primarily on the observing conditions and the weather pattern. The expectation is that during the course of the survey, the project will detect over one million Ly$\alpha$ emission lines. The large-scale clustering of the LAEs will then provide the cosmological parameters sought by the experiment.

This paper presents the observational design of the HETDEX survey in Section~\ref{sec:setup}, the survey requirements to reach the target cosmological constraints in Section~\ref{sec:requirements}, the instrument layout in Section~\ref{sec:IFUnames}, our dither strategy in Section~\ref{sec:dithers}, the reduction procedures needed to reduce and calibrate the spectra in Section~\ref{sec:reductions}, line and continuum detection algorithms in Section~\ref{sec:detection}, simulations for completeness in Section~\ref{sec:data-modeling}, and our method of line identification in Section~\ref{sec:line-identification}. HETDEX will make its catalogs and data public; at present, all data releases have been internal. The latest internal catalog is called HDR2 (for the second HETDEX Data Release).

\begin{figure*}
\includegraphics[clip, height=0.47\textwidth]{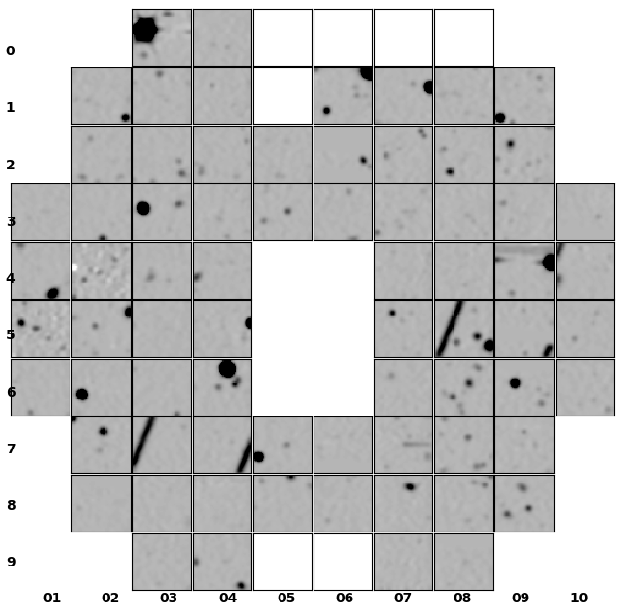}\hspace{20pt}
\includegraphics[height=0.47 \textwidth]{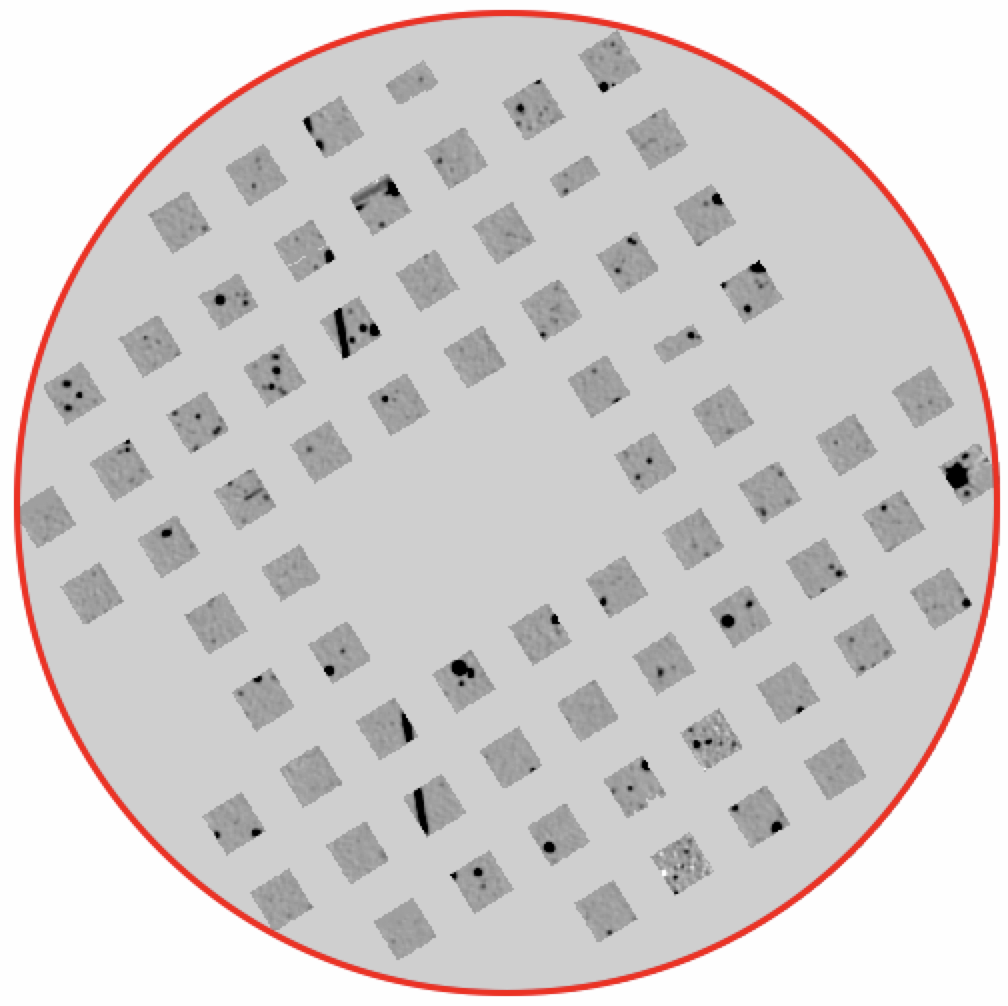}
\caption{Left: a reconstructed image showing the layout of IFUs on the HET's focal plane for data taken on 16 June 2020. This image is formed by collapsing the 4400-5200\,\AA\ spectral region of the VIRUS spectra from a science observation.  The image shows the IFUs as being adjacent to each other in order to save space; each IFU is 51\arcsec\ on a side. The open squares on the outsides of the array denote inactive IFUs; the open rectangle in the middle shows space occupied by other HET instruments.  The streak seen in five of the IFUs (037, 047, 085, 095 and 104, where the first two digits represent horizontal numbering and the last digit gives vertical position) is from a moving object, most likely an asteroid or satellite.  At the time the frame was taken, 71 IFUs were operational.  Right:  The same science observation showing the locations of the IFUs on sky. In this image north is up and east to the left; the IFUs are arrayed on a 100\arcsec\ grid. The red circle has a diameter of 18\arcmin.  Note that the IFUs are orientated in the direction of the parallactic angle, and will therefore change with the azimuth of observation.  The moving object is now obvious.}
\label{fig:focalplane}
\end{figure*}

\section{Observational Setup}
\label{sec:setup}

The science goal of achieving  $<1$\% uncertainty in the cosmological distance measures sets the requirements for the depth and area of the HETDEX experiment.  One needs to survey a large enough volume of space to limit the contribution of sample variance, and observe deep enough so that the number density of galaxies is sufficient to minimize Poisson shot noise. Thus, the project's exposure times and survey area are defined by our current knowledge of the LAE luminosity function in the redshift range $1.88 < z < 3.52$ \citep[e.g.,][]{gronwall+07, ouchi+08, ciardullo+12}, the estimated bias of the LAE population \citep{gawiser+07, guaita+10, kusakabe+18, khostovan19}, the number of IFUs mounted on the telescope, and the expectation for the allocation of observing time. For reasonable telescope usage, we set the base exposure time to 18-minutes, which is split into three 6-minute dithers with an overhead of 2 minutes for read time and dithering.  Under nominal observing conditions, these exposures should detect $\sim 2.5$ LAEs per IFU\null.  
In cases of non-optimal observing conditions, we use real-time estimates of the image quality, sky transparency, sky brightness, mirror illumination, and target availability to adjust the exposure time.

HETDEX produces a significant amount of data. As of August 1, 2020, the date of the HETDEX Data Release 2, 32\% of the planned survey area had been observed. This dataset consists of over 3100 telescope pointings containing 160,000 IFU observations (the number of IFUs per pointing changed with time as we added in more units), 215 million spectra (160,000 IFU observations $\times$ 448 fibers/IFU $\times$ 3 dithers), and 100TB of data storage. Completion of the full survey is scheduled for 2024.

The Hobby-Eberly Telescope is not a fully steerable telescope:  while it can rotate to any azimuth, it can only observe at a fixed elevation of $55^\circ$.  Consequently, observations are only possible when an object passes through the primary mirror's $12^\circ$ diameter field-of-view.  This means that fields can only be observed twice a night, once with one focal surface orientation towards the East transit and another when the orientation is towards the West transit.  Figure~\ref{fig:focalplane} shows the final layout of IFUs on the HET's focal surface; this semi-hexagonal pattern ensures that either of the two tracks provides an optimal tiling on the sky. To maintain the regular tiling \citep{chiang+13}, we populate the hexagon centers (telescope pointings) on the flat rectangle (Figure~\ref{fig:footprint}) and project them onto the survey area (Figure~\ref{fig:fields}) by using an area-preserving map \citep{tegmark96}.  However, because the 74 VIRUS IFUs were installed over a 5-year period, the early HETDEX observations, which were taken with an incomplete IFU array, had unique on-sky footprints.  This necessitated the use of modified field centers in order to optimize sky coverage.  Consequently, data taken during the first three years of the HETDEX project do not have a regular sky tiling. These asymmetries do not impact the measurement of large scale clustering, since the window function of the observations is known very accurately. However, the irregular tiling did produce a significant amount of overlap in the IFU pointings, which allowed us to tune our detection algorithms and improve our understanding of the flux calibration. Only in the spring of 2020 did the number of active IFUs become large enough to use the pre-planned tiling pattern. The window function in the radial (i.e., wavelength) direction is similarly well known, and both directions are considered in the flux limit estimates discussed below.

\section{Science Requirements}
\label{sec:requirements}

\citet{shoji+09}, \citet{chiang+13} and \citet{farrow+21} discuss the forecasts for the cosmology measures. In order to reach our target accuracies of less than 1\% for $H(z)$ and $D_A(z)$, the HETDEX project has a set of science requirements.  These are summarized below.

\vskip8pt
\noindent{\bf Survey size:} The volume and configuration of the survey determine the number of spatial modes that can be used to define the large scale clustering.  Since the HET can only access fields that are $35^\circ$ from the zenith, the footprint of the survey is a compromise between optimizing field observability by having a narrow strip in right ascension, avoiding areas of the sky with significant Galactic extinction, and keeping the shape wide enough so that large scale modes can be adequately sampled.  The fields shown in Figure~\ref{fig:fields} sample 10.9~Gpc$^3$ of space between $1.88 < z < 3.52$. The shortest axis of the spring and fall fields are 7 and 5 degrees, respectively. This width is set in order to allow adequate sampling of the largest clustering scales of interest.

There is a trade between survey depth, survey area, and survey duration. The chosen values for survey volume (10.9~Gpc$^3$) and depth (discussed below) mean that the contributions to the correlation function uncertainties from cosmic variance and shot noise are equal.  For a specified survey duration, going deeper would decrease shot noise while increasing noise from cosmic variance. A more shallow survey would have the inverse effect. The survey duration is a more subjective choice, and is based on our top-level goal of measuring the $z=2.4$ values of $H$ and $D_A$ to an accuracy of 0.9\% and 0.8\%, respectively.  One obvious contingency is to increase the survey duration, and therefore the survey footprint, if needed.

\vskip8pt
\noindent{\bf Number of sources:} The accuracy and precision of the galaxy power spectrum depends on the number of LAEs detected, the false-positive rate, and the fraction of mis-classified sources. We distinguish between false positives and galaxy mis-classifications, since these two errors have different effects on the correlation analysis. For false positives, we are referring to noise or pixel defects that manifest themselves as an apparent emission line. This error should primarily produce white noise, and thus lower the signal-to-noise of our measurement.  (This assumption will be thoroughly tested.) The mis-classification of galaxies is a  larger issue, especially when \OII\ emitters are designated as LAEs.  In this case, the clustering signal of the \OII\ galaxies will leave an imprint on the clustering of the LAEs.

The translation of the galaxy power spectrum into cosmological distance estimates further depends on how the LAEs represent the large-scale clustering of the underlying dark matter distribution.  This factor, which known as the galaxy bias, is a physical quantity that we do not have control over, whereas the other properties depend on our observations and software. For the following analysis, we assume that, for our redshift range of $1.88<z<3.52$, the linear bias parameter is between 1.8 and 2.2 \citep{gawiser+07, guaita+10}; the higher the bias, the more accurate the cosmological measurement, as the power spectrum signal is proportional to bias squared while the Poisson shot noise stays constant. For the volume described above, the HETDEX goal is to identify 1.1 million LAEs with a false positive rate of less than 10\%, and a mis-classification rate (of foreground \OII\ galaxies as LAEs) of less than 2\%. Since the Poisson shot noise is given by the LAE number density within the whole survey volume (instead of the physical number density of LAEs), a finite amount of observing time translates into trade off between depth and area \citep{chiang+13}.  In terms of statistics, this trade is equivalent to the limitations imposed by Poisson shot noise and cosmic variance. 

A sample of 1.1 million sources over 10.9~Gpc$^3$ provides a density of $1.1 \times 10^5$~galaxies~Gpc$^{-3}$. This density optimizes the trade off between cosmic variance and shot noise for measurements of the clustering strength, as discussed in \citet{chiang+13}. The specification of a 10\% false positive rate keeps the white noise effect below the statistical limits of our measurements; the stringent specification of the 2\% contamination limit minimizes the imprint of the \OII\ galaxy clustering signal onto the clustering signal of LAEs \citep[e.g.,][]{pullen+16, leung+17, grasshorn+19, addison+19}. \citet{farrow+21} discuss the implications of having larger or smaller contamination fractions, and redshift dependent contamination fractions. Again, this 2\% limit is designed to keep the systematic uncertainties below the statistical uncertainties.

\vskip8pt
\noindent{\bf Minimum spatial scale:} HETDEX uses the full power spectrum, and does not rely solely on measuring the scale of the baryonic acoustic oscillations \citep{shoji+09}.  This means that in order to reach its cosmological specifications HETDEX must probe down to scales of 5\,$h^{-1}$~Mpc. Calibration down to these scales must include the significant non-linear effects. \cite{jeong2006}, \cite{jeong2009}, and \cite{mccullagh2016} show that we can use scales below 5\,$h^{-1}$~Mpc in our analysis and we utilize these numerical studies. Additionally, on the observational side, the window function has to be accurately determined within the focal plane on similar scales. \S\ref{subsec:flux-calibration} outlines our ability to reach the required specification on the accuracy of the window function on these small scales.

\vskip8pt
\noindent{\bf Wavelength and Redshift accuracy:} Errors on the redshifts can impact the measurement of large-scale clustering by washing out redshift space distortions, which are a powerful tool for cosmological studies. Ly$\alpha$-based redshifts have both a systematic offset and a random scatter about the true systemic redshift of the galaxy. This is due to the physics of radiative transfer, and both the offset and scatter have amplitudes of about $200$~km~s$^{-1}$ \citep[e.g.,][]{shapley+03, shibuya2014, trainor+15, byrohl2019, muzahid+20, gurung2021}. Since we use redshift space distortion in the cosmological studies, any smearing of the redshifts, either from physical or systematic effects, will affect our results.  Thus, to avoid any further increase in the uncertainty of our redshift determinations, we require that the precision of our redshift measurements be less than 180~km~s$^{-1}$. We note that HETDEX redshifts are significantly more accurate than this, with typical uncertainties below 100~km~s$^{-1}$. The precision of these measurements is discussed in \S\ref{subsec:wavelength}.

\vskip8pt
\noindent{\bf Flux limit accuracy:} Any study of the galaxy power spectrum requires measuring the effect that observational selection has on the observed distribution of sources. Thus one needs to know the flux limit versus wavelength at each location in the survey. The required accuracy for these flux limits can be estimated using the expected number of sources per field. The requirement is to not have the flux limit uncertainty be larger than the uncertainty arising from Poissonian errors. For each HETDEX observation we expect about 200 LAEs, which translates into a 7\% variation from the Poissonian noise. We note that this uncertainty limit is for a fully-populated IFU array, and is larger for a focal plane containing fewer IFUs. Thus, we set a limit of 5\% on the flux limit, averaged over the whole focal plane, for one observation; at this level, the uncertainty on the flux limit would only modestly increase the uncertainty on the expected number of sources per field. This 5\% limit translates then directly into the requirements for the measurement of system throughput. The flux limit and throughput accuracy are discussed in \S\ref{subsec:throughput} and \S\ref{subsec:flux-calibration}. We show that our precision on the throughput measurement and therefore the flux limit is, on average, around 2\%. Thus, we are meeting the requirements for flux limit accuracy.

\vskip8pt
\noindent{\bf Dither accuracy:} The emission-line detection algorithm relies heavily upon having an accurate knowledge of an observation's point spread function (PSF), and this model can only be determined from a precise measurement of each frame's dither position.  The positions of the dither offsets are also important for a proper uniform sampling of the sky.   Our specification on the accuracy on a dither position is $0\farcs 4$. This accuracy is discussed in \S\ref{sec:dithers}.

\vskip8pt
\noindent{\bf Astrometric accuracy:} Knowledge of the astrometric accuracy of the HETDEX frames is important for matching emission-line spectra with target lists produced by imaging surveys. Such matchings are used to help discriminate Ly$\alpha$ emission of a high-$z$ source from the \OII\ flux of a foreground galaxy.  The astrometric accuracy of individual sources is discussed in \S\ref{subsec:astrometry},
and is much better than our specification of $0\farcs5$.

\vskip8pt
\noindent{\bf Imaging survey:} HETDEX does not need to select targets beforehand: all the objects within our survey's footprint are observed.  Consequently, we do not need an imaging survey to identify high-$z$ galaxies; instead we require images to assist with line identification.  Over our 2000\,\AA\ spectral range, most HETDEX sources have just a single emission line, produced primarily either by Ly$\alpha$ or \OII\ $\lambda 3727$. (There are other features that may appear in the spectra but these are the dominant lines.) Furthermore, the low resolving power of the VIRUS units does not allow us to split the \OII\ doublet nor resolve the skewed line profile common to Ly$\alpha$ \citep{runnholm+21}. Thus, without additional information, these two lines can be confused, thereby imprinting the power spectrum of foreground \OII\ sources on top of that of the LAEs (and vice versa). 

By measuring the continuum of an object, using either its HETDEX spectrum or its flux in a broadband image, we can estimate an emission-line's equivalent width. This single piece of additional information is extremely useful for helping to discriminate the unresolved \OII\ emission of a foreground galaxy from unresolved Ly$\alpha$ at high-$z$ \citep[e.g.,][]{rhoads+00, gronwall+07, leung+17}. The rest-frame equivalent width distribution for Ly$\alpha$ at $z \gtrsim 2$ is quite different from that of nearby \OII\ galaxies, and the $(1+z)$ boosting that occurs in the observer's frame greatly increases this offset \citep[e.g.,][]{gronwall+07, ciardullo+13}.  As a result, a comparison of emission-line strength to continuum flux density can produce a clean separation of the two lines.  For this calculation, we do not remove the contribution of emission lines to our broadband flux density measurements, since their effect is generally small and our calibration of the LAE/\OII\ galaxy discriminant is empirical.  For the uncertainty in the continuum measurement to not dominate that of the HETDEX emission line, the reference broadband images must reach a limiting magnitude of $g \sim 25$. 

\begin{figure*}
\hspace*{-1cm}\includegraphics[width=350pt, trim={4cm 2cm 0 4cm}, clip]{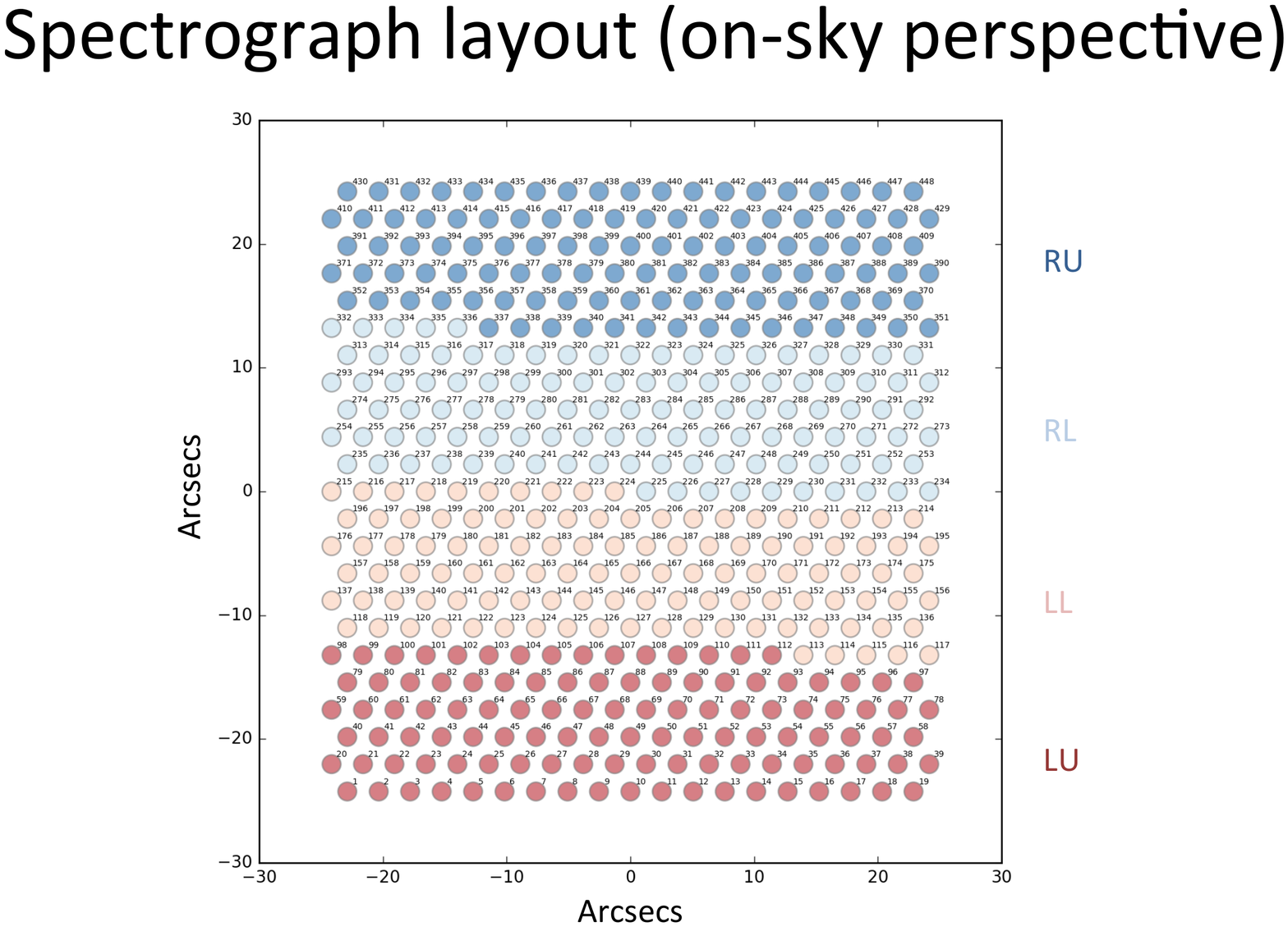}
\hspace*{-4cm}\includegraphics[width=350pt, trim= {0 1cm 0 4cm}, clip]{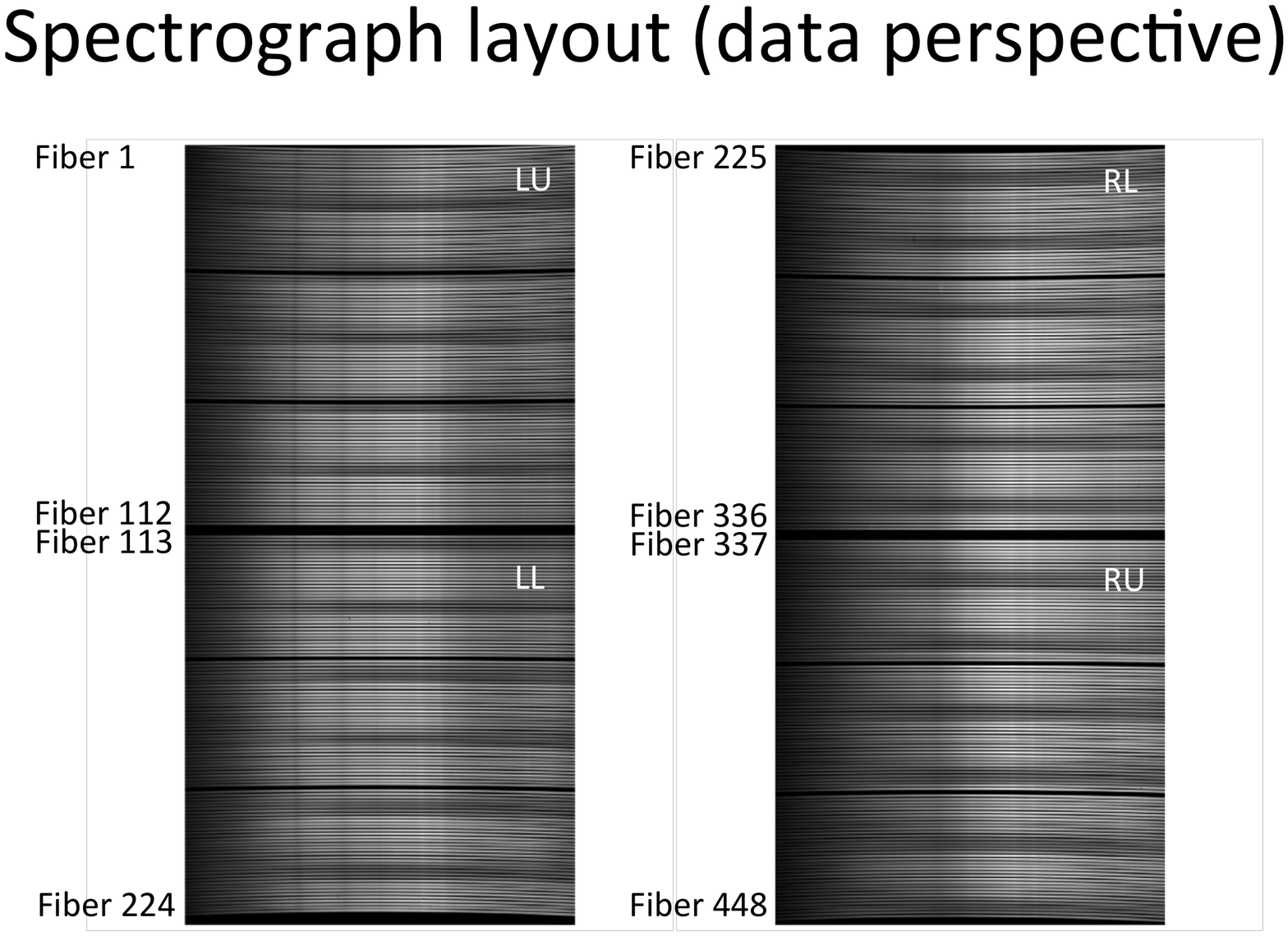}
\caption{The layout of the fibers associated with each IFU\null.  The left-hand panel displays the nominal fiber arrangement in the focal plane and their corresponding detectors and amplifiers. The fibers are numbered 1 to 448. This image represents the standard alignment, though there are a handful of IFUs with different layouts.  The right-hand panel shows the layout of the spectra from the fibers on the detectors. The horizontal is the wavelength direction and the vertical is the fiber direction. Each spectrograph is split into two sides, each with its own detector, ``L'' and ``R'', and each detector (or side) has two amplifiers, designated ``U'' and ``L''.}
\label{fig:IFU-fibers}
\end{figure*}

The imaging surveys and their flux limits are presented in Davis et al.\ (2021 in preparation) and will be described in detail there. The imaging comes from a variety of sources, including our own observations with HyperSuprimeCam (HSC) on the Subaru telescope and the Mosaic II camera on the Mayall 4-m telescope, and archival data from the WFC3 and ACS imagers of the \textit{Hubble Space Telescope} \citep{koekemoer+11}, the Dark Energy Survey  \citep[DES;][]{des2021a}, the Dark Energy Camera Legacy Survey \citep[DECaLS;][]{DECaLS}, the Canada-France-Hawaii Telescope Legacy Survey \citep[CFTHLS;][]{CFHTLS}, and the SDSS \citep{york+00, SDSS-7}.  When multiple imaging surveys are available for the same source, we use the data which gives us the best depth and image quality.  For the vast majority of fields within HETDEX, at least one of the images reaches our specification of $g=25$. The HETDEX spectra themselves, when collapsed over the $g$-band region, typically reach about $g=24$; while this is not quite deep enough for our flux-density requirements, it is useful as a comparison benchmark for the products of the imaging surveys, and allows us to confirm that the imaging data have a common photometric zeropoint.

\vskip8pt
\noindent{\bf Setup time:} In order to reach specifications for observing efficiency, the instrument setup time, which we define as the interval between the end of one 3-dither sequence and the start of the next, must be no more than four minutes. As shown in Hill et al.\ (2021 submitted), HETDEX observations are close to this specification.

\vskip8pt
By meeting the science requirements above in an observing program consisting of about 460,000 IFU observations (baselined originally with over 6000 observations), HETDEX expects to produce a combined distance measure (i.e., a spherically averaged distance, typically called $D_{\rm V}$) of 0.8\% in the $1.88 < z < 3.52$ universe. Given normal weather statistics, instrument stability, and expected observing time allocations, the survey is expected to be complete in 2024. 

\section{IFUs and Detector Designations} 
\label{sec:IFUnames}

The focal surface of the Hobby-Eberly Telescope is capable of housing a grid of 78 IFU fiber arrays, each with 448 fibers covering $51\arcsec \times 51\arcsec$ on the sky. (References for the telescope and instrument can be found in Hill et al.\ 2021, submitted.)  To reach the nominal specification for HETDEX, we require that 74 of these potential IFUs be operational.  The individual IFUs are designated by a 3-digit \texttt{ifuslot} in the focal surface, with the first two digits representing the column number and the last digit defining the row (see Figure~\ref{fig:focalplane}).  Except for the slots near the center of the focal surface, which are reserved for use by other HET instruments, the center of each IFU is separated from that of its nearest neighbor by 100\arcsec, thus leaving 49\arcsec\ of unused space between the fiber bundles.  As a result, in a fully-populated focal plane, the VIRUS IFUs take up $\sim 22\%$ of a circle where the diameter of the circle is defined by the corners of the outermost IFUs\null. (The actual HET focal surface is larger than this, as the telescope's guide cameras extend over a region further out.)

Each IFU in the above array feeds its own spectrograph unit, which is designated by a \texttt{specid}.  Each unit has two spectral channels or sides, each with its own detector, designated ``L" and ``R".  Finally, each detector (or side) has two amplifiers,  designated ``U" and ``L".  Thus, during each exposure, four separate detector images are generated by each VIRUS spectrograph; for example, a file containing the sequence 073RU is from the ``U"-amplifier of the ``R"-side CCD of the spectrograph fed by the IFU in column 7 and row 3. A single VIRUS exposure with 74 IFUs generates $74 \times 4=296$ data files.  Figure~\ref{fig:IFU-fibers} displays the nominal layout of the fibers for each IFU, along with their CCD and amplifier. A few IFUs have slightly different alignments, which we handle on an individual basis.  The fiber numbers run from 1 to 448 and are labeled in the figure.

\section{Dither Sequence} 
\label{sec:dithers}

Since the center of each $1\farcs 5$ diameter fiber is separated from its nearest neighbor by $2\farcs 543$, a dither sequence is needed to fill the gaps between fibers.  These dither offsets are performed by shifting the fiducial position of a star on the focal surface guide camera, thereby forcing the telescope to move accordingly. Each HETDEX field is observed at three different positions, for a total exposure time of 18 minutes.  The commanded dither pattern is triangular: dither 2 is offset from dither 1 by $1\farcs 27$ in $x$ and $0\farcs 73$ in $y$, while dither 3 is offset from dither 2 by $1\farcs 46$ in $y$.  The resultant overlap in dither positions is thus very small, about 2\% in area. This dither sequence provides complete spatial coverage over the $51\arcsec \times 51\arcsec$ region of each fiber array.

\begin{figure}
\includegraphics[width=242pt]{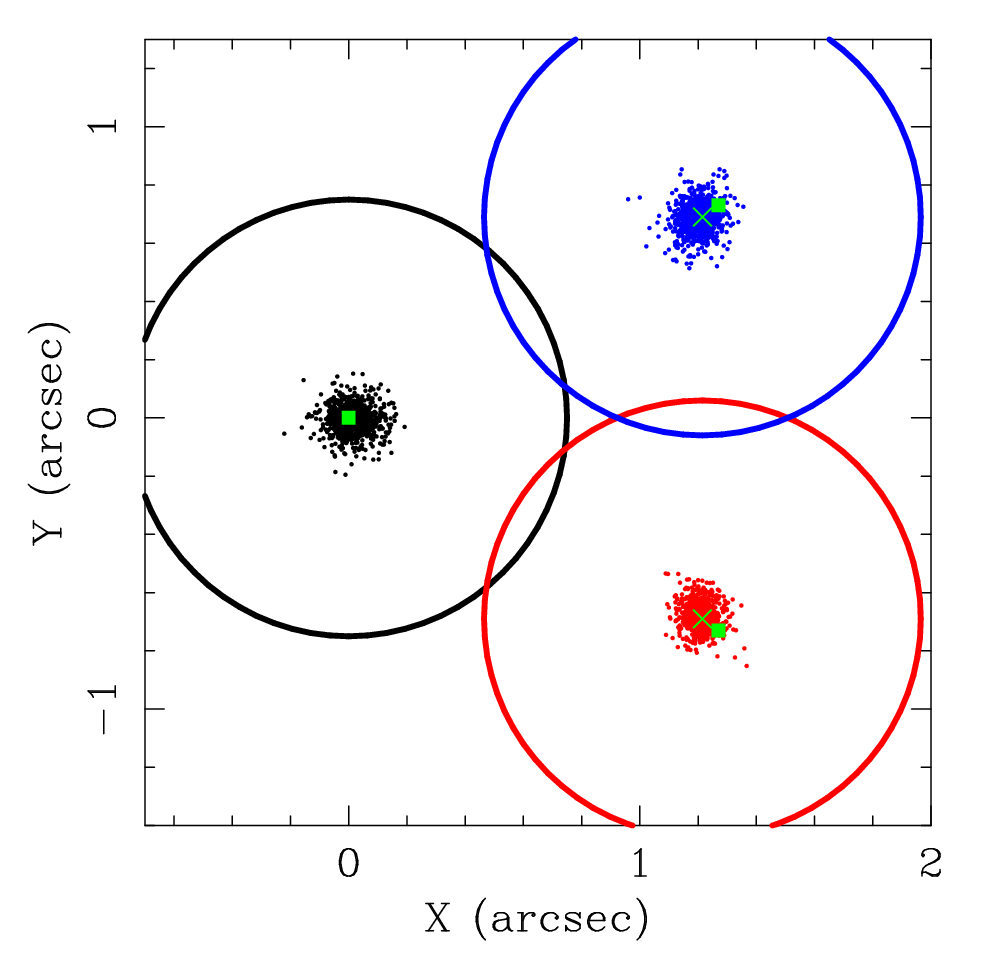}
\caption{The measured locations of the dither offsets derived from all continuum sources in fields observed between 1 Jan and 30 June 2020. The large black, blue, and red circles represent the size of a fiber and the assumed position in the standard 3-point dither. The black, red, and blue points are the measured positions for dithers 1, 2 and 3, respectively, where we have centered the three dithers on their average position. The light green crosses show the centroids of the dithers 2 and 3, while the light green squares illustrate the desired center; the offset between these positions is likely due to a small difference in plate scale.  For HETDEX reductions, it is not important that the commanded and actual dither positions match, but it is critical that we know what the actual dither position is.  After accounting for an overall scale factor, the measured offsets are consistent with the requested centers, and the measured scatter of the dither centers is the same as our measurement accuracy.  This agreement allows us to use the requested dither offsets in our data reduction; the measured dither positions are only used as a check on the data.
}
\label{fig:dithall}
\end{figure}

Figure~\ref{fig:dithall} demonstrates the consistency of the dither pattern by illustrating the dither positions for all HETDEX datasets taken between 1 Jan 2020 and 30 June 2020.  To create the figure, the locations of every continuum source on the VIRUS frames are compared to those derived for the observation's other dithers.  Over the entire sample, the mean difference between the commanded and measured dither position is 6\%, i.e., the mean offset is $0\farcs 08$ smaller than the expected value of $1\farcs 47$.  After accounting for this scale difference, the scatter of the measured positions, $0 \farcs 15$, becomes equal to the measurement uncertainty of the individual dithers, which is also $0 \farcs 15$.  This consistency implies that the true position of the telescope can be measured to a similar accuracy. All subsequent data processing uses the commanded dither offsets while accounting for the small reduction in scale. Thus, dither 2 is assumed to be offset from dither 1 by $1\farcs 215$ in $x$ and $0\farcs 70$ in $y$, and dither 3 is considered to be offset from dither 2 by $1\farcs 40$ in $y$.

The large circles in Figure~\ref{fig:dithall} represent the size of the fibers and the location for the assumed 3-point dither sequence. Ideally, there would be no overlap, but the dither sequence shows a few percent overlap. This small overlap is taken into account in our flux calibrations.

\section{Basic reductions} \label{sec:reductions}

As mentioned above, the HET can only observe objects that are $35^\circ$ from the zenith.  The telescope has a 11.1-m spherically shaped primary mirror, which consists of 91 separate hexagonal segments. To observe an object, the telescope structure rotates to the appropriate azimuth and the tracker at the top end of the telescope follows the object in $x$, $y$, and $z$ as it moves across the primary mirror's $12^\circ$ diameter field-of-view.  To correct for the primary mirror's spherical aberration, this tracker contains a wide-field corrector (WFC) capable of improving image quality over a $22\arcmin$ diameter field-of-view.  Since the WFC's pupil diameter is 10-m,  the VIRUS fibers see different sets of mirror segments at different times during a track.  Moreover for long exposures, the effective aperture of the telescope changes, especially near the ends of the track, since part of the WFC's pupil falls off the projection of the primary. More details about the tracker and WFC can be found in Hill et al.\ 2021 (submitted). 

For HETDEX, we need to take into account differences in mirror illumination for all exposures, and use observations that go to the edge of the track.  This means we need to measure the effective integrated throughput during each exposure. Similarly, there are also slight changes in the telescope plus instrumental resolution in the spectral direction. As discussed in \S\ref{subsec:efit}, we compensate for this by not using a fix line-width in emission-line detection algorithm.  Thus small changes in the spectral shape do not affect our ability to detect emission-lines.

Because of the HET's dynamic nature, the calibration of its data products is tricky.  HETDEX has a complicated data model, as it is important to calibrate all of its $\sim 33,000$ fibers to an accuracy of a few percent and to keep the residuals associated with the sky subtraction to less than 3\%.  The latter systematic is especially important, since the primary targets of HETDEX have line fluxes well below that of the sky. For example, a typical $z \gtrsim 2$ galaxy may have the flux from its Ly$\alpha$ emission-line distributed over $\sim 150$ pixels obtained from $\sim 5$ different fibers associated with the observation's three dithers. Since a detection signal-to-noise of 5 corresponds to a total count level of around 300 ADU spread over about 100 individual detector pixels (3 dithers, and including the spectral and fiber direction on the detector), this means controlling pixel level issues to a precision below one count per detector pixel. For comparison, the background counts on a dark night range from 10 to 30 per detector pixel away from sky lines. Below, we describe all aspects of the spectral reductions including overscan removal, bias subtraction, flat-fielding, sky subtraction, flux calibration, astrometric calibration, and object detections for both continuum and emission line sources.

\subsection{Data Processing Requirements}
\label{subsec:processing}

All HETDEX reductions and data storage use resources at the Texas Advanced Computing Center (TACC\null). The raw data coming off the telescope are transferred from the HET to the TACC as soon as the CCDs are read out.  While quick-look quality checks are performed at the telescope upon readout, the primary data reductions are run on a monthly cadence on the TACC computers. The cpu resources are significant, but given the excellent facilities at TACC, the project can process 4 years of observations in 1 to 2 weeks using $\sim 200$ processors simultaneously. A full processing of the data is generally performed multiple times, as with each iteration our understanding of the behavior of the detectors improve.  Although nearly all aspects of the reduction scripts on TACC are automated, the large number of files associated with HETDEX reductions, the wide variety of instrumental differences across the spectrographs, and the lack of stability in some of the first generation detectors means that a significant amount  of individual attention is needed throughout the process.

\subsection{Reductions to Sky-Subtraction}
\label{subsec:sky-subtraction}

Basic detector characterization uses pixel flats, bias frames, dark frames, twilight sky exposures, and the sky background of science exposures. The twilight frames provide the primary calibration for the fiber profile, spatial trace maps, spectral trace maps, fiber-to-fiber normalizations within a given IFU, and IFU-to-IFU normalizations across the focal surface.  (These frames also allow us to measure the resolving power of the instrument.)  This information is then slightly modified using the data of each individual science frame:  since for a typical observation, about 50-70\% of the fibers are looking at blank sky, these on-sky data allow us to make small adjustments to the calibration. 

The VIRUS spectrographs are mounted on the side of the HET and are very stable, with calibrations that do not change significantly over many months to years (Hill et al.\ 2021, submitted). Given this stability, calibrations frames averaged over a month are superior to those taken daily.  Moreover, the tight packing of fibers on the CCDs means that there is overlap in the fiber profiles at the level of a few to 10\%, and being able to measure this well requires a large number of datasets.  As a result, individual calibrations are not as robust as monthly averages;  this improvement is noticeable on the sky-subtracted frames.

The sequence of reduction steps for the twilight sky frames is overscan subtraction, bias subtraction, pixel flat correction, background light model subtraction, measurement of the fiber traces, measurement of the fiber profiles, derivation of the relative wavelength solution, fiber extraction, and the assignment of the full wavelength solution. The reduction order for night-time science frames is overscan subtraction, bias subtraction, pixel flat correction, background light model subtraction, adjustment of the fiber trace position, fiber extraction, adjustment of the full wavelength solution, sky measurement, sky subtraction in both the 2-D and extracted 1-D spectra, and flux calibration.

\subsection{Bias Frames}
\label{subsec:bias-frames}

The HETDEX project takes 11 bias frames every day, or about 330 bias frames per month.  An interesting feature of these frames is the presence of a low-level, temperature-dependent interference pattern in some of the detectors. Left unaccounted for, this pattern can create false positives for emission line detections. 

Unfortunately, the interference pattern is not stable enough to remove via bias frames taken at different times of the night. Thus, our master bias must preserve the broad-scale features present on the individual biases, but not include the transient small-scale interference patterns. To do this, we smooth over the bias pattern on the individual biases using a $1 \times 9$ pixel boxcar average.  The dimensions are designed to not mix the bias from individual columns (i.e., the $1\times$) and to remove the interference pattern (i.e., the $9\times$).   Then, on every detector readout throughout the night, the interference pattern (if any) can be measured and recorded.  If the pattern is present, the 1 to 2 count noise increase associated with its presence can be incorporated into the analysis.  We also keep track of the bias patterns and trace their behavior with time.  Since these patterns reflect the matched pairing of a controller and a detector, any change to the pattern may indicate a failure in the controller. There are some detector amplifiers that have an unstable interference pattern, and these must be removed from the analysis.

\subsection{Dark Current}
\label{subsec:dark-current}

Our array of $74\times4$ amplifiers have a range of dark currents.  Although we take dark exposures every day, the daily variations are large enough to preclude the use of daytime darks for night-time observations.   Instead, the daily darks provide quality checks, which alert us to issues with individual amplifiers.  We then fold the dark frame information into the background light analysis discussed below.

\subsection{Pixel Flats}
\label{subsec:pixel-flats}

Given that a single weak emission-line may be spread over about 150 pixels on three dithered frames, we require excellent knowledge of the response of each individual pixel.  Thus an important aspect of the data reduction is the application of pixel flats for each detector. The initial step in deriving these flats is to examine the high signal-to-noise flatfield frames acquired in the lab before the CCDs are installed at the telescope.  These detector flats cannot be used on their own, since the devices are temperature cycled before installation, but the lab flats are useful for identifying the locations of hot pixels on the CCDs.

Pixel flats generated with the instrument on the telescope are the most important component for the flat-field correction.  These are produced using de-focussed spectra, as it is important to illuminate those pixels in between the fibers on the detector and still maintain the spectral dispersion. To de-focus the light, we use a set of spacers that increase the separation of the IFU head attachment and the spectrograph. We then take a set of images from a laser-driven light source (LDLS) with an integrating sphere. This setup allows light to easily reach into the fiber gaps, enabling the creation of an accurate set of flatfield frames. From start to finish, this procedure takes a few hours of daylight time per detector. Since this is a time-consuming process, we only perform these observations upon detector installation, and once every 12 months thereafter. We have looked carefully at pixel flats taken over three years of operation, and the flats are remarkably stable, to better than 0.1\% on average of the pixel flat value. When significant changes are found, they are all traceable to instrument maintenance, and we monitor these changes with new flats.

We reduce the flatfield frames by first dividing each row by a smoothing spline, and then repeating the procedure for each column.  The resultant frame residuals are then examined for pixels lying more than $3\,\sigma$ above or below the predictions of the spline; when such pixels are found, they are masked and a new spline is generated.  This process is repeated until convergence is achieved for all the individual pixels.  The result is a highly-accurate pixel flat along with a variance frame, as determined from the individual exposures.  In general, the uncertainties associated with our pixel flats are below the 1\% level.

Many VIRUS CCDs have significant features, including large dust spots, many charge traps, and a ``pox'' contamination where the quantum efficiency of individual pixels can be suppressed by 10-40\%. These issues were quite common on the first generation of detectors and still present in a few of the later units. This ``pox'' tends to be located on the corners of the detectors, and is particularly difficult to deal with. The HETDEX project has been removing badly-affected detectors, and slowly building a set of CCDs that do not have significant pox. While we will never have a completely pox-free dataset, the worst units are being addressed. We have run extensive simulations for object detections, including regions affected by the pox, and, as expected, the pox regions have a higher flux limit.

The pixel flats come from the LDLS, which provides high-count level observations and correspondingly high signal-to-noise data.  However, the night time science data are always in the low-count regime, and flats generated with low-light levels do not exactly match the bright-light LDLS flats, especially for pixels with relative throughput values below 0.6.  We therefore flag the low-throughput pixels and do not use them in subsequent analyses. Including cosmic rays and flagged pixels, we generally exclude about 3\% of the pixels in any exposure.

\subsection{Background Light}
\label{subsec:backgroundlight}

There is extra light in most detectors that needs to be modeled. This background light has contributions from the extended wings of the PSF, scattered light, errors in modeling the wings of the fiber profiles, bias counts not included in the master biases, and controller issues.  Because these effects involve a mixture of additive and multiplicative sources, their individual contributions are not easily modeled, and we do not attempt to measure the relative importance of each component.  Instead, we rely on an empirical approach that combines all the effects. This procedure is not exact, but it allows us to reduce the background subtraction residuals to below a few percent. If one requires background light removed at a level below this, then an additional correction is likely needed.

Our background modeling uses all the twilight sky and night-time science exposures for a given month; this sums to roughly 500 twilight frames and $\sim 1000$ night sky observations. After subtracting the overscan and bias from each frame, we locate regions on the detector that should not contain any light from fibers: these are gaps at the bottom and top edges of the detector and 2 or 3 gaps in the middle. We measure the light in the gaps as a function of wavelength for every exposure. This background light allows us to calculate a full-frame background model via interpolation. Thus, every science exposure and twilight exposure provides information for the background light model.

The background model is not linear with count level. To deal with the non-linearity, we use three different flux levels for the science and twilight frames. For each level, we compute the biweight average \citep{beers90} from the hundreds of frames taken that month.  This gives us a set of three background models and their corresponding flux levels for each detector.

To apply this correction, we find the average count level in the science frame using a specified region of the chip. We then interpolate this count level using the three background models to determine the appropriate background to apply. For a typical science exposure, the background light is about 2 counts per pixel, with a variation that depends on the sky brightness and the brightness of nearby continuum sources.

\begin{figure*}
\includegraphics[width=0.83\textwidth]{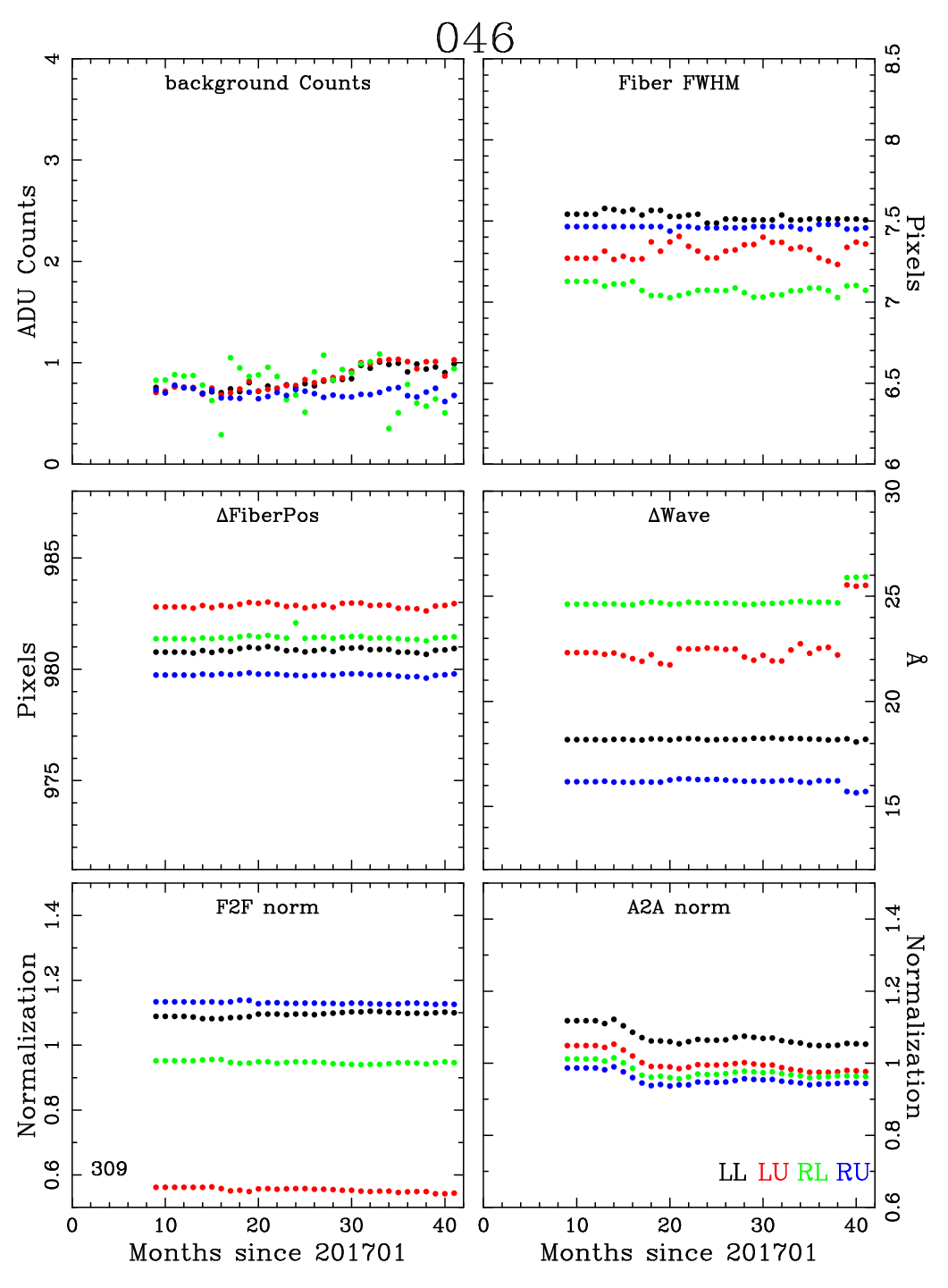}
\caption{Variations in the calibration of one typical IFU over three years of operation.  Plots like these enable us to look for trends in the calibration of all IFUs in the VIRUS array. This set of data is for IFUslot 046, which feeds spectrograph 309. The six checks we apply are: background count level (top-left), FWHM in pixels across the fiber profile (top-right), separation in pixels between the first and last (112th) fiber on an amplifier (middle-left), change in the wavelength solution (in \AA) from the corner fibers of the IFU (middle-right), normalization differences between the ratio of the corner fibers of the IFU (bottom-left), and the overall normalization of the particular amplifier compared to the average of all the amplifiers in all the spectrographs (bottom-right). The colors represent the four amplifiers of the IFU, with LL in black, LU in red, RL in green, and RU in blue.
}
\label{fig:qa_calib}
\end{figure*}

The top-left panel of Figure~\ref{fig:qa_calib} shows the month-to-month variation in the background count level of the models for IFUslot 046. The colors correspond to the four different amplifiers used in this IFU.

Our background models only provide a full-frame correction, and do not account for scattered light that is local.  Photons from bright objects, such as stars, galaxies, and meteors, can scatter off a variety of surfaces, such as the IFU plate, the optical and mechanical elements of the telescope, and the optics of the spectrographs themselves.  We see evidence for all three types of scattering in the data, and there are likely more. We do not apply a local scattered light model in the current data, and it is a subject to be explored in future analyses. 

The loss of light due to scattering is automatically included in the flux calibration. If a bright continuum source creates excessive scattering over the whole detector that detector is removed from consideration. For the emission line detections, we exclude from consideration those sources close to bright objects. Since our focus for the cosmology is high-redshifts objects, we only remove sources that happen to be projected in the same region of sky as a bright source. (For science focused on nearby galaxies, one should rely on a separate catalog for emission lines near bright sources.) These bright sources will create holes in the survey and these are included in the window function. For calibration, the brightest sources are not used since they exceed the acceptance criteria for extraction from 2D to 1D spectra. Thus, while scattered light from local sources is not included in the analysis, it does not create problems for the calibration of the spectra or the detection of objects.

\subsection{Fiber Trace}
\label{subsec:fiber-trace}

In order to perform our spectral extractions, it is crucial that the fiber traces be as accurate as possible, and that they routinely reach a precision of 0.02 pixels in their centering.  To achieve this level of accuracy, we rely on exposures of the twilight sky. As a first pass for the fiber trace, we use the initial measures for the fiber positions, which were determined when the spectrographs were first installed on the telescope.  We then use the twilight sky data to compute the location of the peak flux of each fiber in 10 spectral bins by weighting the summed-pixel positions in the spatial direction by the square of the total flux. This calculation begins at one end of the spectrum, and then increments to the next spectral bin, using the previous bin's centroid as a starting point.  When the full detector field has been measured, we fit the resultant array of fiber centroids versus wavelength with a smooth spline.  Finally, using this initial fiber trace as a starting point, we repeat the procedure, this time weighting the points by the fiber profile itself, instead of the simple square of the flux weighting. This iterative method is extremely accurate:  based on the measured wavelength-to-wavelength variations, which we track over periods of months, the centroids of our fiber traces are known to a precision of $\sim 0.02$~pixels.

Although the shapes of fiber traces are extremely stable, there are nightly shifts in their position:  the trace positions have zeropoint offsets and a breathing mode, where the fiber separations expand and contract with time. To handle these variations, we first compute the biweight average of all twilight trace positions taken over a month, assuming that each fiber trace is the same except for a possible zeropoint offset in its position.  Then, for each individual nighttime exposure, we take these fiducial fiber positions and adjust the traces with a linear model.  (In other words, we apply a zeropoint offset and a scale factor.)  These nightly motions can be up to 0.1 pixel in amplitude, and must be included for a proper sky subtraction.

The middle-left panel of Figure~\ref{fig:qa_calib} shows the evolution with time of the positional difference between the first fiber and the last (112th) fiber on one of the amplifiers for a CCD in IFUslot 046.

\subsection{Fiber Profile}
\label{subsec:fiber-profile}

A key element required to achieve an accurate sky subtraction is knowledge of the fiber profiles. We rely on the twilight images exclusively for this information. The shape of a typical profile is flat-topped with steep wings, and, despite many attempts, we were unable to robustly describe the profiles analytically with a limited number of parameters.  Thus, we  model the fiber profiles in both the spectral (along the fiber) and spatial (across the fiber) directions without having to rely on an analytic form. The most robust approach is to fit each profile non-parametrically using 13 spatial bins across the fiber.  At each spectral location along the fiber, we use our knowledge of the fiber trace to fit the fiber profile with a smooth spline \citep{wahba+90} using a 10 pixel bin size in the wavelength direction.  After performing $\sim 100$ of these fits along the fiber, we interpolate through the individual spectral bins to create a set of $13\times1032$ entries for each fiber. The 1032 is the number of pixels on the detector array in the wavelength direction. Doing this for all fibers provides the full frame fiber profile, with $13\times1032\times112$ entries per amplifier. For further refinement, we use the residuals from these fits and fit a spline to the residuals, adding the results back into the models. We then iterate our fiber profile with the fiber trace, and repeat both measurements. The results typically converge in two iterations, but we perform five iterations for all solutions. This procedure is done on all twilight frames and all amplifiers for a month. Finally, we compute the biweight of the hundreds of twilight frame fiber profiles to provide the month's master fiber profile.

We note that because of the high density of fibers on the CCDs, there is overlap of the fiber profiles between adjacent fibers. We do not explicitly take this overlap into account when deriving the fiber profile. Instead, we rely on the background model to correct for the fiber-to-fiber contamination. The overlap in counts is typically 2-3\%.

The top-right panel of Figure~\ref{fig:qa_calib} shows the variation in fiber full width at half maximum (FWHM) in pixels for IFUslot 046 at a single location on the detector. 

\subsection{Fiber Extraction}
\label{subsec:fiber-extraction}

The fiber extraction algorithm uses a weighted sum across each wavelength bin. Using the fiber trace position as the centroid, and the fiber profile as the weights, we perform an optimal spectral extraction as outlined by \citet{horne86}. For each resolution element, we also track the root-mean-square and the $\chi^2$ deviation from the fiber profile. These values are used directly in the object detection algorithms. Most importantly, the reduced $\chi^2$ measurements allow us to accurately remove cosmic rays.  Any resolution element with $\chi^2 > 10$ is assumed to be affected by a cosmic ray hit, and is flagged for subsequent removal in the analysis. Furthermore, the $\chi^2$ value provides a method for the identification of low-level charge traps. The VIRUS detectors suffer from a number of such traps, which, if not identified as such, can result in false emission-line detections.  The sum of the $\chi^2$ values over all the data provides a clean way to find these defects.

\subsection{Wavelength Solution}
\label{subsec:wavelength}

Traditionally, one would use arc-lamp exposures to determine both the wavelength solution and the spectral resolution of an instrument.  In our case, the integrating sphere has both vignetting and spectral features that create differences between the wavelength solutions based on arc lamps and solutions produced by on-sky calibrations.  As a result, arc-lamp calibrations cannot attain the accuracy required to reach the HETDEX specifications.  We therefore rely on the twilight sky to define the wavelength scale.

The wavelength calibration of HETDEX spectra is performed in two steps. First, we adjust each fiber to a local calibration for its individual amplifier. Second, we fit all the spectra over the full CCD to a standard calibration, thereby placing the full set of fibers on the same system.

We begin by defining the sum of the central three fibers on an amplifier as the fiducial spectrum from the frame.  Each extracted fiber spectrum is compared to this fiducial over 10 equally-spaced wavelength bins between 3500 and 5500\,\AA\ in the twilight sky. To determine the wavelength offset between each fiber and the fiducial, we use the full wavelength range in those 10 bins and find the offset that gives the minimum root mean square between the two spectra. These wavelength bins define a two dimensional plane of wavelength offsets for all 112 fiber spectra on the amplifier. We then interpolate this low-resolution map onto individual 2\,\AA\  wavelength bins.  For the edges of the detectors, we perform an extrapolation based on the slope derived from the last few bins. We create this map for all twilight frames taken over a month and then average the maps (after adjusting for a zeropoint) to create a master relative wavelength solution for the amplifier.  Wavelength differences from this master solution, measured on adjacent corners of the amplifier, then give us a measure of the stability of the solutions. 

The second step uses each of the 112 fibers to build a highly-sampled twilight sky spectrum.  We compare this spectrum to that of the \citet{kurucz+84} KPNO solar spectrum  (convolved down to the resolution of the VIRUS spectrographs) by fitting the wavelength offsets of the 10 bins, and interpolating those solutions onto each 2\,\AA\ pixel.  This results in residuals to the solar atlas that are roughly 0.3\,\AA\ in amplitude. The wavelength solution is similar for all the spectrographs: it is very linear between 3500 and 4800\,\AA\ before experiencing significant curvature at the reddest end of our spectral coverage. There are differences between the solar atlas and the twilight sky, but the solar light dominates enough that the comparison is robust. Thus, each observation for a night is tied to that particular evening's twilight.

We routinely check whether the wavelength solutions obtained from the twilight exposures adequately represent those for the night-time observations by inspecting the sky-subtracted images for unexplained residuals. We observe no systematic residuals over all amplifiers.   In addition, as an end-to-end test, we compare the radial velocities derived for HETDEX field stars with higher-precision velocities obtained from the SEGUE \citep{yanny+09} and LAMOST \citep{luo+15, xiang+17} surveys.  As shown by \citet{hawkins+21}, the resultant 30~km~s$^{-1}$ rms of the VIRUS spectra is within our specifications, as are the small systematic offsets compared to LAMOST ($\sim$13 km/s) and SDSS (a few km/s).

We do not apply a heliocentric correction to the final spectral database, although that information is provided in \cite{hawkins+21}.  However, that correction will be included in the catalog paper of Mentuch Cooper et al.\ (2021 in preparation).

The middle-right panel of Figure~\ref{fig:qa_calib} shows how the absolute value of the wavelength difference between the corners of the 112 fibers for IFUslot 046 changes with time. For this dataset, there is a jump in early 2020 which corresponds to when the spectrograph was taken off the the telescope and then returned. This jump is included in the data reductions. We note that these twlight calibrations are used as starting points for the nightly reductions, and we refine each science frame calibration as small modifications to the twilight calibration. Thus, small changes such as those seen in the wavelength shift are incorporated.

\subsection{Instrumental Resolution}
\label{subsec:instres}

There are two aspects for the instrumental resolution that we consider:  the line-width in the spectral direction and the source profile in the spatial direction. In detector pixels, these should be similar, but we note that the focus in the spectral direction tends to be slightly better than in the fiber direction. We discuss both of these and their implications.

The line-width in the spectral direction is not critical for the HETDEX project, as the parameter is not used by our detection algorithms.  As outlined below in \S\ref{sec:detection},  emission-line detections are initially performed using a spectral sum over 3--4 pixels in wavelength and is then refined using a fit where one of the free parameters is the line-width.  Thus, the instrument's spectral resolution is incorporated into the HETDEX data products via a cataloged line-width.  If a subsequent study requires knowledge of a source's intrinsic line-width, then the instrumental values will need to be considered. Hill et al.\ 2021 (submitted) show that the instrumental resolution of VIRUS is fairly constant as a function of wavelength, with a FWHM around 4.7\,\AA, giving a resolving power that varies from 750 to 950, blue to red. There are some spectrographs that have a larger change in the spectral resolving power over the fibers. We track these units for possible re-focussing in the future.

More important is the VIRUS instrumental resolution in the spatial direction. Since the fiber packing of each IFU is relatively tight with 10-20\% overlap, poor focus in the spatial direction will cause point sources to spread into neighboring fibers, thereby lowering the signal-to-noise. Our quality control involves examining the FWHM in the spatial direction of every fiber in every IFU\null.  As can be seen in the top-right panel of Figure~\ref{fig:qa_calib}, the spatial FWHM for fibers in IFUslot~046 is about 7.5 pixels; the full range of FWHMs for all the IFUs extends from $\sim 6.5$ to $\sim 8.5$ pixels, although some individual fibers can have larger values. We use the cumulative distribution function for the instrumental fiber FWHM for all fibers in an IFU to help determine whether a unit needs upgrading. The fiber profiles are stable with time to within the measurement uncertainty.

\subsection{Fiber-to-Fiber Flats} 
\label{subsec:F2F}

The twilight-sky frames also provide a measure of the fiber-to-fiber relative throughput over the full HET field. Once again, we break this measurement into two steps. First, we measure the fiber-to-fiber variation over an individual amplifier, and then we measure the amplifier to amplifier variation over the full field of all IFUs.

In the first step, we use the extracted spectra of every twilight sky exposure. We initially scale each of an amplifier's 112 fiber spectra using the biweight average of the pixel values between 4300 and 4900\,\AA\null.  We then use the 112 spectra (all of which have slightly different wavelength solutions) to make a single, highly-sampled sky spectrum that has $112 \times 1036$ elements, and take the biweight average over 11 pixels to make an array with approximately $112 \times 1036/11$ elements. The 1036 comes from the wavelength re-sampling, which covers 3470 to 5540\,\AA\ in steps of 2\,\AA\null. This becomes our fiducial scaling profile.  The original individual fiber spectra are then divided by the fiducial to produce a measure of the fiber-to-fiber variation as a function of wavelength. This procedure is then iterated five times to produce a fiber-to-fiber map for the twilight spectrum under consideration.  Finally, we take the biweight of all the twilight fiber-to-fiber maps created over a month to produce our final fiber-to-fiber map. The bottom-left panel of Figure~\ref{fig:qa_calib} shows the variation in the relative fiber normalization between the first fiber on the blue edge of the amplifier and the 112th fiber on the amplifier's red edge for IFUslot 046.


After normalizing all the fibers with a single amplifier, the next step is to calculate the relative normalization for each amplifier in the array of VIRUS units. In a similar fashion to that described above, we generate a master sky spectrum using the results from all the amplifiers. With a fully-populated IFU array, this involves $\sim 33,000$ spectra. The fiber-to-fiber map is divided by the master spectrum of each amplifier to provide the relative amplifier normalization.  As above, we iterate the solution using the new normalization, and compute our final estimate as a function of wavelength. The biweight of the amplifier normalizations found from every twilight exposure over a month then gives the master profile.

The bottom-right panel of Figure~\ref{fig:qa_calib} shows the variation in the relative amplifier normalization compared to the average of the full field for IFUslot 046. Our analysis of these ratios suggests that the amplifier normalizations of our older units decrease with time.  We attribute this change to better performance from the newer IFUs and possible accumulation of dust on the older spectrographs. Any change over time in the overall performance of the observations is included in the calibrations and source simulations (presented in Section 8).

\begin{figure}
\hspace*{-25pt}\includegraphics[width=310pt]{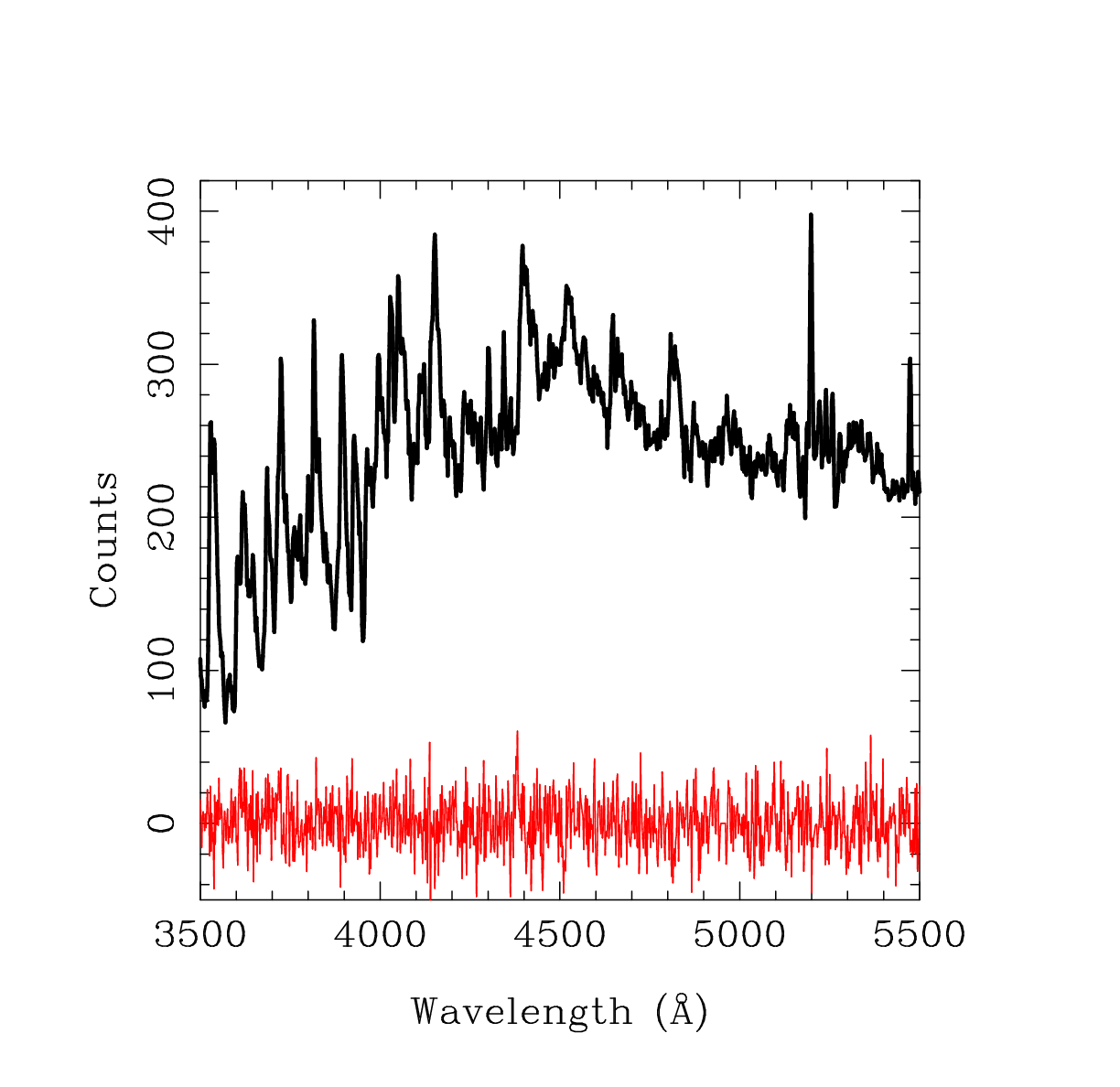}
\caption{Sky model and sky-subtracted residual (in counts per pixel) for fiber 55 on amplifier RU of IFUslot 093 (spectrograph 507).  The sky model, in black, comes from the local-sky estimate, and is based on all 112 fibers of the amplifier. The sky-subtracted residual is in red, and is derived by subtracting the sky model from the extracted fiber data.  The spectra are from the first dither sequence of exposure number 016 on 2021-07-31. The variation in the residual is equal to that expected from the counting statistics of the sky counts.}
\label{fig:skyres}
\end{figure}

\subsection{Sky and Background Subtraction}
\label{subsec:sky-extraction}

One of the more critical reduction steps for the HETDEX program is sky and background light subtraction. At very low signal-to-noise, any residual introduced by sky subtraction will reduce our ability to find faint objects, and create a significant increase in the number of false detections. All previous reduction steps are therefore tuned to make background light subtraction as robust as possible over the entire range of detector properties.

The majority of the background light comes from the night sky, but there are other sources to consider, especially when trying to measure down to levels of less than one count per pixel. Some of these other effects include scattered light, dark current, unaccounted for wings in the telescope point spread function, unaccounted for wings in the fiber profile, and stray charge. 

The fiber extractions produce 1D arrays of flux versus wavelength.  After this step, we estimate a local sky for each individual amplifier by using our knowledge of the fiber fluxes to identify and exclude all discrete continuum sources in the fibers. We start with a background light subtraction in an amplifier, where there are 112 fibers. We compute the biweight average over all fibers and flag any source that is more than three times the biweight scale from the average as a continuum source. For the remaining fibers, we assume that faint continuum sources may still be hiding within the data. This hypothesis is confirmed via deep \textit{HST} imaging; for HETDEX the fields with \textit{HST} overlap, we find that about 10\% of the fibers that made it through the initial continuum cut have faint sources in the \textit{HST} images. Thus, to correct for this systematic, we also remove 10\% of the remaining fibers with the highest count-rates. This approximation will likely introduce a very small residual zeropoint in the background, and studies that are very sensitive to the background should consider applying an additional contribution to the zeropoint. After this last cut, the biweight average is used to determine the sky value at each wavelength.  To ensure the robustness of our sky measurement, we require at least 30 fibers be involved in our background light estimation; for most frames, 80 to 90 fibers are used. 

For the sky-background estimate, we do not attempt to remove spectral regions in fibers that have detected emission lines. This can be a problem for large objects,  where the emission covers most of all of an IFU (at least $13\arcsec \times 50\arcsec$).  Large Ly$\alpha$ blobs, diffuse nebulae, and the halo regions of nearby galaxies can, in theory, fall into this category.  For these objects, the problem can be mitigated by performing a full-field sky subtraction derived from the entire array of IFU spectrographs (see below).  While we have not seen emission lines affect the sky-subtraction of our frames, we realize that this might happen at a low level and we continue to monitor such issues.

In cases where an object extends over a substantial fraction of full IFU array, then there is no easy solution for sky subtraction, and the observation is excluded from the cosmological analysis.  The area of the HETDEX spring field that includes M101 falls into this category.  

Figure~\ref{fig:skyres} shows the sky and sky-subtracted residuals for one fiber from one exposure. The black line is the sky model using the local estimate that is based on the 112 fibers of the amplifier. The residual spectrum in red is the sky model subtracted from the extracted fiber spectrum. The variation in the sky residuals is consistent with that expected from the counting statistics of the sky.

In general, our sky subtraction technique is extremely robust and accurate to the photon noise level. Aside from cases where the data are affected by controller problems, the main failure mode is when a bright star or a large nearby galaxy extends over a substantial fraction of an IFU\null. With our minimum criteria of 30 fibers required for a sky estimate, the inclusion of object flux over so many fibers causes the background to be overestimated. Even then, our detection algorithms (described below) can still pick out emission lines on top of the over-subtracted, negative continuum.

The excellent sky subtraction is partly a product of having a very robust measure of the fiber profile and the fact that we are working in a spectral regime with no strong sky lines.  In fact, the strongest line visible in our sky spectra is from mercury vapor at 5461\,\AA\null.  Sky lines do create correlated residuals in the spectral dimension for some of our amplifiers. These residuals are included in the noise estimation, with the noise being increased in those regions.

In addition to creating a sky estimate for each amplifier based on its own sky fibers, we also produce a full-frame sky subtraction based on the combined data of all the amplifiers.  This global sky estimation, which is especially useful for removing the effects of bright stars and extended galaxies, is produced in a manner very similar to that of the local sky estimate. We take all fibers, in this case up to $\sim 33,000$, flag the fibers containing continuum sources, and generate a background light spectrum using the 90\% of the remaining fibers with the lowest flux (generally 20k to 25k fibers).  Both the full-frame and local sky subtraction are kept and used in subsequent analysis.

\subsection{Astrometry} 
\label{subsec:astrometry}

We require that the astrometric positions derived from the HETDEX spectra match those from imaging catalogs to $0\farcs 5$. This specification is driven by our need to separate the Ly$\alpha$ emission from $1.88 < z < 3.52$  objects from the emission lines of foreground contaminants. Because the VIRUS spectrographs only cover the spectral range from $3500\,\textrm{\AA} \lesssim \lambda \lesssim 5500$\,\AA, virtually all LAEs and most \OII\ galaxies between $0.13 < z < 0.47$ will have only a single emission line.  (At redshifts below $z < 0.13$, H$\beta$ becomes a second confirming line, and at $z < 0.10$, \OIII\ shifts in our instrument's spectral range.)   To discriminate between the cases where only one line is detected, we need to compare the line fluxes obtained from the spectra with continuum flux densities measured from broadband imaging \citep[see][]{leung+17}. Given the large ($1\farcs 5$ diameter) size of the fibers, and the fact that some of our imaging comes from the {\sl Hubble Space Telescope,} the astrometric uncertainty of VIRUS sources will always be larger than that obtainable from imaging.  However, simulations demonstrate that with three dithers, we can centroid any detected source to $\sim 0\farcs 5$. (Doing better would require sub-sampling of the dithers, and even then, the improvement would be marginal at best, to $0\farcs 4$.)  This is sufficient to allow robust matching for most of the detected emission line objects.  

We note that Ly$\alpha$ from high-$z$ sources can be offset from the centroid of their host galaxy \citep{shibuya2014, bond2009, lemaux2021} although this offset is generally much smaller than $0\farcs 5$. This scatter, along with the astrometric error, is taken into account when establishing the probability of a counterpart. Moreover, on those occasions where a HETDEX emission line has more than one possible counterpart, we can employ a Bayesian decision algorithm similar to that used in the HETDEX Pilot survey \citep{Adams2011}.

Obtaining $0 \farcs 5$ precision on VIRUS sources requires that the uncertainty associated with each field's global astrometric solution be negligible compared to the centroiding error of any individual source.  There are multiple parameters that affect our ability to create such a solution.  We consider four aspects that drive our astrometric analysis:  1) our knowledge of the field center, 2) the accuracy of the IFU seat positions, 3) our understanding of the dither offsets, and 4) our ability to reconstruct source positions for astrometric reference stars in the field.  In fact, it is this last issue that limits our ability to determine the other three parameters. Specifically, in order to achieve $0\farcs 5$ astrometric precision for individual sources, we need to define each field center, IFU seat position, and dither offset to better than $0\farcs 2$. We typically reach $0\farcs1$ precision in these measurements.

To obtain an astrometric solution for each field, we first average each fiber's counts between 4400\,\AA\ and 5200\,\AA, and spatially interpolate those counts over the three dithered exposures to construct a pseudo ``image'' of sky.  An example of such an image is shown in Figure~\ref{fig:focalplane}.  We then use the PSF-fitting routines of DAOPHOT \citep{stetson87, stetson90} to measure the IFU positions and fluxes of all the continuum sources, both on the interpolated image and on the individual dither frames.  These lists are fed into the DAOPHOT routines {\tt master} and {\tt match} to determine the astrometric offset of each dither (thereby confirming the dither pattern), and best-fit $(x,y)$ locations of the continuum sources on the HET's focal surface.  Finally, the derived $(x,y)$ positions of stars on the IFUs and the nominal ($x,y$) locations of the IFUs in the focal plane are compared to the objects' equatorial ICRS coordinates in the SDSS DR15 \citep{SDSS-7}, \textit{Gaia} DR2 \citep{Gaia2018}, or PanStarrs \citep{ps1, flewelling+20} catalogs.

\begin{figure}
\hspace*{-0pt}\includegraphics[width=240pt]{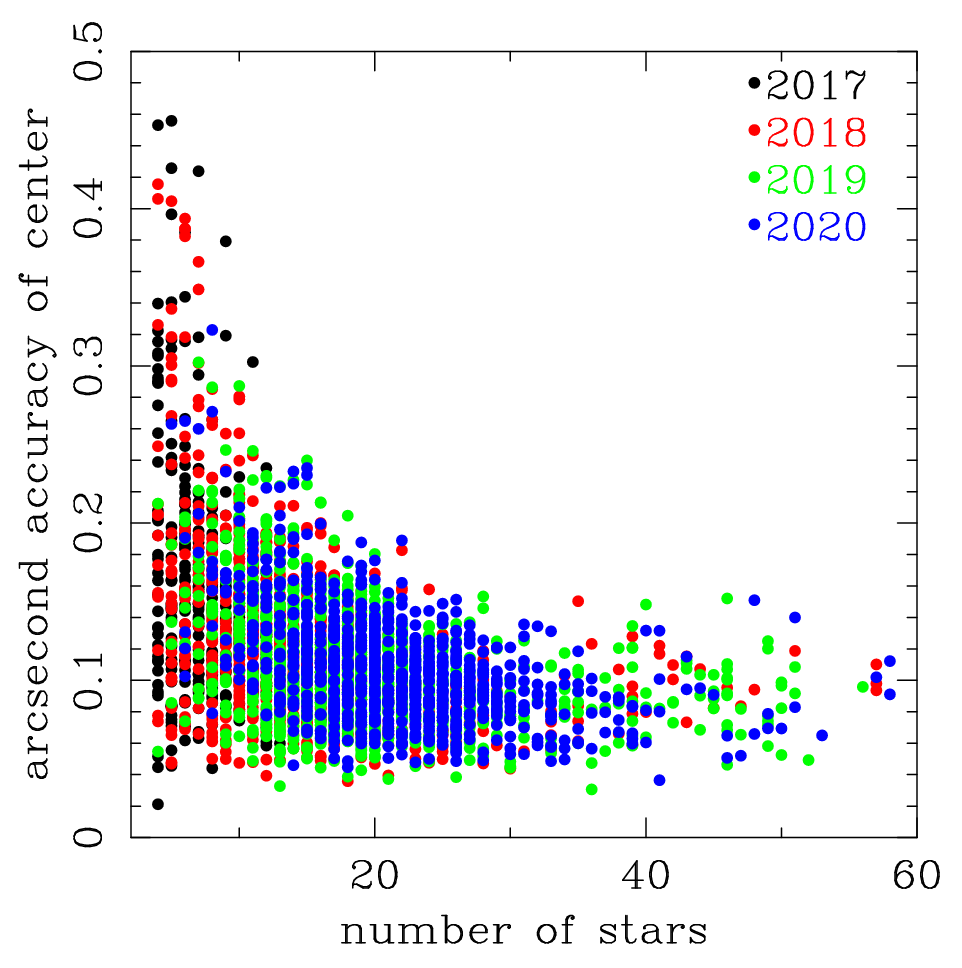}
\caption{Accuracy in arcseconds of the focal plane center against the number of stars used for the astrometric solution. These data are for HETDEX observations for all of HDR2\null. The color corresponds to the year, where black is from 2017, red from 2018, green from 2019, and blue from 2020. The 2017 data had far fewer units installed and therefore fewer stars, which leads to the increased astrometric uncertainty.}
\label{fig:astromacc}
\end{figure}

High galactic latitude observations generally contain around 30 stars bright enough ($g \lesssim 22.5$) to use in an astrometric solution.  With this number of stars, the global solution is usually accurate to better than $0\farcs 2$. Figure~\ref{fig:astromacc} demonstrates this by showing the accuracy of the astrometric solutions for all the HETDEX fields in HDR2 plotted against the number of stars used. The colors represent the year, with black from 2017, red from 2018, green from 2019 and blue from 2020. The older data had far fewer active units and therefore fewer stars to use for the measurement of the focal plane center. For the majority of the observations, we are meeting the required specification for our astrometric accuracy.

As noted above, to perform this measurement, we need to know the relative positions of each IFU in the HET's focal plane.  The approximate IFU seat positions are known from laboratory measurements and the strong expectation is that these relative positions do not move. Since the IFUs and spectrographs have been added steadily over the years, we regularly have the chance to test this assumption by re-measuring all the IFU seat positions using on-sky observations.  Our data confirm the constancy of the IFU seats.

\begin{figure}
\hspace*{-10pt}\includegraphics[width=250pt]{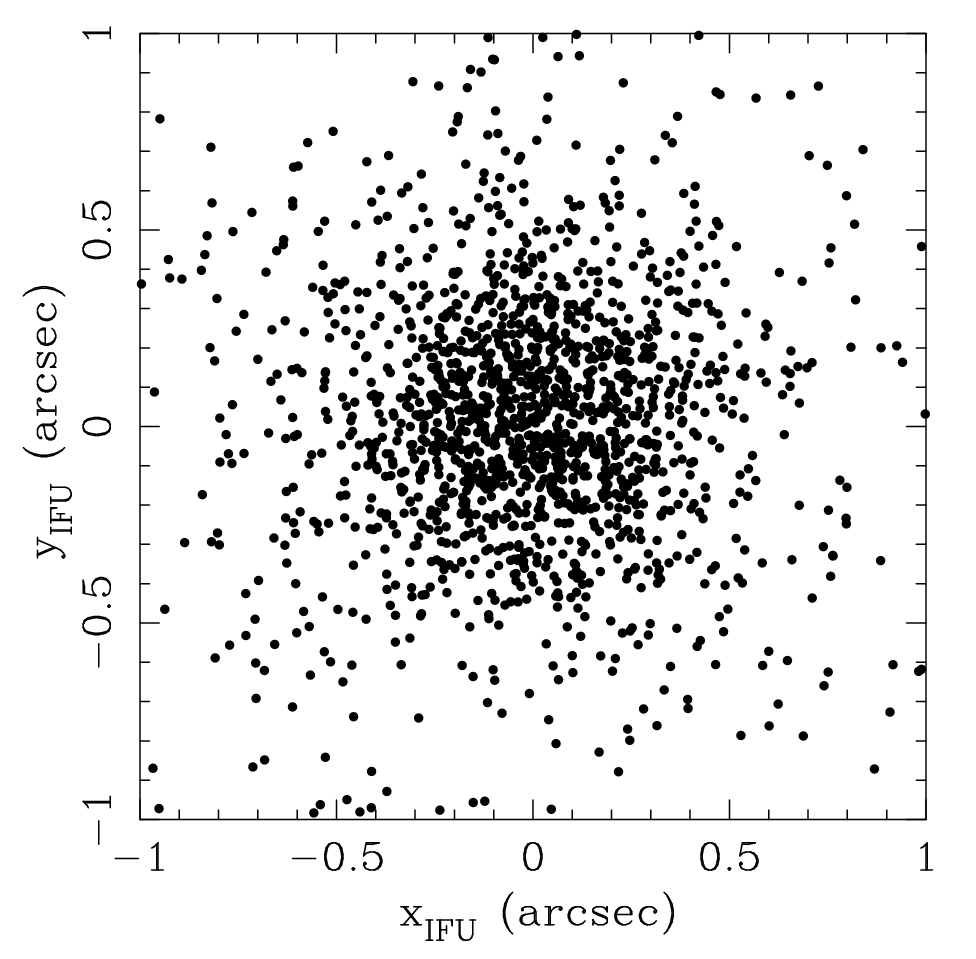}
\caption{Offsets in arcseconds for stars selected for the astrometric measurement of the IFU seat position for IFUslot~046. There are 1800 stars in this figure covering all HDR2 fields, with magnitudes ranging from 14 to 20 in the $g$-band. The rms is $0\farcs 35$.}
\label{fig:046seat}
\end{figure}

Our knowledge of the relative IFU seat positions improved over the years using the ensemble of HETDEX observations.  A typical HETDEX field contains between 0.5 and 1 star per IFU in the magnitude range 
$14 < g < 20$.  By tracking the astrometric offsets between the stars' cataloged positions (primarily from \textit{Gaia}) and the positions found from our nominal astrometric solutions, we built up an offset map for each IFU\null. Figure~\ref{fig:046seat} shows an example data set for IFUslot~046, where each point represents the offset of a HDR2 star in ($x,y$) coordinates.  Since the focal plane center is determined using star positions on all the other IFUs, any change in the relative position of one affects the relative positions of all the others.  Convergence therefore requires about three iterations, and at present the IFU seat positions are known to better than $0\farcs 05$.

This same procedure also allows us to track our ability to measure a stars' positions in the absolute sense. As illustrated in Figure~\ref{fig:046seat}, our overall astrometric solutions have an rms accuracy of $0\farcs 35$.  This number can be broken down into its component terms as follows:
\begin{enumerate}
    \item Fiber positions within an IFU: $0 \farcs 05$
    \item Equatorial position of field center: $0 \farcs 1$
    \item Focal surface position angle: $0 \farcs 2$. This assumes a $0 \fdg 05$ uncertainty in our knowledge of the field rotation, which is determined from the astrometric solutions for each observation at a mean radius of 4\arcmin.
    \item IFU seat positions: $<0\farcs 05$
    \item IFU seat rotations: $<0\farcs 08$. This assumes $0\fdg 2$ rotation error at a radius of 19\arcsec\ within an IFU.
     \item Statistical: $0\farcs 2$ for a 3-pt dither sequence on an individual star.  This depends weakly on S/N of the continuum sources.  
    \item Cataloged star positions: $0\farcs 15$. The main sources of scatter for this term are visual binaries/optical doubles and non-stellar sources.
    \item Source proper motions: $< 0\farcs 05$
\end{enumerate}
Each of terms 1-7 has been measured with the current data, while the mean proper data come from the \textit{Gaia} DR2 catalog \citep{Gaia2018}.  The sum of these terms in quadrature results in a formal accuracy of $0\farcs 35$, in excellent agreement with the results shown in Figure~\ref{fig:046seat}.

\subsection{Image Quality} 
\label{subsec:imagequal}

In order to properly extract the spectrum of a point source, we need to measure the image quality of an observation, quantify how image quality changes with wavelength, and model the effects of differential atmospheric refraction.  All of these quantities are derived from the VIRUS data themselves.

The image quality for each observation comes from modeling the distribution of light expected in the fiber array from a point source. To do this,  we use the bright stars in the field; typically $\sim 30$ objects are suitable for analysis.  For each star, we integrate the spectral flux between 4500 and 5000\,\AA, and fit for the star's equatorial coordinates, total flux, and best-fit PSF FWHM using a Moffat function with a fixed $\beta=3.0$. This beta value is an average best-fit from multiple exposures where we have determined the minimum $\chi^2$ to the stellar profiles; we keep $\beta$ fixed for robustness, although we expect the true value varies by a small amount. The PSF fit itself is performed via a grid search where we integrate the model PSF over the face of the fibers and then calculate the amount of light expected in each fiber. The best fit PSF is the one which minimizes the $\chi^2$ between the model and the data.  We then use the PSFs of all $\sim 30$ stars to determine the mean FWHM and scatter for the frame. Once the FWHM is known, we re-measure all the positions and fluxes for the stars, and use the mean FWHM in all subsequent analyses.

We also use the brightest stars to measure the behavior of the FWHM with wavelength.  By repeating the above procedure within ten 200\,\AA\ wide spectral bins from 3500\,\AA\ to 5500\,\AA, and comparing the FWHM values, we recover the standard Kolmogorov turbulent atmosphere result, FWHM $\propto \lambda^{-0.2}$ \citep{roddier81}. This relation is included in all subsequent analyses.

The Hobby-Eberly Telescope does not have a atmospheric dispersion corrector, and we must account for the shift of an object's position as a function of wavelength. The same bright stars used for the PSF analysis provide a very accurate measure of this differential atmospheric refraction (DAR\null). Our data demonstrate that from 3500\,\AA\ to 5500\,\AA, a source position moves by $0\farcs 95$, a value in agreement with the calculations of \citet{filippenko82}.  Given the large fibers, the large separation of the fibers, and the amplitude of this offset, it is essential to consider this systematic in any spectral extraction. In fact, in order to accurately determine an object's flux, this criterion demands that a source's position be known to a precision of at least $0\farcs 2$.

\begin{figure*}
\includegraphics[width=510pt]{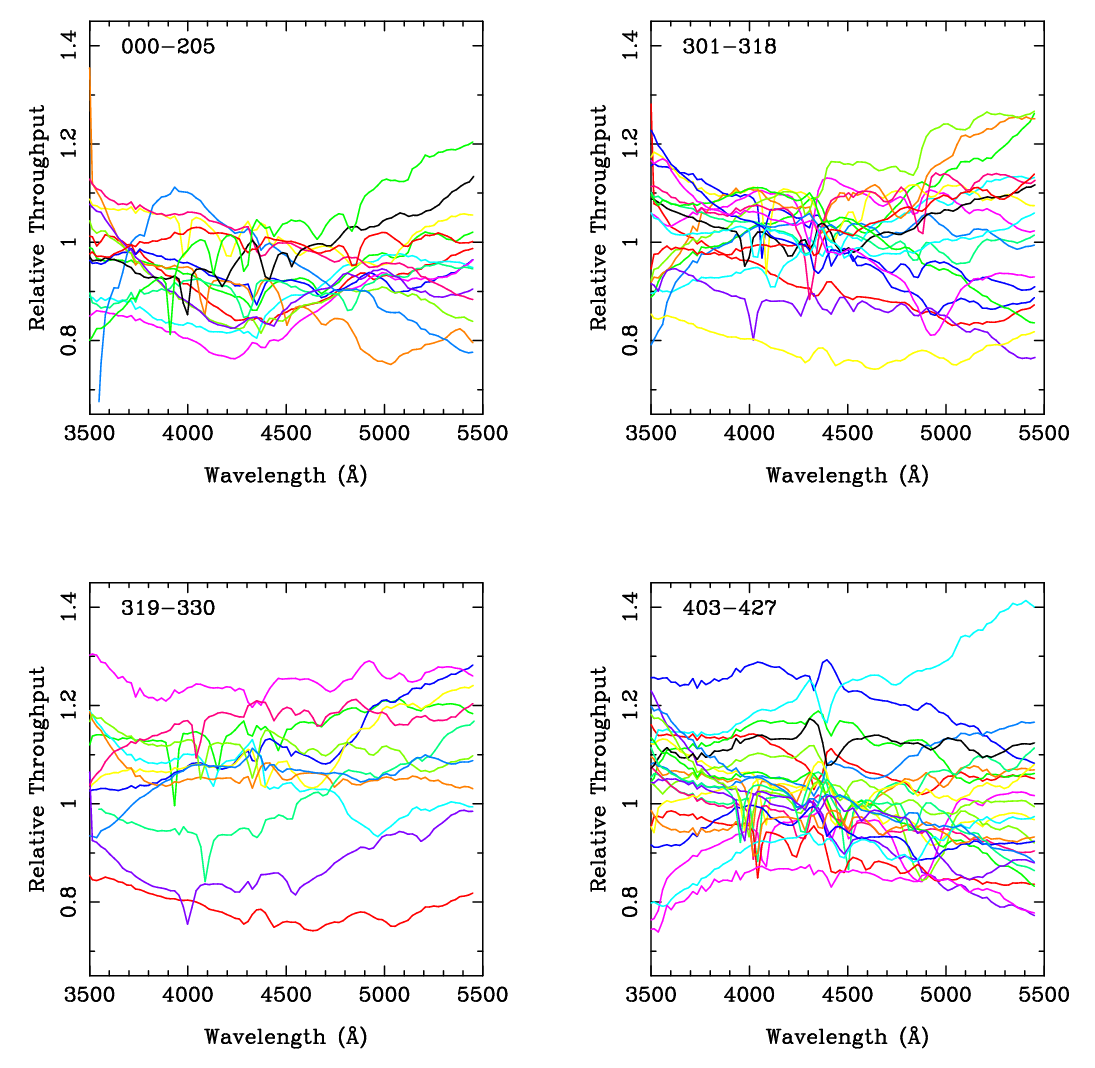}
\caption{Relative throughput measures of amplifier LL versus wavelength for each spectrograph. Each panel presents data from a different set of spectrographs, where the spectrograph identification is given in the label near the top. The spectrographs with higher numbers are those installed at a later time and generally have higher relative throughput by about $10\%$. Each curve is based on the average of 112 fiber-to-fiber profiles within the amplifier. We use these curves to determine the relative throughput calibration over all IFUs.}
\label{fig:amp2amp}
\end{figure*}

\subsection{Throughput} 
\label{subsec:throughput}

In order to measure the power spectrum of $1.88 < z < 3.52$ LAEs over 540~deg$^2$ of sky, the flux limits of each observation must be known to high precision. Our target goal for this limit is 5\%, averaged over the full field and all wavelengths. For each pointing, we must make sure that the uncertainty in throughput does not dominate over the error associated with the field's counting statistics. Given that the HETDEX survey consists of over 6000 individual pointings taken over many years, this is a formidable task. 

Our definition of throughput is based on a 50~m$^2$ clear aperture and a set of 3 dithers, each with a 360s exposure taken in clear weather; for exposures through the best conditions, this value is $\sim 0.16$ at 5140\,\AA\null. Since our goal is to collect a dataset that is as homogeneous as possible with respect to this throughput, we adjust the HETDEX exposure times to compensate for transparency variations, image quality changes, primary mirror illumination fraction, and sky background. In practice, we cannot obtain a constant flux limit over the full survey, due to limitations imposed by field availability and rapid changes in the observing conditions.  Thus, for purposes of our experiment, we define ``throughput'' by treating each exposure as if it were 360s long through a 50~m$^2$ clear aperture. We also refer to the throughput curve as the response function, as it allows for the translation of counts to flux units.

The 5\% value on our flux limit precision is derived from the expected number of sources and from simulations. Our goal is to keep the error associated with the flux limit below that caused by Poisson noise of an observation, and is based on the expectation of detecting 150 to 200 LAEs per field.  A flux precision of 5\% is good enough to produce only a modest increase in the overall error. This is confirmed by our simulations which show that a 5\% uncertainty on the flux limits has a negligible effect on our ability to measure the cosmological distance indicators. 

We approach the flux and throughput calibration in two ways. First, for the scales within the focal plane, we measure the variation of a star taken in multiple frames, after removing an overall relative normalization. Second, for the uncertainty of the absolute normalization of a given field, we use the variation of all stars within that field. The ultimate test of the procedures then come from comparisons with externally calibrated spectra.

For scales smaller than $23\,h^{-1}$~Mpc, we rely on different tests for the spatial and spectral variations. Since this scale is within the focal plane, we first remove the overall normalization of the measurements and look at the spectral variations of a star taken in multiple exposure sets.  This is done by observing the same star on many IFUs, and measuring the rms scatter in the count rate between observations, while taking into account atmospheric diffraction in the spectral extraction process.  This test typically produces rms scatters of 2\% in 2\,\AA\ bins, and is consistent over many different stars and over time.  This is well within our specification.  Thus, the limiting accuracy for our spectral calibration is determining the absolute flux calibration and the integrated flux limit. 

Our calibration of spatial scales within an individual observation is performed using flux limit cubes created using the measured noise properties of the data. These cubes are calibrated using input source simulations specific for the observation, and are especially important for addressing issues such as edge effects, transitions over amplifiers, fibers with little or no transmission, and the effect of bright stars and nearby bright galaxies on our measurements. The cubes are created with scales of 2\arcsec\ per pixel, but this size can be adjusted for specific cases, as the only limitation is computer time.  These cubes allow us to measure  spatial variations within a given observation on scales similar to those for the spectral direction. The limiting accuracy for both spectral and spatial calibration is determined by the overall absolute flux limit. For the remainder of this section, we focus on obtaining the broadband absolute flux calibration of our spectroscopic data, integrated over a full frame and over all wavelengths.

\subsection{System Flux Calibration}
\label{subsec:flux-calibration}

The Hobby-Eberly Telescope's design complicates the derivation of an absolute flux calibration for its spectra.  Since the HET is set at a fixed zenith angle of $35^\circ$ with a range of $\pm6^\circ$, differential atmospheric attenuation varies by at most 20\%, as all observations are taken between 1.14 and 1.32 airmasses.  (In fact, the variation is generally much smaller than this, since most HETDEX observations are taken close to center of the track.) However, during an observation, an object's movement across the telescope's 11.1\,m primary mirror may cause the effective aperture of the system to change, as different segments of the primary (with potentially different reflectivity coefficients) come into view. As a result, while observations of spectrophotometric standard stars might be used to infer the nominal wavelength dependence of the throughput curve, the applicability of these data to any single HETDEX observation is uncertain.  Fortunately, the HETDEX spring and equatorial survey regions have been well-characterized by the Sloan Digital Sky Survey \citep{SDSS-7}. Since the flux calibration of bright SDSS field stars ($g\le 18.5$\,mag) is good to $<$1\% in \griz\ and $<$2\% in $u$ \citep{padmanabhan+08}, we can treat these stars as in-situ standards. By integrating their stellar counts in 3\arcsec\ radius apertures, modeling the effective aperture of the telescope, and taking advantage of the stars' SDSS photometry (along with photometry from other wide-field surveys), we can obtain an accurate characterization of the system throughput for each individual HETDEX pointing.

\subsubsection{Methods of Flux Calibration}
\label{subsubsec:flux-methods}

There are a number of ways to exploit the availability of SDSS photometry for HETDEX flux calibration. Below we describe three methods for performing this calibration, and the results of various tests of their robustness.  We note that, since each method shares a dependence on the SDSS $g$-band magnitude, the three calibrations are not completely independent.  Nevertheless, the techniques are sufficiently different so that each procedure provides a crosscheck on the other two methods, and the combination of all three yields a robust measure of our throughput curves.


Each method relies on generating the expected spectral energy distribution (SED) of field stars.  First we choose stars from a catalog formed from a combination of \textit{Gaia} \citep{Gaia2018}, Pan-STARRS \citep{ps1}, SDSS \citep{SDSS-7}, and USNO \citep{USNO} astrometry.  We extract the spectrum of each star using a $3\arcsec$ radius aperture (see \S\ref{subsec:extraction}) and generate a throughput curve using the techniques described below.  By combining the throughput curves derived from all the stars present in a field, we obtain a first-iteration mean response curve for the observation, along with a measure of the star-to-star variance.  

After deriving these mean response curves for all the observations, we normalize the curves to a common throughput and combine the data to produce a final global response function versus wavelength. We then compare this global response function to the curve derived for each individual observation and find the best-fit normalization and slope that transforms the individual curves into the global curve.  This final fit is what we use in our analysis. Thus, our model for the system response is based on a two-parameter model (normalization and slope) fitted on top of a global response curve. Below we provide the details for the three possible methods for this flux calibration. 

\begin{figure}
\includegraphics[width=240pt]{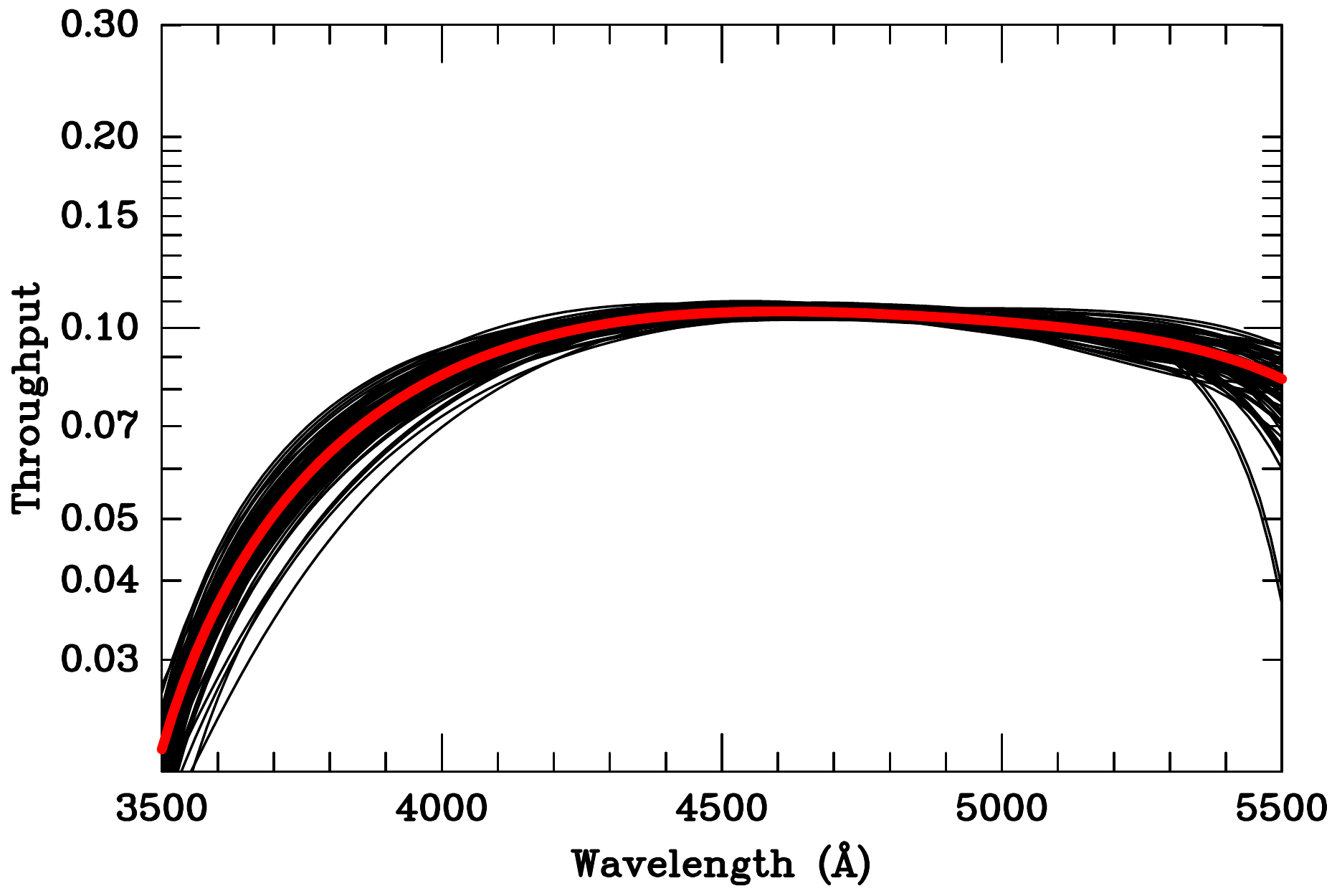}
\caption{Individual response curves derived from 89 observations of spectrophotometric standard stars between 7 Jan 2018 and 2 Nov 2018.  The mean response curve is shown in red.  All the curves have been normalized to a common $g$-band filter response.}
\label{fig:stdstar_throughput}
\end{figure}

\vspace{\baselineskip}

{\bf System throughput from $g$-band normalization.}

On fully-steerable telescopes, the response curve for most spectrographs is derived through the observation of spectrophotometric standard stars. In the case of the HET, where objects track onto and off of the different mirror segments, this simple procedure is inadequate.  However, a slight modification of this standard procedure can produce a first-order estimate of the system response.

Observations of faint ($m_V \gtrsim 15$) spectrophotometric standard stars \citep{massey+88, oke90, allende-prieto+09, bohlin+14, narayan+19}, if performed using the same three-position dither pattern employed by HETDEX, can be used to derive the nominal wavelength dependence of the instrument's throughput curve (see Figure~\ref{fig:stdstar_throughput}).  Measurements of SDSS field stars can then be used to re-normalize this flux calibration curve to each individual pointing.  More specifically: since the $g$ filter's bandpass falls entirely within the spectral range of the VIRUS spectrographs, one can compensate for a change in the effective telescope aperture or a grey shift in sky transparency by simply multiplying the $g$-band filter's transmission curve \citep{doi+10} by the fluxed-calibrated spectra of SDSS field stars, computing the stars' expected $g$-band magnitudes, and comparing these magnitudes to those listed in the SDSS catalog \citep{SDSS-7}.  Any offset can then be applied to the nominal throughput curve derived from the spectrophotometric standards.

The advantage of this approach is that the method can be applied to field stars as faint as $g \sim 23$, ensuring that more than enough objects are always available, even at high galactic latitude.  The disadvantage is that the calibration assumes that any change in the system response is achromatic.  Such an assumption may be problematic, as the aerosol content of the air may change during the night. (Ozone absorption may also vary significantly on a timescales of hours, but its effect largely lies outside the wavelength range of the VIRUS spectrographs. For a discussion of the various atmospheric components, see \citealp[][and references therein]{hayes-latham75}.)  Moreover, since the effect of aerosols is most important in the blue and ultraviolet, observations taken through clouds may introduce a redshift-dependent systematic error into the HETDEX analysis. 

\vspace{\baselineskip}

{\bf System response from stellar absorption lines.}

A second way of obtaining a full-spectrum estimate of system response is by modeling the absorption lines of individual SDSS stars.  Objects with $g \lesssim 17$ have sufficient counts in their VIRUS spectra to allow a detailed comparison to libraries of stellar spectra, such as those of ATLAS9 \citep{kurucz70, kurucz79, castelli+04} and MILES \citep{cenarro+07, falcon-barroso+11}. These fits, along with the stars' precision SDSS magnitudes, can then yield an estimate of the objects' effective temperatures, metallicities, reddenings, and, most importantly, their expected flux distributions above the Earth's atmosphere.  A comparison to the observed spectra then yields the system response at each wavelength. 

The weaknesses of relying on individual stellar absorption lines include the relatively low resolution of the VIRUS spectra, which results in the blending of several key spectral features, and the low sky density of bright stars at high galactic latitude.  Specifically, for most HETDEX observations, there is only $\sim 0.01$ $g < 17$ star per IFU\null.  Thus, even with the fully complement of 74 spectrographs, there is typically only $\sim 1$ suitably bright comparison star per telescope pointing, and many fields contain no such stars.  

Another major issue associated with using a single calibration star is the systematic error associated with its spectral extraction.  The large fibers, the large fiber spacing, the mean $1\farcs 8$ FWHM seeing of the survey, and need to compensate for atmospheric dispersion, all contribute to the overall shape of the extracted spectrum. As a result, even a small ($0\farcs 05$) error in the assumed stellar position can produce a many percent shift in the blue-to-red normalization of an extracted spectrum. This systematic is further complicated by the fact that, even if the astrometric position of the star were known perfectly, we do not know the relative position of the individual fibers to requisite accuracy. Thus, the uncertainty in the inferred response curve will be dominated by the systematics caused by our imprecise knowledge of the stellar centroid and the effects of atmospheric dispersion, rather than simple counting statistics. As a result, throughput curves that rely on measurements of a single star are problematic, and cannot produce a robust calibration.

\vspace{\baselineskip}

\begin{figure*}
\includegraphics[width=260pt]{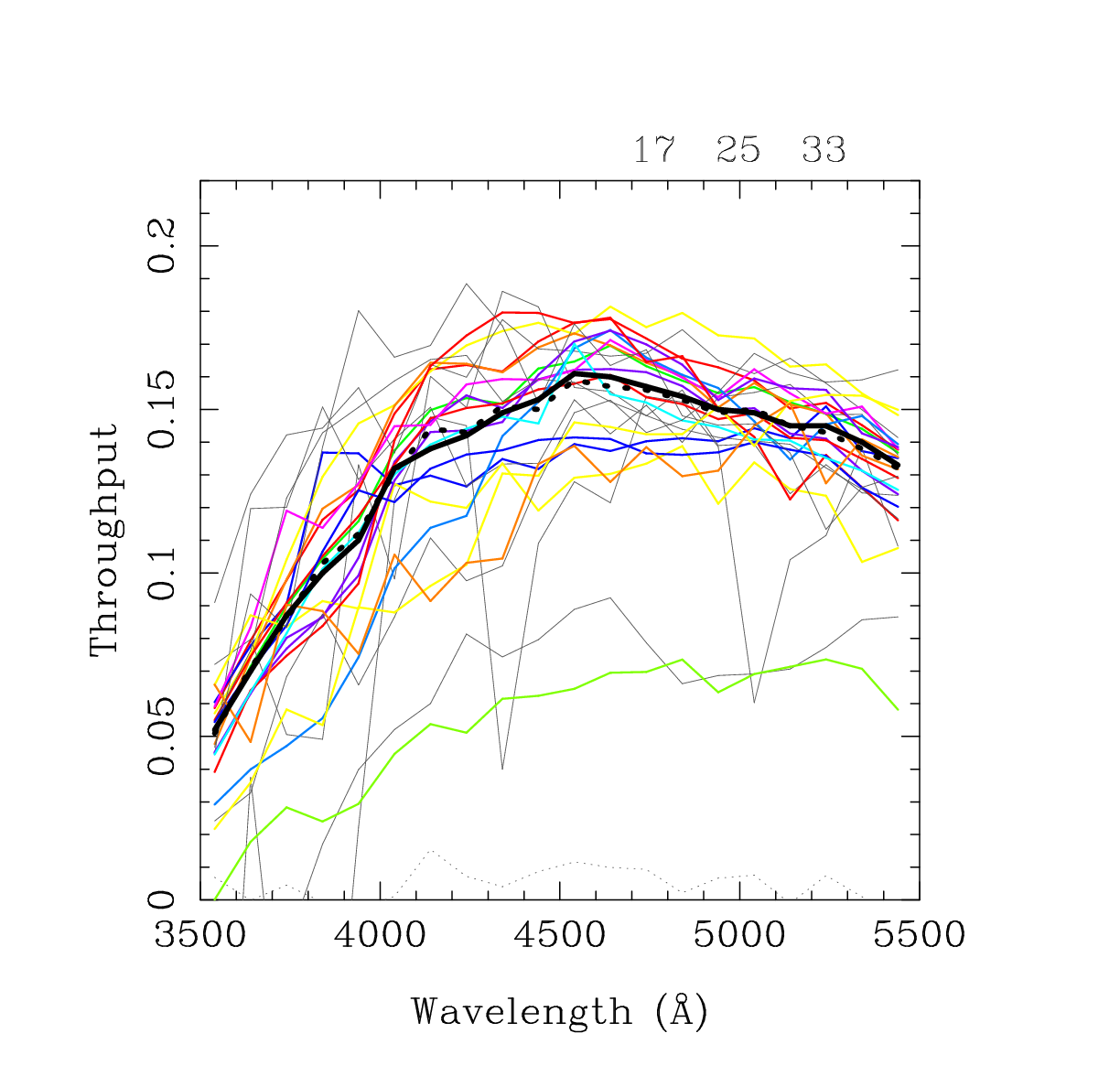} 
\hspace{-11pt}
\includegraphics[width=260pt]{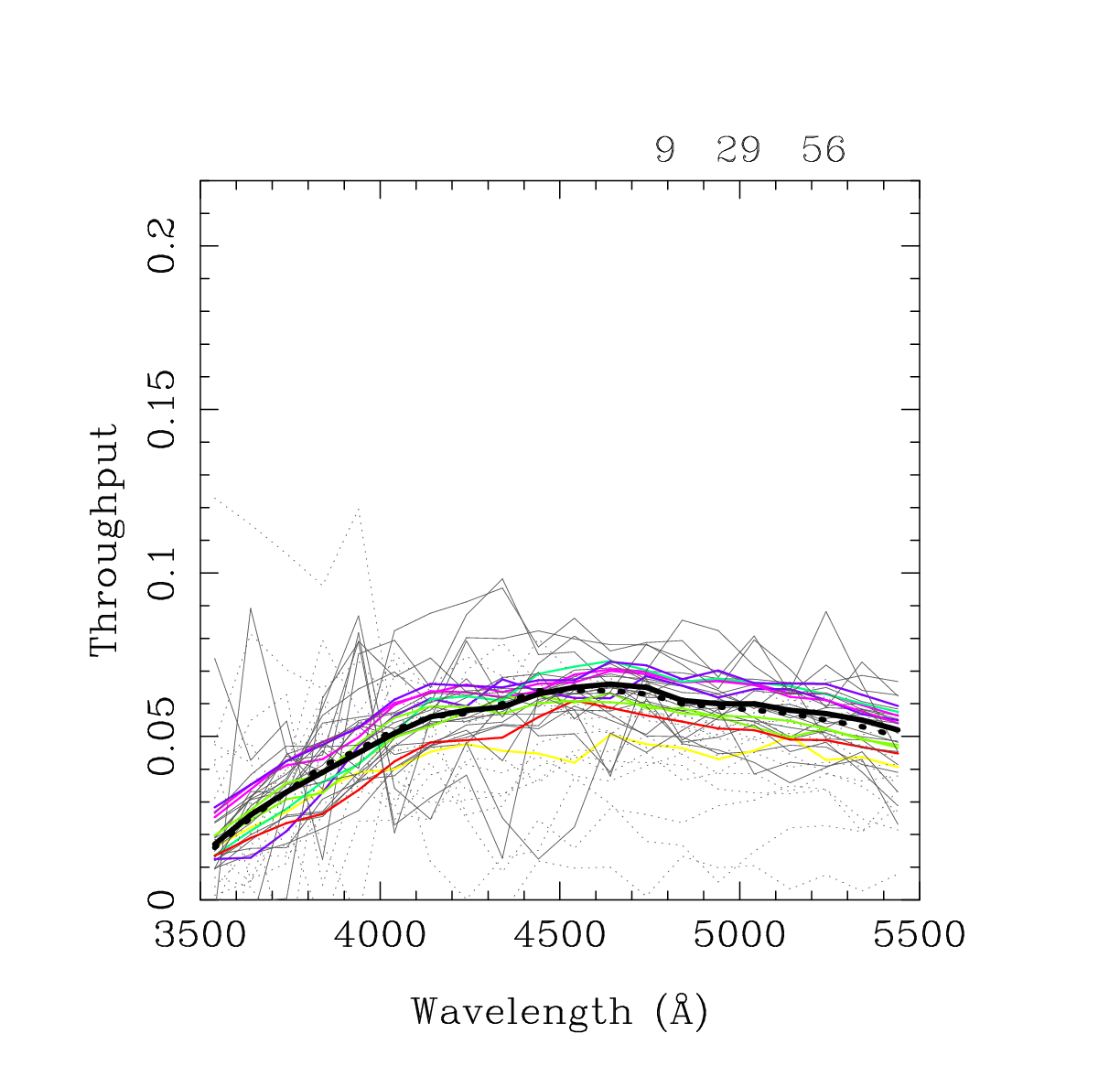}
\caption{Two examples of the multi-band photometry method for absolute flux calibration. The plot on the left is from observation 017 on the night of 10 June 2020; these data were taken through conditions typical for the survey.  The plot on the right is from observation 021 on 14 June 2020, which was taken through clouds. The thin lines are the throughput curves derived for individual stars using the multi-band fits: the thin grey lines, solid or dotted, show curves which have been rejected either due to poor extractions or poor predictions as to their intrinsic flux distribution, while the lines with color display throughput curves that passed our quality criteria. The numbers on top, from left to right, are the number of continuum sources used in the final fit (lines with color), the number of continuum sources including those rejected based on their rms (lines with color and thin solid grey lines), and the total number of detected continuum sources, which includes stars, galaxies and sometime meteors (all thin lines). The thick solid black line is the average derived from the accepted stars, while the thick dotted line is an average which includes the rejected sources.  The agreement between the two means demonstrates the robustness of the fitting.
}
\label{fig:broadband-fit}
\end{figure*}

{\bf System response from multi-band stellar photometry.}
The third approach to HETDEX flux calibration involves the creation of synthetic VIRUS spectra of field stars based on their available multi-band photometry and prior knowledge of the distribution of stellar luminosities, temperatures, and metallicities along the line of sight.  These data can be used to predict the spectra of the SDSS field stars, which can then be compared with their actual VIRUS spectra.  
 
To create the synthetic spectra of the field stars, we begin by generating a set of 1024 template stars with high-quality predictions for their photometric and spectroscopic properties.  Our choice of stars is dictated by the Galactic model of \citet{green+14} along two lines of sight taken to be representative of the HETDEX spring and fall survey regions.  Using this model, we can predict the distribution of absolute magnitudes ($M_g$) and metallicities ([Fe/H]) of $g < 24$ stars in the HETDEX field.  These predictions can be translated into model spectra using PARSEC isochrones and luminosity functions \citep{bressan+12, marigo+17}, a grid of MILES stellar spectra \citep{cenarro+07, falcon-barroso+11}, and a foreground reddening.

Our analysis uses a constant value of $E(B-V) = 0.02$ across the field. This is, of course, an 
approximation, and there are reddening maps that could be used to infer extinctions to individual sources.   However, these maps are for sources outside of the Galaxy \citep{schlegel+98, schlafly+11} or are limited by the low density of field stars at high galactic latitude \citep{capitanio+17}.  More importantly, because the reddening in the HETDEX fields is already small and our procedure is probabilistic in nature, improved reddening estimates do little for the overall quality of our solutions.  As shown below, we reach our specification on the flux calibration without the use of more complex, three-dimensional reddening maps.  

Using our grid of stellar spectra, we compute
the most-likely spectrum for each field star based on its observed SDSS $ugriz$ magnitude \citep{SDSS-7}, its \textit{Gaia} distance \citep{Gaia2018}, and our assumption of foreground reddening.  By comparing the predicted spectra for an ensemble of field stars to their observed spectra, we can derive an estimate of throughput versus wavelength.

There are two obvious improvements that can be made to this procedure. First, it is fairly straightforward to include other data sets in the calculation of the spectral calculation. For instance, \grizy\ photometry from PanSTARRS-1 \citep{ps1}, NUV and FUV data from GALEX \citep{galex1, galex2}, near-IR $JHK_{\rm s}$ measurements from 2MASS \citep{2MASS}, and 3.4 and $4.6~\mu$m flux densities from WISE \citep{WISE} can all be used to help identify the most likely entry in our spectral grid. Second, rather than using just two sets of priors (one for the spring field and one for the fall field), the priors on $M_g$ and [Fe/H] could be recomputed for each individual line-of-sight.  Our simulations suggest that variations in the priors over the HETDEX fields are minimal, but this flexibility would be required for observations at lower Galactic latitude.


The multi-band stellar photometry approach can be applied to stars as faint as $g \sim 23$, which at high galactic latitude corresponds to about 0.35 stars per IFU, or $\sim 30$ stars per pointing for the final array of 74 IFUs. An example of a calibration curve derived via this method is shown in Figure~\ref{fig:broadband-fit}. We show example fits for two different nights in 2020, one with average throughput and one affected by clouds. The individual response curves derived for each source are shown as either lines with color, which represent sources used in the fitting, or thin grey lines which indicate objects that were excluded from the analysis via our flux calibration software.  Most of the rejected sources are either galaxies, stars with poor extractions due to their proximity to the edge of a detector, bright, saturated objects, or detector artifacts. In the early 2017 data, when only a small fraction of the IFUs were installed, generally only 5 to 10 stars were available for this calibration and the flux calibration of these fields has a relatively large error.  By 2019, when the IFU array was nearing completion, a typical calibration involved 20 or more objects, and produced results that were extremely robust.

It is this multi-band photometric method  that is used to flux calibrate the vast majority of the data generated by the HETDEX survey.  Of the 3300 fields observed in HDR2, less then 20 do not have a sufficient number of stars for the multi-band photometric approach; for those frames, we relied on $g$-band normalization.  The expected calibration error on these frames is illustrated in Figure~\ref{fig:stdstar_throughput}.


\subsubsection{Final Flux Calibration}
\label{subsubsec:final-flux-calib}

\begin{figure*}
\includegraphics[width=260pt]{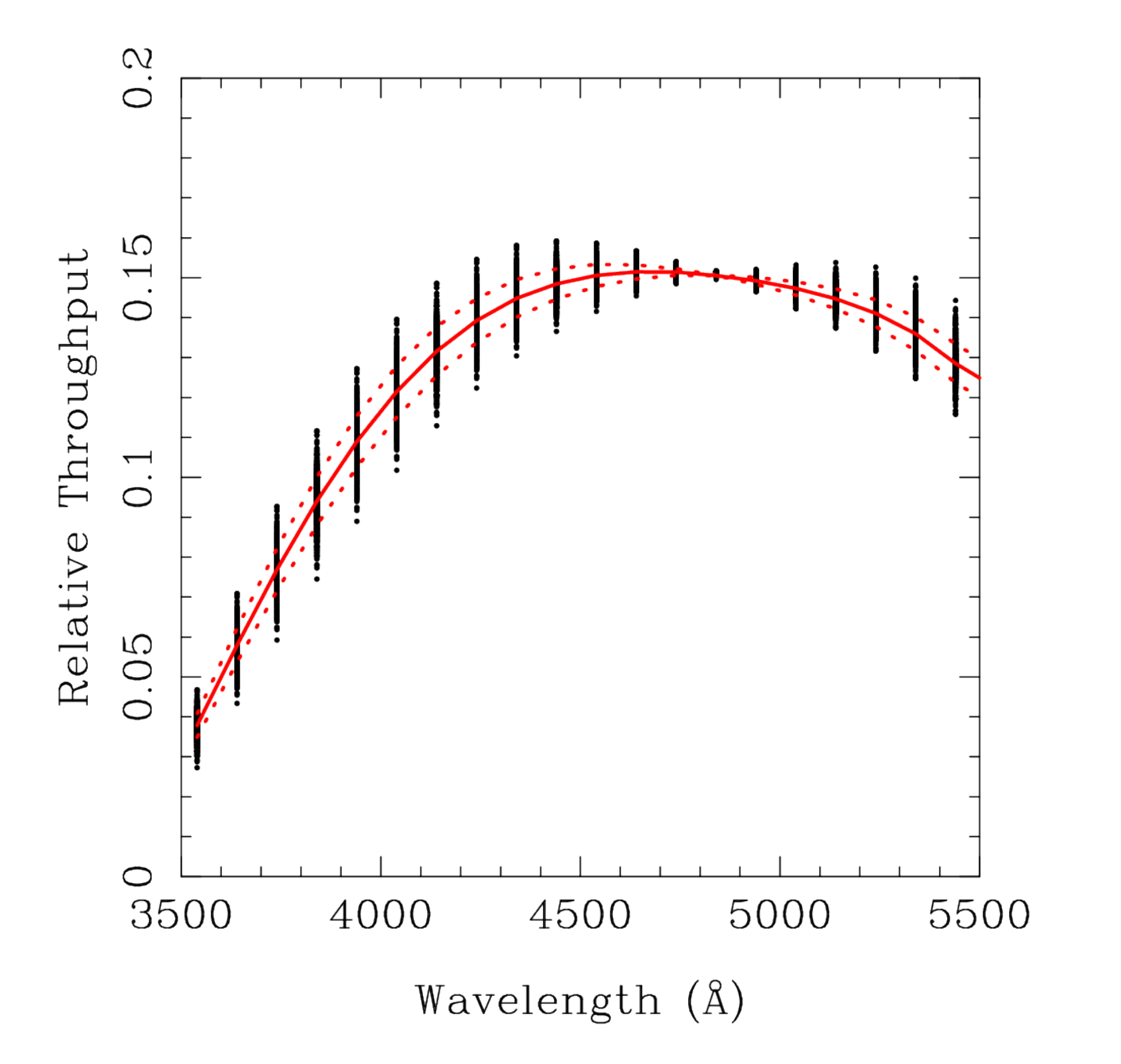}
\hspace{-9pt}\includegraphics[width=260pt]{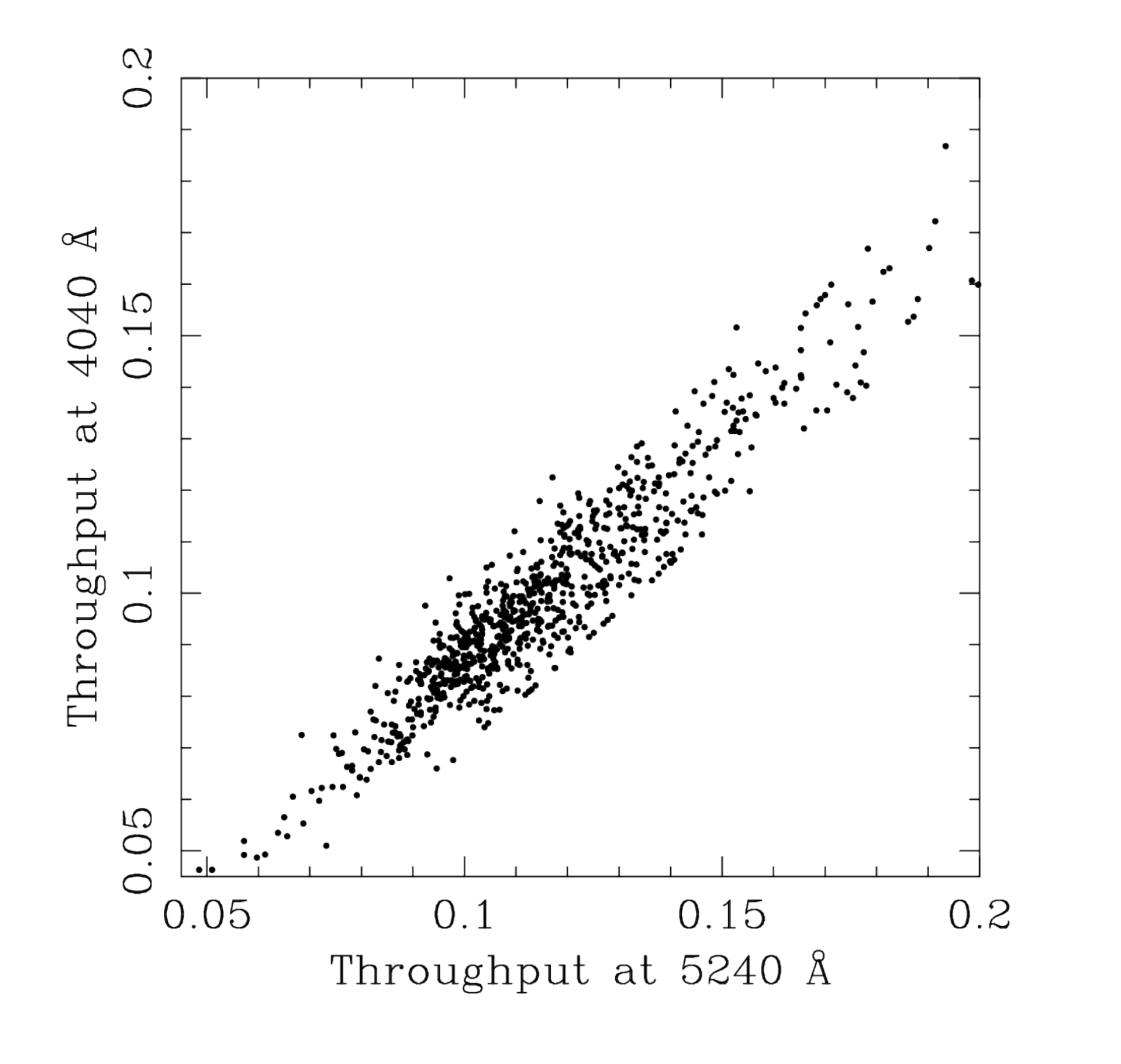}
\caption{Throughput measurements derived using multi-band stellar photometry. The left-hand figure normalizes each curve to 0.15 between the wavelengths 4600\,\AA\ and 4800\,\AA\ and shows the median 4-th order polynomial fit (solid line) and the region occupied by the central 68\% of the curves (dotted lines). The individual measurements that went into creating the curves are also plotted as the black points.  The right-hand panel compares the throughput at 4040\,\AA\ to that at 5240\,\AA. These data come from all the HDR2 observations in 2020.}
\label{fig:rel_throughput}     
\end{figure*}

\begin{figure}
\includegraphics[width=242pt,trim={40pt 10pt 60pt 60pt}, clip]{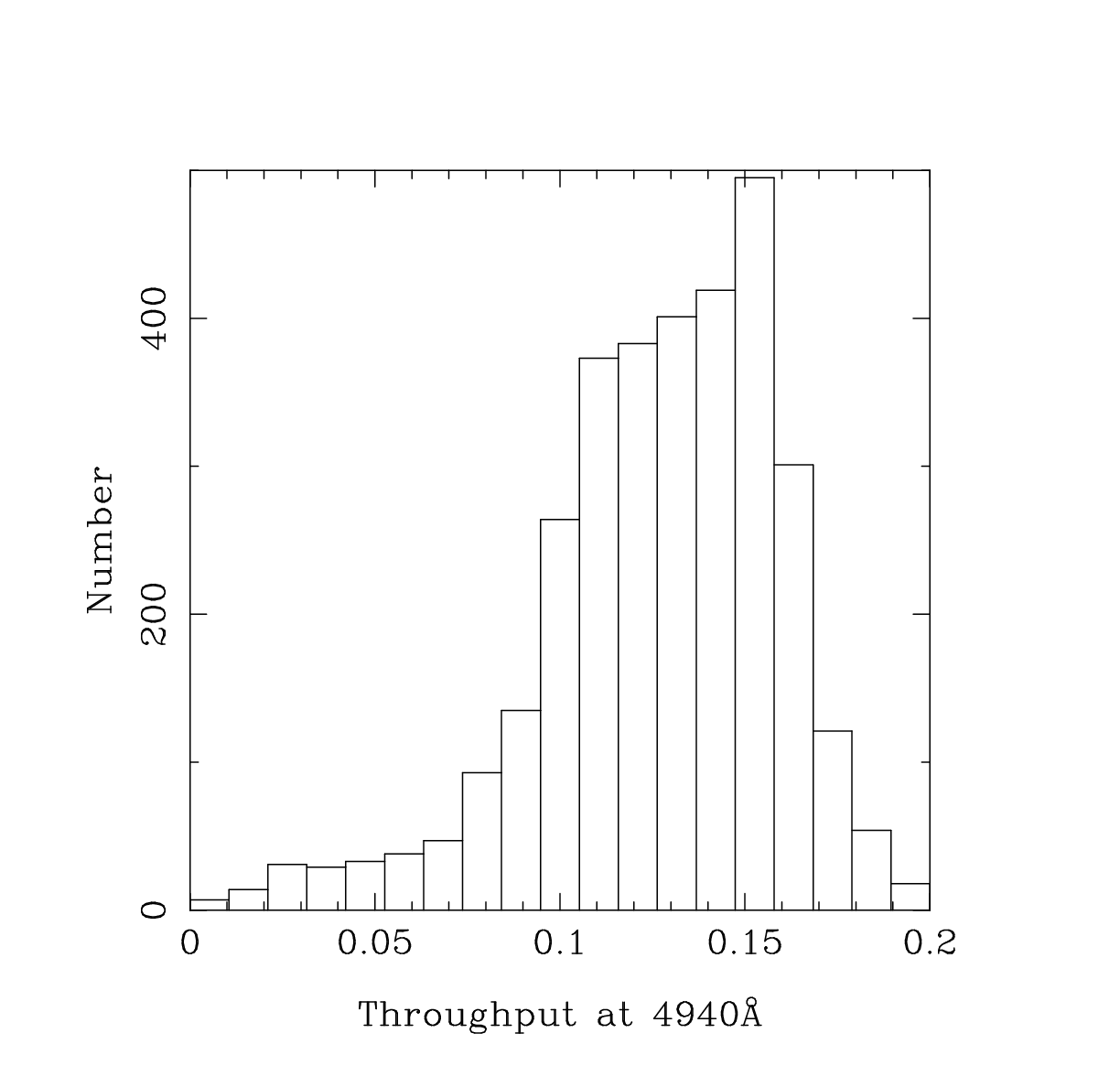}
\caption{The effective response (in flux per count) at 4940\,\AA\ for all HETDEX observations in HDR2. The data assume a standard 3-dither sequence with 360\,s exposures and a telescope aperture of 50\,m$^2$. This distribution is close to our predictions from the pilot survey \citep{Adams2011}.  Observations with throughputs below 0.08 are not used in the final catalog.}
\label{fig:abs_throughput}
\end{figure}

As discussed above, the throughput curve derived from $\gtrsim 20$ individual field stars (i.e., Figure~\ref{fig:broadband-fit}) provides an initial estimate of an observation's response function.  The next step takes advantage of the fact that, over small portions of the VIRUS spectral range, we expect the relative response curve of the instrument to be approximately constant over time.   Our final flux calibration therefore comes from the analysis of all observations, and applies only a normalization and a linear gradient to our survey's response curve; the latter corrects for any large-scale trend in the blue-to-red throughput that might be caused by atmospheric aerosols or systematic error in our spectral extractions.

The results of combining all the throughput curves derived from multi-band photometry are shown in  Figures~\ref{fig:rel_throughput} and \ref{fig:abs_throughput}. To create the figures, the stellar count rates of each frame's field stars were measured in a series of 100\,\AA-wide bandpasses and compared to their predicted flux distributions based on the field's priors for stars of the given color and absolute magnitude. We then took the biweight average of these normalized throughput curves, and fit a fourth-order polynomial through the data. This polynomial serves as the global throughput curve, and defines the instrument's nominal blue-to-red response. We then divide each individual response curve by the global response function and fit the residuals with a linear model. The resultant normalization and slope were then applied to the global curve to yield the observation's final response function, and this function was applied to all the extracted spectra.

The left-hand panel of Figure~\ref{fig:rel_throughput} shows the general behavior of the solutions by normalizing the individual throughput curves for 894 fields observed in 2020 to the response between 4600\,\AA\ and 4800\,\AA\ and then scaling the curve to the mean throughput value of 0.15.  The solid line shows the median of the fits, the dotted lines display the region containing 68\% of the solutions, and the points illustrate the individual measurements that went into creating the curves.  The right-hand panel uses these same data to compare the throughputs at 4040\,\AA\ to those at 5200\,\AA\null. This figure confirms that, although discrepant points are relatively common and there is considerable noise in the individual 100\,\AA\ wide data points that go into making the response function, the fitted curves derived from the ensemble of many field stars are relatively robust.  Although the data were taken over a six month period under a variety of observing conditions, the throughput ratio has little dispersion and the flux calibration at any wavelength is stable to better than $\sim 5\%$. 

Figure~\ref{fig:abs_throughput} uses the throughput curves found from the multi-band photometric approach to examine the system throughput for all 1442 HETDEX observations obtained between 1 Jan 2017 and 9 Feb 2019.  The figure displays a histogram of the effective system response at 4940\,\AA, assuming a nominal 3-dither sequence of 360~sec exposures and a telescope collecting area of 50\,m$^2$. Under most conditions, the effective throughput of the atmosphere+telescope+fiber+spectrograph system is greater than 13\%; for reference, the HETDEX survey requires the effective throughput at 4500\,\AA\ to be greater than 10\%.  When combined with Figure~\ref{fig:rel_throughput}, the data demonstrate that even in the extreme blue, the throughput of the VIRUS spectrographs is generally greater than 5\%.

\subsubsection{Evaluating the Flux Calibration and Uncertainties}
\label{subsubsec:flux-testing}

We use a variety of measures to estimate the uncertainties for the full system throughput. The discussion below provides the error associated with each component of the measurement.

To investigate the relative instrumental throughput versus wavelength over the focal surface, we use all the twilight frames taken throughout a month. We create two masks from the data, one showing the relative fiber-to-fiber variation, and the other displaying the detector-to-detector consistency.  Based on the standard deviation of these $\sim 100$ frames, our relative calibration is accurate to 0.5\%, with slight increases (up to 1\%) at the edges of the field or detectors. This 0.5\% variation represents a floor to our instrumental calibration, and is significantly smaller than our measurement accuracy.

The measurement of the overall absolute throughput is affected by two different issues:  the accuracy of the fits to the stellar spectral energy distributions and the errors associated with our spectral extractions. Both of these issues are significant.

To assess the accuracy of our throughput curves, we use the individual response functions derived from each star on a given frame and determine the standard deviation of the mean of those curves.  In other words, since a field's response curve is a mean value based on SED fits to $\sim 20$ field objects, we use the response estimates from each individual star to measure the dispersion about this mean, and therefore the error associated with the mean value.  This simple analysis suggests that the overall accuracy of flux calibration is generally between 2-5\%, with a small tail extending up to 10\%. (The 10\% values are from the very early data when only a few field stars are available).  In fact, the overall accuracy is better than this, since our fits model the response curve as a fourth-order polynomial, and are therefore insensitive to high-frequency changes in the throughput.  This analysis therefore represents the systematic floor to our throughput accuracy and is the dominant source of noise in our determination of absolute throughput.

\begin{figure}
\includegraphics[width=240pt]{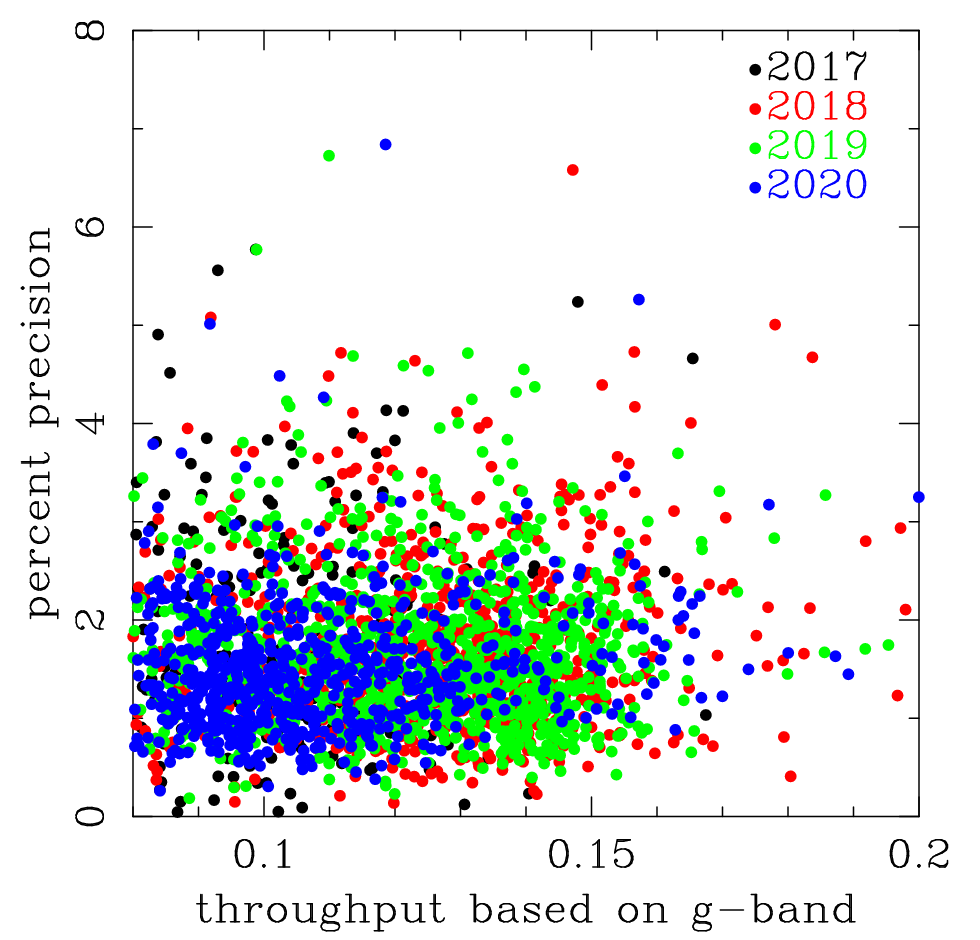}
\caption{Precision of the throughput measurements versus the value of the throughput at 4940\,\AA\ as determined from the $g$-band normalization of stars in the frames. The points include all HDR2 fields; black is from 2017, red from 2018, green from 2019, and blue from 2020.  The throughput is measured by scaling the individual stars' fluxes to their $g$-band magnitude, and the precision is the uncertainty on the mean value (i.e., variation of the individual stars divided by square root of the number of stars). Our target accuracy is 5\% on average, with individual observations going up to 10\%. In terms of average absolute accuracy, we are meeting the project's specification.}
\label{fig:tpaccuracy}
\end{figure}

Another way of seeing this result is to plot the standard deviation of the mean response at 4940\,\AA\ against the total $g$-band throughput for all the HETDEX frames in HDR2.  This is done in Figure~\ref{fig:tpaccuracy}  where the color corresponds to the year of observation (black is from 2017, red from 2018, green from 2019 and blue from 2020) and the mean throughput is estimated from simply scaling the stellar flux densities to match the stars' SDSS $g$-band magnitudes.  Since the target accuracy for the absolute flux calibration is 5\% on average, the plot suggests that we are well within this specification.  

It is difficult to improve on this systematic floor to our flux calibration.  For any individual source, extraction errors dominate:  these errors include local pixel issues (charge traps, etc), inaccuracies in the PSF, edge effects, image crowding, astrometric uncertainties, and, in particular, small uncertainties in the relative fiber positions. (Concerning this last issue, we could improve our knowledge of the relative fiber positions by using images of the fiber array taken in the lab.  However, since we are currently within our overall specification, this improvement has not yet been implemented.)  Based on repeat observations of bright sources located on or near the same region of our detector, our spectral extractions have a minimum variation of $\sim 5\%$.  For random sources in the field, this number increases to a mean value of 11\%, with amplitudes of $\sim 14\%$ in the blue and $\sim 10\%$ in the red. Thus, for each exposure, it is important to average over many stars in order to reduce the errors associated with individual spectral extractions.

\begin{figure*}
\includegraphics[width=510pt]{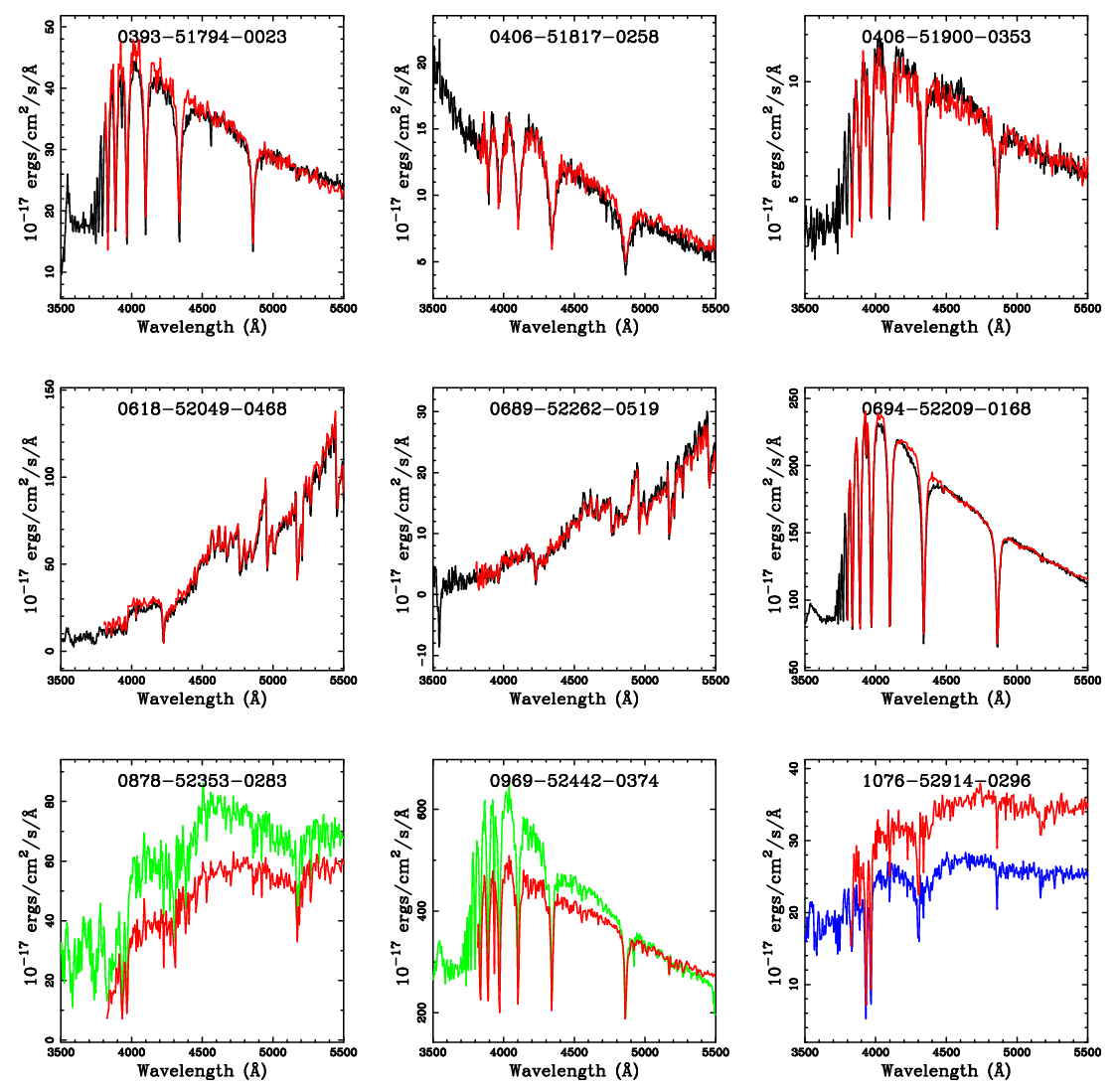}
\caption{Example comparisons between SDSS spectra (in red) and spectra obtained by HETDEX (in black).  The spectra in green represent data from sources at the edge of an IFU; the spectrum in blue shows a source where the SDSS and HETDEX results disagree significantly. We do not track the reasons for the disagreements, as we have enough stars for a robust calibration. The name at the top of each panel is the SDSS ID (plate, MJD, fiberid). The stars in the bottom row are included in the measure of our uncertainties for the overall calibration for a single object, and we are within specification. Since we use a calibration averaged over many sources, we can tolerate individual sources being off by up to 20\%.}
\label{fig:sdsscomp}
\end{figure*}

\begin{figure}
\includegraphics[width=242pt]{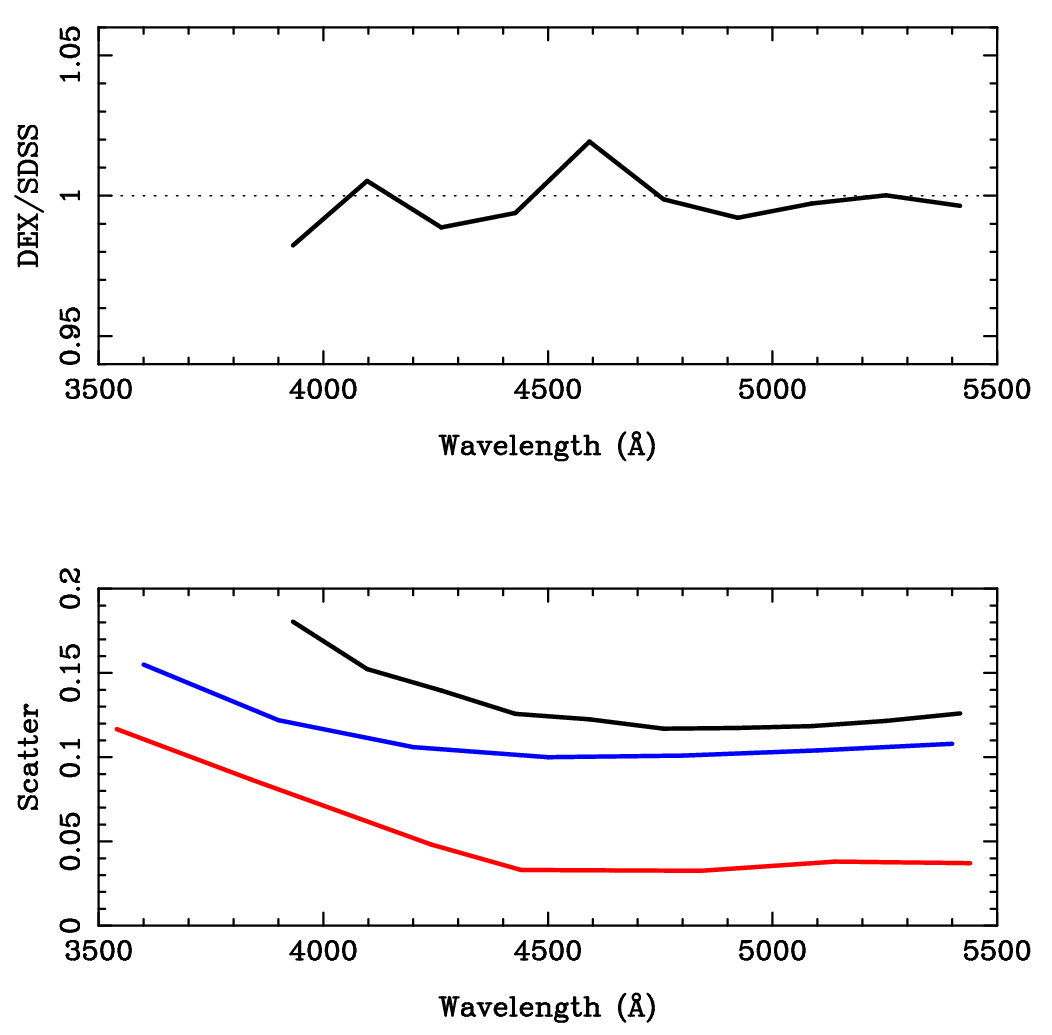}
\caption{The top panel displays the average flux in a HETDEX spectrum divided by a corresponding measurement in the SDSS\null.  This curve is based on 3363 sources observed between 1 Jan 2017 and 6 June 2020.  The bottom figure is the scatter of the flux calibrations produced by our different methods.  The red line is the average accuracy as a function of wavelength, measured by performing SED fits to the SDSS $ugriz$ photometry of field stars; the blue curve is derived by normalizing the response curve found from observations of spectrophotometric standard stars to the counts of SDSS field stars in a synthesis $g$-band filter. The black line is derived by using field stars with SDSS spectroscopy as in situ flux standards.  The multi-band photometric technique, which relies on observations of an ensemble of stars, is more accurate than methods that rely on a single spectral extraction.}
\label{fig:dex_sdss}
\end{figure}

\begin{figure}
\includegraphics[width=242pt]{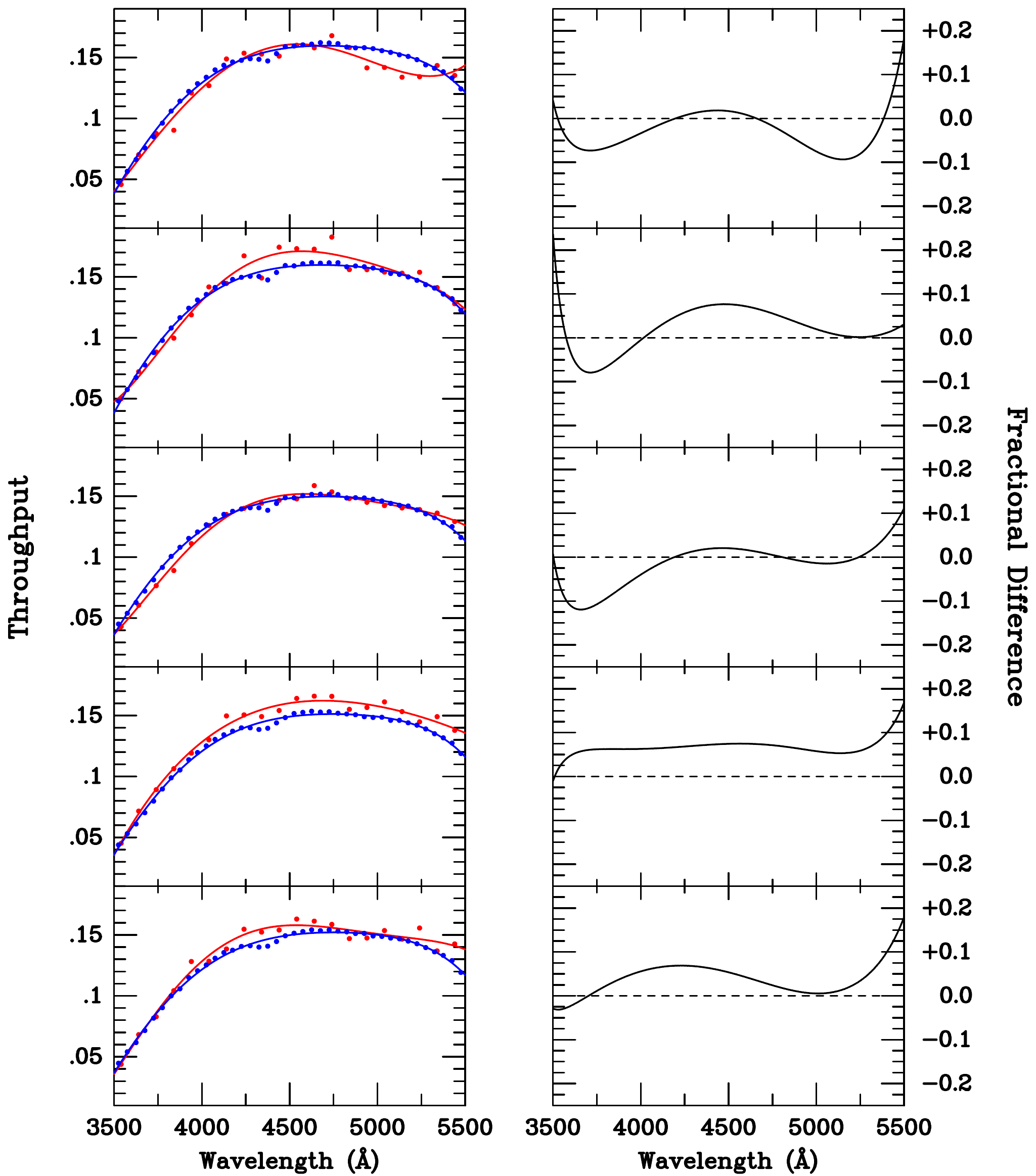}
\caption{The left panels show the HET+VIRUS throughputs derived from observations of the spectrophotometric standard star GD193-74 (blue) and the field stars surrounding the standard via multi-band stellar photometry (red).  The right-hand panels show the ratios of the curves.  The measurements were performed over five nights in Jan 2018.  This comparison is limited by the errors associated with the spectral extraction of the standard star, but the agreement is still good to better than $\sim 7\%$ over most wavelengths.}
\label{fig:compare_stdstar}
\end{figure}

An obvious external test of the accuracy of our flux calibration is to compare the stellar flux densities derived by HETDEX to similar  measurements from the SDSS \citep{SDSS-6}. In HDR2, there are over 3300 field stars with a high enough signal-to-noise to make this comparison. Figure~\ref{fig:sdsscomp} shows 9 such objects, with the SDSS spectra displayed in red. In the bottom row, the first two stars are at the edge of (or even just outside) an IFU, making our flux calibration suspect. The last panel on the bottom row shows an example where there is a significant discrepancy between the two spectra; this is usually due to image crowding or a bad CCD \citep[although, in a few cases, it is due to stellar variability - approximately 3\% of SDSS sources have been found to be variable;][]{bhatti+10}. We do not attempted to identify the objects that deviate significantly from their expected SEDs, as we have enough good stars in each observation to mitigate their contribution. As can be seen from the top two rows of the figure, most of the comparisons are quite good, both in a relative sense, and in terms of the absolute throughput.

Figure~\ref{fig:dex_sdss} summarizes the result of this test. The top panel displays the average ratio between HETDEX and SDSS spectra for 3363 sources observed between 1 Jan 2017 and 9 Feb 2019. To within the precision of the comparison, the relation is flat, demonstrating that we do not need to apply any post-reduction correction.  The black curve in the bottom panel of Figure~\ref{fig:dex_sdss} shows the dispersion between the HETDEX and SDSS spectra. This curve represents the expected flux accuracy for using any single star as a flux standard. The red curve represents the precision obtained by our adopted procedure of calibrating HETDEX using the SED fits of $\sim 20$ stars with $ugriz$ photometry, while the blue curve is derived by using the response curve found by scaling the mean throughput derived from spectrophotometric standard stars to match the $g$-band photometry of SDSS field stars. Note that the two procedures that rely on individual stars, i.e., the curves based on the observations of spectrophotometric standards and the stars with SDSS spectroscopy, have a higher uncertainty.  This is as expected, due to the systematic issues discussed above. Our SED-fitting procedure mitigates this problem, as it averages the results of many stars in the field. Since a typical HETDEX exposure contains at least 20 stars bright enough to be fit, we should expect about a factor of $\sim 4$ reduction in the uncertainty in the flux calibration.  This is close to what is measured.

Yet another way of examining the robustness of the multi-band photometry calibration is to observe a field containing a spectrophotometric standard star and compare the field's throughput curve, as derived from the multi-band stellar photometry, with that found directly from the standard star.  The results of this experiment are displayed in Figure~\ref{fig:compare_stdstar}, where we plot the response curves derived for observations of the \citet{oke90} standard star G193-74.  Five nights of data are shown, with the left-hand panels comparing the response curve derived from the analysis of G193-74 itself to the results from the field stars.  Both curves are represented by fourth-order polynomials.  The right-hand panel shows the ratio of these two curves.  At almost all wavelengths, the multi-band procedure generates a response curve that is within 10\% of that inferred from the spectrophotometric standard.  There are some systematics associated with the fit, as the low-order polynomial used for the fitting does not completely capture the true shape of the curve.  (The worst deviation is at the extreme red end of the spectrum.)  Also, since G193-74 is at a slightly lower Galactic latitude than the HETDEX fields ($l = 166^\circ$, $b = 31^\circ$), the HETDEX spring-field priors used for the calculation are not strictly appropriate for the observation.  Nevertheless, the comparison does suggest that the field-star colors provide enough information to flux-calibrate HETDEX data to better than $7\%$ precision. 

We note that it is possible to combine the results of multi-band SED-fitting, stellar absorption-line analysis, and spectrophotometric standard star observations to produce a set of throughput curves that are more accurate than those found from the multi-band photometry alone. While this may be an option for future HETDEX analyses, it is not needed at the present time, as our current procedure is already meeting or exceeding survey requirements for the vast majority of pointings. Since the multiband photometry method also yields quantifiable uncertainties, it is currently our baseline technique for calibrating HETDEX frames.

\subsection{Galactic Extinction}
\label{sec:galext}

All HETDEX spectra are flux calibrated, but are not corrected for foreground Galactic extinction. Such a correction is employed in our flux calibration process, as our SED matching algorithm needs to know the stars' true colors.  However, the factor is small (we assume $E(B-V) = 0.02$ throughout), and we do not correct our observed spectra for this effect.  In the final HETDEX catalogs, estimates of Galactic extinction will be included in the database.

\subsection{Post Corrections}
\label{sec:post-corrections}

Since the fiber-to-fiber flatfield calibrations are primarily based on twilight sky frames, and since these frames may need different corrections for illumination and scattered light, there could be frame-to-frame differences in their calibration.  Our data suggest that the residuals from these effects are no more than 5\%, and the worst cases are identified during the reduction process.  However, there are regions of the HETDEX spectra that can be systematically off by a few percent. One reason for this is scattered light in the blue; another is associated with an absorption feature near 4350\,\AA\ produced by the VIRUS optical coatings.  Both of these effects are difficult to remove in the pipeline reduction.  To correct for these systematic offsets, we compare the HETDEX and SDSS spectra, and average the fractional residuals over time, matching our spectra to the SDSS calibration. Specifically, we take all the data acquired over a year (over 1000 exposures) and compute the fractional residual correction that must be applied to all science frames to remove this small (few percent) systematic.  Since the residual maps are stable over years, we can apply the same map to all HETDEX data. As the top panel of Figure~\ref{fig:dex_sdss}, illustrates, there is little need for this post correction in HDR2.

\subsection{Noise Model}
\label{subsec:noise-model}

An understanding of the 
element-by-element noise in our sky-subtracted spectra is essential for our line-detection algorithm and the generation of expected source counts. To acquire this knowledge, we compare three different noise models and use the differences between the results to identify potential reduction issues. The three noise models come from 1) the propagation of photon noise throughout the system, including sky subtraction errors, 2) an examination of local noise based solely on spectral and spatial variations within and across fibers, and 3) the analysis of a global model which tracks the variance of each spectral element over time. The third technique is by far the most robust, while the classical model of propagating photon noise is the least reliable. Consequently, we rely on the third model for our HETDEX analyses, and use the measurements of the local noise (model 2 above) to help flag reduction issues. We discuss each model below.

The classical approach to measuring noise is straightforward: one simply propagates the photon noise through the reductions and tracks the additional uncertainties which come with each step. The final noise model is then based on the counting statistics of the object, the sky, and the calibration frames, with the largest noise contributor coming from the sky subtraction.  We assume the fiber profile is uncertain to 2\% and we include an additional component of 2\% of the sky in counts, i.e., $\sigma^2 = \sigma^2_\mathrm{classical} + 0.02*\mathrm{sky}$.  Though mathematically simple, this approach tends to underestimate the actual noise in our spectra. The most likely cause of the underestimation is a combination of scattered light, dark current, sky-subtraction errors, fiber profile errors in the wings of the PSF, and detector controller issues that are not captured in the bias. We use these errors as a starting point when we track expected uncertainties over time (discussed below).

The second approach is empirical as it relies on measuring the local variations in the sky in both in the spectral and spatial directions. We rely on this technique to help flag data reduction issues.  This procedure utilizes 100 pixels in the spectral direction and the 7 nearest fibers to inform the uncertainty of each pixel in the survey. We run this analysis box across each pixel for an observation, and define the noise to be the greater of two estimates: 1) the biweight scale over the 100 spectral pixels for a given fiber, or 2) the biweight scale using the 7 nearest fibers with 2 spectral pixels each. For bright continuum sources, this model overestimates the noise, since absorption lines and other continuum features increase the scale; for these objects, we simply use the noise found from counting statistics. This local empirical model provides a very robust measure of the statistical noise, but it can miss systematic noise introduced by sky subtraction errors. Thus, even though this method is a significant improvement over the classical model, it underestimates the noise around bright night sky lines.  We therefore cannot rely on this approach for object detection, and only use it to help track down reduction uncertainties.

The third approach to noise estimation, and the one we adopt, is to merge the noise determined from photon counting with a global noise model. For every observation, we produce a sky subtracted frame with the expected uncertainty based on the propagation of photon statistics. For each exposure and each extracted spectral element, we then examine the ratio of the sky residuals to the sky measurement, which we call the residual ratio.  Finally, we measure the variance of each spectral and fiber element by combining the residual ratios for a year's worth of observations (i.e., many thousands of exposures).  The result is a map of the actual variance compared to the expected variance for all $\sim 33,000$ fibers and 2000 spectral elements. 

This noise ratio map provides an estimate of how well an element reflects it expected noise properties. For each pixel in an exposure, we multiply the noise from photon counting by the noise ratio map to generate an estimate of the true noise for the pixel.   For most of the pixels of most of the detectors, the ratio is $\sim 1.0$. However, there are some detectors that show overall offsets, some that display blue to red variations, and some that exhibit large offsets around (relatively) bright sky lines. Most of these variations are understood (i.e., due to scattered light, a systematic issue with the wavelength solution, or errors in the fiber or fiber profile), and we simply use the ratio as measured in all subsequent analyses. 

Unfortunately, some of the HETDEX detectors go through phases when the background and amount of electronic interference increases. The root cause of these excursions is likely an instability in the controller temperature.  In these systems, the variance measured across a year's worth of observations will overestimate the true noise, as most of the observations have noise values that are much lower than that of the ensemble average.  The occasional noise increases in the problematic detectors are easily seen when comparing our adopted noise model to the local noise model.  That is, there are some emission line detections which have a reduced $\chi^2$ (from a Gaussian fit in the spectral direction) much lower than one, reflecting an overestimation of the uncertainties. We do not correct these cases, and instead adopt the conservatively large uncertainties. The flux limits are derived from the same noise model, so the higher flux limits are included in our source simulations.

\subsection{Flagging Pixels, Amplifiers, Detectors, and Bad Data}
\label{subsec:pixel-flags}

The final design of HETDEX uses 74 IFUs, 33,000 fibers,  148 spectrographs, 592 amplifiers, 592 fits frames,  0.62 billion pixels, and will include about 30,000 science exposures. Each component can create its own particular problem, and, given the size of this experiment, if a problem is possible, HETDEX will experience it. Thus, we must maintain a high level of quality control, especially down to the pixel level since our primary targets are only a few pixels across. Pixel defects can manifest as detected sources. 

We maintain a barrage of measures to flag bad or compromised data. Most are based on statistical analyses of the individual frames, including tracking the number of detections on a CCD amplifier versus time and versus the number of detections on other amplifiers.   

For each frame and each fits image (i.e., an image from a single amplifier), we track the sky level relative to all other fits images, the pixel rms after sky subtraction relative to the values on all the other frames, and the number of object detections. We then track this information over time, look for obvious trends, and, if necessary, remove the data of individual amplifiers based on this statistical analysis. Some amplifiers can come and go over time, and we maintain a database for whether one should be used.

We start our analyses with the pixel flats, where we identify bad pixels via their low throughput (a relative value generally $<0.6$) or ``hot'' response above a threshold (generally $>1.2$). Another significant problem are charge traps, and we generally remove the whole column when a charge trap is identified. The existence of charge traps on a detector range can remove anywhere from 0 to 20\% of the area. There are also detector regions that are excised due to extreme pox or poor performance.

Some amplifiers have a low level (a few counts per pixel) interference pattern on the detector. This pattern is a few pixel across and can cause significant a number of false positives. We understand this interference to be due to poor grounding within a given spectrograph, and in some cases we remove the whole spectrograph from the analysis. This obviously creates holes in the survey that are included in the window function.

Individual fibers can be bad for a few different reasons, such as being broken, having low or variable throughput, or just being poorly calibrated. These bad fibers are identified and flagged as well.

Another issue is that, on occasion, the  HET's guider will vignette a corner of one of the IFUs. This happens about 1\% of the time, and we flag the data from that IFU for subsequent analysis.

In our baseline 6~minute integration, stars brighter than $g = 13$ will generally saturate.  A small amount of saturation is not a problem as it will just result in a  stellar spectrum that is not usable in any analysis. But brighter stars can cause significant issues when trying to determine the local sky value.  For example, the light from a $g=11$ star can result in an entire  amplifier being flagged as bad, and the scattered light from stars brighter than this can affect multiple IFUs.  In fact, since the HETDEX spring field goes through Ursa Major, there are cases where the bright stars of the constellation have caused the majority of the IFUs to be flagged as bad.  Bright meteors can have a similar effect, and cause much of an observation to be lost.

Concerning this last point, our HETDEX database contains a large number of meteors, observed at a very high signal-to-noise.  Since the spectrum of a meteor is basically pure emission with very little continuum, these objects create false positives and are often initially flagged as LAEs.  However, since we know the average spectrum for a meteor, we can flag these objects in software.  Once alerted to a possible meteor, we can visually inspect the full frame and remove the whole trail. The width of a typical flagged meteor trail is about 10\arcsec, though this number can be larger in extreme cases.

All of the effects above are flagged and removed in the subsequent analysis.

\section{Object Detection}
\label{sec:detection}

The above reduction steps produce a set of calibrated fibers with flux, flux uncertainty, a set of flagged pixels and the fiber's equatorial coordinates on the sky.  We search this dataset for both emission-line and continuum sources using a model based on the PSF, which is measured from the data itself (see \S\ref{subsec:imagequal}).  The procedures used for the emission-line and continuum source detections are very similar;  we highlight the differences below.

The primary targets of HETDEX are LAEs between $1.88<z<3.52$.  Since the typical image quality for our observations (FWHM=$1\farcs 8$) corresponds to about 15 kpc at these redshifts, we assume that the spatial profile of most LAEs will be consistent with that of a point source.  This is good assumption:  although galaxies at $z \gtrsim 2$ do have extended Ly$\alpha$ halos, the surface brightness of a typical $z \sim 3$ LAE will e-fold more than 3 times within the radius of our median PSF \citep{ouchi+20}, and a $3\arcsec$ extraction radius (see \S\ref{subsec:extraction}) will contain virtually all the Ly$\alpha$ flux of a typical object \citep{wisotzki+16}.  Moreover, as shown in \cite{herenz2017} the detection signal-to-noise for a faint LAE is virtually unaffected by a modest template-mismatch:  we will still detect our target galaxies at close to the maximum S/N using a point-source model.  While some of the brightest LAEs may have more extended emission \citep[e.g.,][]{ouchi+20}, the use of the point-source assumption should not introduce any biases for the clustering analysis.

HETDEX does not rely on images to inform the placement of spatial apertures for extraction. We instead search all spatial and spectral resolution elements in the survey for emission-line or continuum flux. Object detections involve a three step process of generating an initial sample of targets of interest, refining the selection around those targets, and culling the sample to generate a final catalog. Simulations inform each of these steps. For the emission lines, the properties that determine whether a resolution element is retained are its signal-to-noise, the $\chi^2$ of a Gaussian fit in the spectral direction, and the fitted line-width. For continuum sources, presence in the final catalog is simply a function of signal-to-noise. We discuss each step in turn.

\subsection{Emission-line Grid Search}
\label{subsec:gridsearch}

The emission-line detection algorithm begins by searching all spectral and spatial resolution elements associated with each IFU\null.  At each fiber location, we spatially extract a spectrum based on the PSF and attempt to fit an emission-line at each location in the spectrum.  This is initially done using a grid that is $0\farcs 5$ in the spatial direction and 8\,\AA\ in the spectral direction. Based on simulations where we attempt to recover artificial sources placed on the sky, this spatial and spectral sampling is optimal for the initial search. This grid has $448 \times 3 \times 500 \times 9 = 6$M elements per IFU (where the numbers correspond to the number of fibers, dithers, spectral elements, and spatial elements); since the current data release has $\sim 160,000$ IFU pointings on the sky, our initial object search involves analyzing $\sim 10^{12}$ elements.  We do note that a slightly finer grid would increase our detection sensitivity, but the computation time needed for such grids becomes prohibitive.  Moreover, our numerical experiments demonstrate that $0\farcs 5$ spatial bins produce source lists that differ from the ideal case by less than 1\%.

At each resolution element, we fit a Gaussian line profile with a line-width fixed to the instrumental line-width. We then keep all sources that have S/N$>$4.0 and $\chi^2 <3.0$ for follow-up analysis. These numbers are quite generous and produce a catalog that is about $10 \times$ larger than the final catalog. To produce a more robust source list, we then conduct a secondary search using a $5 \times 5$ raster centered on the position of each candidate, with $0\farcs 15$ spatial steps.  The location within this raster that provides the highest S/N of an emission line is assumed to be the true source position. During this step, the same input source may appear more than once in the output catalog; to deal with this, objects that are within some specified distance of each other, both spatially and spectrally are combined. For our initial catalog, any two sources that are within 3\arcsec\ and 3\,\AA\ of each other are considered to be one object. The position that has the highest S/N is then defined to be the source's location.

For a single IFU, the initial grid search provides about 50 to 100 sources of interest. The secondary grid search generally reduces this number to about 20 sources per IFU\null.  This decrease is due primarily to more aggressive requirements on S/N and $\chi^2$: we prefer a more lenient limit for these properties in the initial search, as our fits generally improve when the spatial location is better centered. (A difference of $0\farcs 2$ can make the difference as to whether a source makes it into our sample.)  A source measured in the secondary search must have a S/N$>4.8$ and $\chi^2<1.2$ to make it into our final catalog (see Mentuch Cooper et al.\ 2022, in preparation). Currently, we maintain a souce list that go down to S/N=4.5, but objects below the S/N threshold of 4.8 have not yet been  vetted to our satisfaction. The translation from S/N to flux changes from field to field, and over IFUs within a field, and we base the individual IFU catalogs on the S/N selection. In the final catalog, we generally have about 5 sources per IFU, coming from an initial search of 6 million resolution elements. 

\subsection{Continuum Grid Search}
\label{subsec:continuum-search}

While the focus of HETDEX is on emission lines, there is substantial ancillary science and calibration improvement that comes from the creation and search of a continuum catalog. Of course, we could rely solely on deep imaging surveys to select regions for the extraction of continuum sources, but there are advantages to providing a continuum catalog based entirely on a grid search of HETDEX data. Some examples include measurements of variable stars,  transient sources, and moving objects, and identification of issues associated with our flux calibration.  In fact, the combination of results from a continuum grid search with the spectral extraction of sources with known positions will ultimately provide the most science. Here we focus on defining a continuum catalog based on a grid search, without any reliance on an imaging catalog.  The first step is such a search is to define a lower bound for continuum detection.  For first pass, we use a conservative cut, corresponding to around $g=22.5$. In future analyses, we will extend our cut to lower fluxes and combine these data with dedicated extractions based on an imaging catalog.

For each of the 448 fibers in an IFU, we measure the detector counts in a 200\,\AA\ window in the blue (from 3700 to 3900\,\AA) and in the red (from 5100 to 5300\,\AA\null). If either region contains more than 50 counts per 2\,\AA\ pixel on average (about $g=22.5$), we flag it as a possible continuum source. As mentioned, this 50 count limit is arbitrary and designed to be conservative; objects can be detected more than magnitude fainter than this limit.  Once we detect a possible source, we then search about the fiber position, using a $15 \times 15$ element raster and $0\farcs 1$ spatial bins.  The spatial location that achieves the lowest $\chi^2$ fit to our PSF model defines the center of source, and a point-source extraction at that position (see \S\ref{subsec:extraction}) generates the spectrum. This peak-up procedure is obviously different from the emission-line peak-up algorithm where we center by using the highest S/N\null. Currently for the continuum source grid search, we only employ point-source model extractions, which is clearly not adequate for resolved galaxies. Refinements for extended sources are planned for the future.

At our current threshold, we generally find about 0.5 continuum sources per IFU, with about 80\% triggered from the red bandpass and 20\% triggered from the blue spectral region. In the current continuum catalog, we have about 80,000 sources. We are currently refining our ability to robustly extract fainter sources, and we expect to extend down to around $g=23.5$ in the near future. Thus, we plan to increase significantly the continuum sources found from a grid search. Once combined with an imaging catalog, we should produce around 3 continuum sources per IFU.

\begin{figure*}
\includegraphics[width=\textwidth]{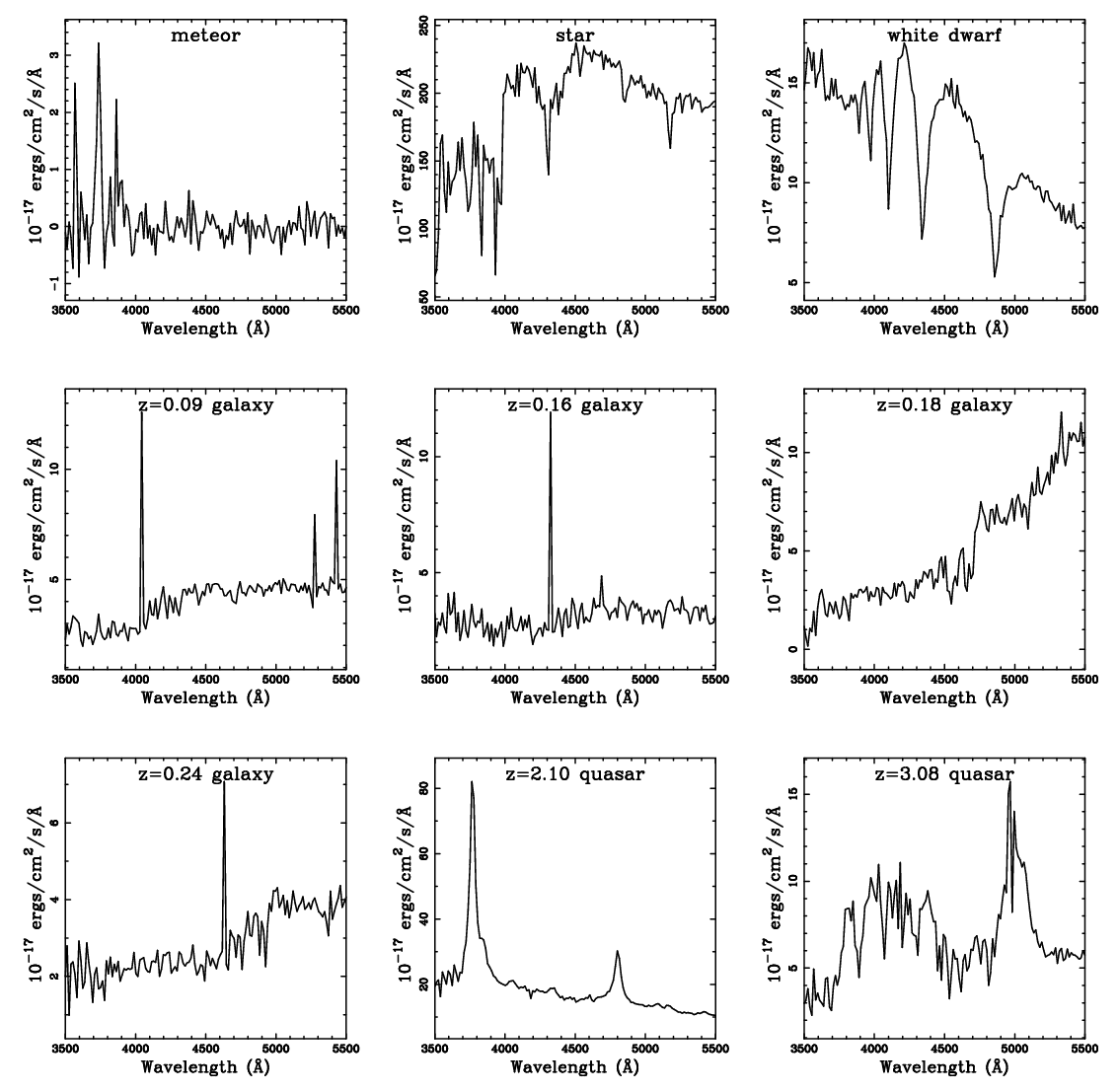}
\caption{Nine examples of continuum sources. The top-left spectrum is that of a meteor; all meteors make it into our emission line catalog and some are detected as continuum sources. The remaining eight sources are stars and galaxies,  with their classification given at the top of each panel.}
\label{fig:example2_spec}
\end{figure*}

As with its emission-line counterpart, the HETDEX continuum catalog does not rely on input positions from an external database.  Obviously, many of the detected continuum sources are stars in the Milky Way:  for example, \citet{hawkins+21} cross-matched the HDR2 continuum source catalog with objects in the \textit{Gaia} DR2 database \citep{Gaia2018} to examine the spectra of stars with a detected proper motions. But the HETDEX continuum source catalog contains much more than just stars: it has galaxies, transients sources, and moving objects, such as asteroids, comets, satellites, and meteors. Figure~\ref{fig:example2_spec} shows typical spectra for continuum-selected objects identified in January 2020.  Some objects, such as meteors, emission-line galaxies, and quasars, can be present in both the continuum and emission-line source catalogs.  Others are pure continuum sources.  

\subsection{Point Source Extraction}
\label{subsec:extraction}

The size of our HETDEX extraction aperture is dictated by the design of the instrument and the typical seeing at the telescope. For the LAEs, we use a point-source model, which will bias the measured fluxes for spatially-resolved sources.  Since the HET does not have an atmospheric dispersion corrector, the offset between an object's position at 3500\,\AA\ and that at 5500\,\AA\ is $\sim 0\farcs 95$ (see \S\ref{subsec:imagequal}).  This value, combined with the typical image quality of the observations ($\sim 1\farcs 8$) leads to our use of a 3\arcsec\ extraction radius. All fibers within 3\arcsec\ of a source position are taken into account in our analysis.  A typical object therefore contains counts from $\sim 15$ different fibers. We note, however, that given the typical PSF of the observations, fibers that are close to 3\arcsec\ away contribute very little flux due to the point-source model used in the extractions. 

The initial step for spectral extraction is to define the weights given to each fiber as a function of wavelength.  This is computationally intensive, as the mean PSF (as computed in \S\ref{subsec:imagequal}) of a typical 3-dither observation has a FWHM similar to diameter of a HETDEX fiber, $1\farcs 5$.  Thus, to determine the proper weights for an extraction, we must use an object's centroid, as determined from our   $0\farcs 15$ step-size raster search, and integrate the PSF across the face of each fiber.  This means that after performing our initial 6 million element object search and our higher-resolution $5 \times 5$ raster search, we must integrate the PSF of each source over the $\sim 15$ fibers contained within the aperture.

After determining the appropriate weight for each fiber, we combine this number with the fiber's noise properties.  Recall that the fibers that go into an extracted spectrum come from 3 dithers and are spread across the detector. Their noise properties may therefore be quite different, with variations up to a factor of $\sim 2$.  Furthermore, about 3\% of all pixels in an image are flagged as bad data for various reasons (e.g., cosmic rays, charge traps, pixel defects, bright star contamination, etc.).  This information modifies the weights defined by the observation's PSF and generates the final weights for each fiber.  These weights are then used to produce the object's one-dimensional spectrum via the optimal extraction algorithm of \citet{horne86}.

\subsection{Emission-Line Fitting}
\label{subsec:efit}

The final step in the detection of emission lines is to examine the one-dimensional extracted spectra and apply a matched filtering algorithm to the data.  For each extracted spectrum, we run a wavelength window over the full spectral range, varying the initial central wavelength and range of the search. By breaking the spectrum into different regions, we can identify multiple lines in the same spectrum and mitigate the effect of local systematics introduced by the data reduction. 

We fit the lines using a step size of 4\,\AA\ and a window of $\pm50$\,\AA\null. The first spectral fit uses an initial guess of 3500\,\AA\ for the wavelength of the emission line, and uses a range that extends from the minimum wavelength for the spectral extraction (generally around 3490\,\AA) to 3550\,\AA\null. Each successive window then increments the red end of the range by 4\,\AA, until the window becomes 100\,\AA\ wide; after that, this 100~\AA\ wide window moves across the spectrum in 4\,\AA\ steps, until it gradually tapers down to 50\,\AA\ at the red end of the spectrum.  The spacing of 4\,\AA\ leads to a large number of multiple hits for the same source; these are trivially combined into a single source at the detected wavelength. We have measured the fraction of missing sources as a function of step size. Increments of 8\,\AA\ and smaller never miss a source, whereas larger step sizes will sometimes cause the fitter to not peak up on an emission line. 

For our initial run, we adopt a Gaussian line-width of $\sigma = 3.0$~\AA ; for subsequent fittings in a refined search, we allow this line-width to float.  The advantage of fixing the line-width is that it improves our ability to detect faint emission-lines matching the line-width value. For example, if we allow the line-width to float initially, a low-level continuum source can be identified as an object with a  wide emission line; in this case, a high line-width fit would be reported, and a potential real source with a smaller line-width might be omitted from the sample. Based on our simulations of input sources with varying line-widths, we have settled on the approach of initially setting the line-width at 3\,\AA, a value that still allows us to recover sources with wider emission-lines.

The detected emission lines are fit with a Gaussian model using Levenberg-Marquardt minimization.  In the initial fit with $\sigma=3.0$\,\AA, there are 3 parameters: continuum level, total line flux, and line centroid; in subsequent fits we treat line-width as an additional parameter. We have experimented with additional terms, i.e., using Gauss-Hermite polynomials and fitting more than one line, but have found that a single Gaussian fit is adequate.  In future work, we may  explore other types of line profiles, such as the asymmetric Gaussian used by  \citet{shibuya2014} or a mean profile determined by from our own data.  However, since the resolution of the VIRUS spectrographs is low ($R=750-950$) and most of our sources have low signal-to-noise, such improvements are unlikely to add much to the detection optimization.  We note that fits with symmetric profiles will be poor if the source has a high signal-to-noise, skewed line profile. For example, active galactic nuclei and quasars will not be well fitted by a Gaussian. And for some extreme LAEs that have very skewed line profiles, the adopted profile may not be adequate. For cosmology measurements, this effect should be insignificant, since we generally care about whether the source is detected and not the precise amount of flux recorded in the emission line.  There is the secondary consideration that we need to know the completeness for any detected source, and there might be a small systematic in the case  of faint, but extremely skewed Ly$\alpha$ profiles. For AGNs and quasars, Liu et al.\  (2021, in preparation) re-fits the lines using the full spectrum and also derives a refined completeness limit.

\begin{figure*}
\includegraphics[width=\textwidth]{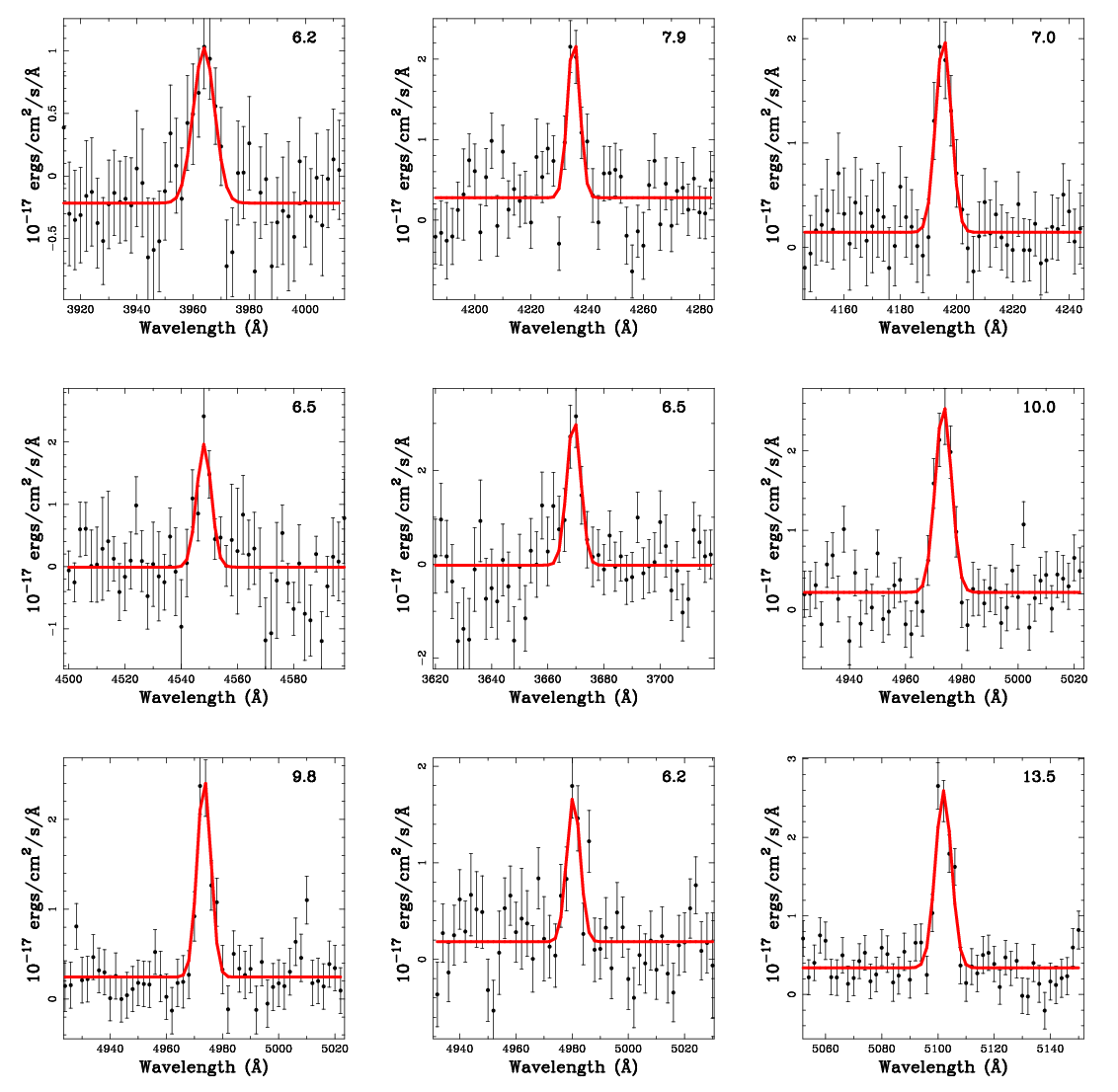}
\caption{Typical spectra and their fits from one exposure set taken on 25 April 2020. The S/N is shown on the top-right of each figure. These spectra were chosen at random from over 400 emission lines present in the observation.}
\label{fig:example_spec}
\end{figure*}

We define the signal for the source using the area under Gaussian fit; we have compared this area to the result obtained by simply summing the flux within an emission line, and the results are nearly identical.  Of the 15 fibers that go into an extracted spectrum, generally three have weights above 10\% and at most five have weights above 1\%. These numbers, of course, are dependent on the PSF of the observations and the angular size of the object. Any object with a high-weight ($w > 10\%)$ pixel within 3\,\AA\ of the line centroid that has been previously flagged as bad is removed from the source list. Cosmological parameters can be sensitive to false positives, and this step helps reduce the number of false detections in our sample.

The uncertainties on the fitted line parameters come from a series of Monte Carlo simulations. We do not rely on the covariance matrix that we generate from the initial fit, although we do compare the predictions of this matrix to those from the numerical simulations in order to look for inconsistencies. (The simulated uncertainties are always higher than those derived from the covariance matrix.)  For our Monte Carlo simulations, we use the input line profile and its uncertainty, which come either from the root-mean-squared of the fit or our noise model, whichever is greater.  Generally speaking, the use of the higher noise value improves the rejection rate of false positives at the expense of missing some real sources. We note that it is not unusual for the noise model to be higher than the root-mean-square: since this model is based on data taken over a year, detectors whose noise properties that change with time will occasionally be better than characterized (see \S\ref{subsec:noise-model}). We track these noise differences to identify variability in the detectors.

The Monte Carlo simulations start with the 4 fitted parameters, generate a Gaussian line profile for the source, use the noise model to create a synthetic spectrum, and then fit the spectrum's emission-line using the same algorithm as for the original measurement.  This process is repeated 100 times to determine 68\% uncertainties on the emission line's location and strength.  This information, along with the line's measured wavelength centroid, flux, line-width, continuum level, S/N, and $\chi^2$ is then entered into the emission-line catalog.

Figure~\ref{fig:example_spec} shows some typical emission-lines and their fits for the night of 25 April 2020. 

\subsection{Detection Rate}
\label{subsec:detection-rate}

There are three science requirements that are essential to track, as these are the most important for reaching the cosmological specifications of measuring $H$ and $D_A$ to better than 1\% accuracy. These are 1) the number of Ly$\alpha$ emitters detected, 2) the number of false detections, and 3) the number of interloping galaxies, such as $z<0.5$ \OII\ emitters that are mistaken for LAEs.

\begin{figure}
\includegraphics[width=240pt]{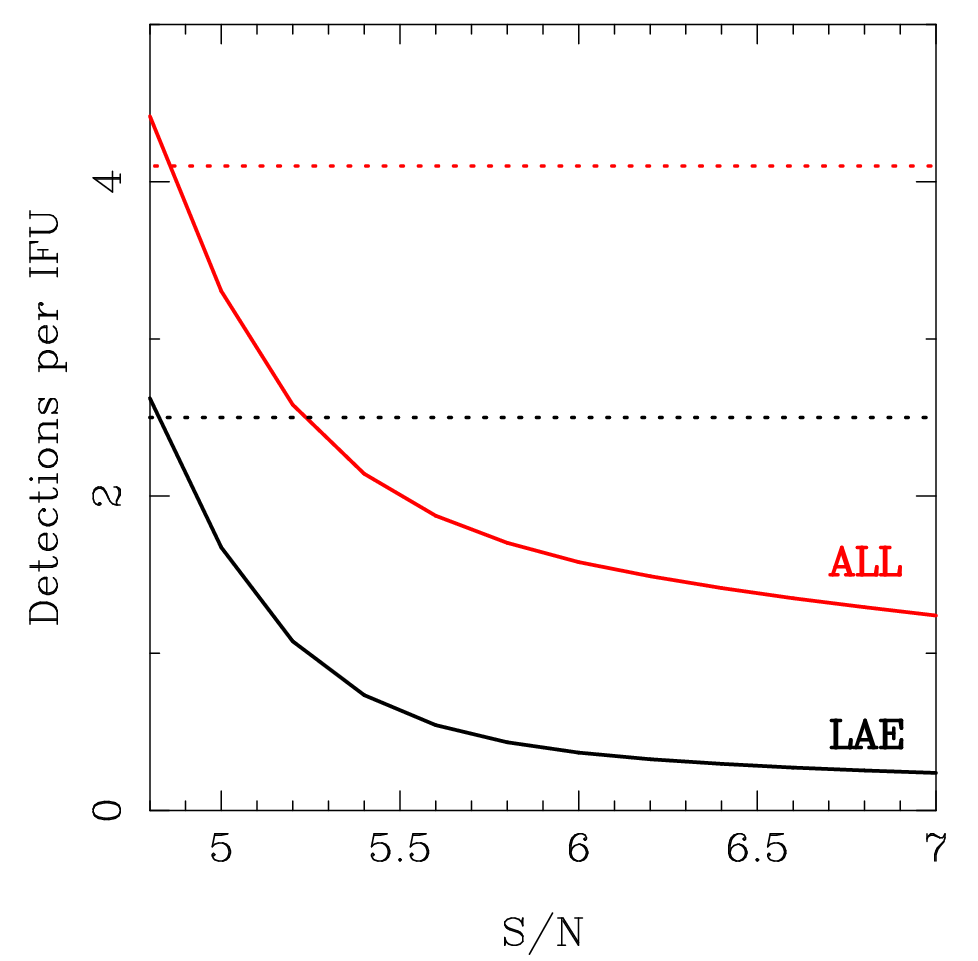}
\caption{Number of emission line sources per IFU for HDR2 versus signal-to-noise ratio. The red solid line is for the total number of emission lines in the catalog, and the red dotted line is the required number for the sum of LAEs and \OII\ emitters. The black solid line is the current number of LAEs using our automated classification software. The dotted black line is the required number of LAEs.  We expect to increase the number of LAEs in the catalog as we study the various cuts that we have applied and refine the classification algorithm. Currently, we are meeting the specification for S/N$>4.8$, but small changes to our procedures may result in significant improvements.}
\label{fig:ndets}
\end{figure}

For the total number of detected LAEs, the best metric to use is the number of emission-line detections per IFU\null. From the science requirements, our goal for this number is 4.1.  Based on the  LAE and \OII\ galaxy luminosity functions \citep{gronwall+07, ouchi+08, ciardullo+12, ciardullo+13} we expect that, on average, 2.5 sources will be LAEs and 1.6 will be \OII\ emitters. (There are sources detected from other lines, but these comprise less than 10\% of the full sample.) Figure~\ref{fig:ndets} compares the results of our HDR2 analysis to these goals. From the figure, is it apparent that to reach specification, we need to work down to S/N $\sim 4.8$. We also include an estimate of the number of LAEs per IFU based on P(LAE)/P(\OII), i.e., the probability ratio that a detected emission-line is Ly$\alpha$ as opposed to \OII\ $\lambda 3727$.  This likelihood is derived primarily by comparing the object's emission-line flux to its broadband flux density, either from the HETDEX spectra itself, or from imaging catalogs \citep{leung+17}.  However, there are multiple additional factors that enter into the probability calculation, including the presence of other emission lines, the size of the source, and the astrometric offset between the emission-line and the continuum. These will be discussed in detail in Davis et al.\ 2021 (in preparation).  In Figure~\ref{fig:ndets} LAEs are identified as having an P(LAE)/P(\OII) $>5$, but we have not yet performed a statistical analysis for how this ratio affects the rate of contamination. The science requirements specify that no more than 2\% of the LAE sample be misclassified \OII\ emitters. We will be evaluating the probability ratio with this 2\% limit in mind (Davis et al.\ 2021, in preparation).

\subsection{False Detections}
\label{subsec:false-detections}

The target false positive rate from the science requirements is below 10\%. This number is chosen so that the effect of this white noise would be well below the statistical noise. This false positive rate is included in all cosmological analyses.

There are several ways to estimate the false positive rate produced by our detection algorithms, including the analysis of inverse frames, use of classical statistical techniques, and repeat observations. The optimal way to quantify false detections is to empirically measure the rate from fields observed more than once, either through serendipitous overlapped regions on sky or through intentionally repeated observations at either nominal depth or deeper. We discuss each method below, and find that the overlapping regions provide the best measures since there are a large number of them and they cover different fibers.

The inverse frame technique, which involves multiplying a sky-subtracted frame by $-1$ and then running the data through the detection algorithms, requires even more attention to the pixel level defects that have been discussed. Our focus for the pixel-level defects has been on the normal frames, where we maintain a list of pixels or regions that trigger detections at a much higher rate than expected. We have not spent the time listing pixels or regions that trigger detections in the inverse frames. An analysis on the inverse frames is still worthwhile, and we will re-address this issue once the pox issues are mitigated a better level. The other sources of detections in the inverse frames that differs from the normal frames are faint continuum objects. On an inverse frame, absorption lines turn into emission lines so one must carefully remove all the faint continuum sources before conducting the analysis. We do automatically generate the inverse frames and run detections on these negative spectra.  In fact, after going through individual inverse detections and removing those features associated with continuum sources and obvious pixel-level defects, we find a rather small number for the amount of contamination, below the 10\% specification. However, we cannot trust this estimate until we refine the inverse frame technique with the more recent data.  Given the importance of the false positive rate for cosmology measurements, this technique will be revisited in future analysies.

The classical approach of counting resolution elements across the field and using Gaussian statistics to estimate the number of false positives is also inadequate. The issue here is that the appropriate noise model does not have Gaussian tails, and since many of our objects have a signal-to-noise of 5, understanding the shape of these tails is critical. While this technique does not work globally (that is, defining the noise per resolution over the full field), we do use the noise shape for individual detections in order to help identify real sources. As discussed below, for an individual source with repeat measurements, we generate the signals from all fibers within an aperture and compare the distribution of negative fluxes to that of positive fluxes. Real sources stand out clearly with this local statistical estimator.

\begin{figure}
\includegraphics[width=242pt]{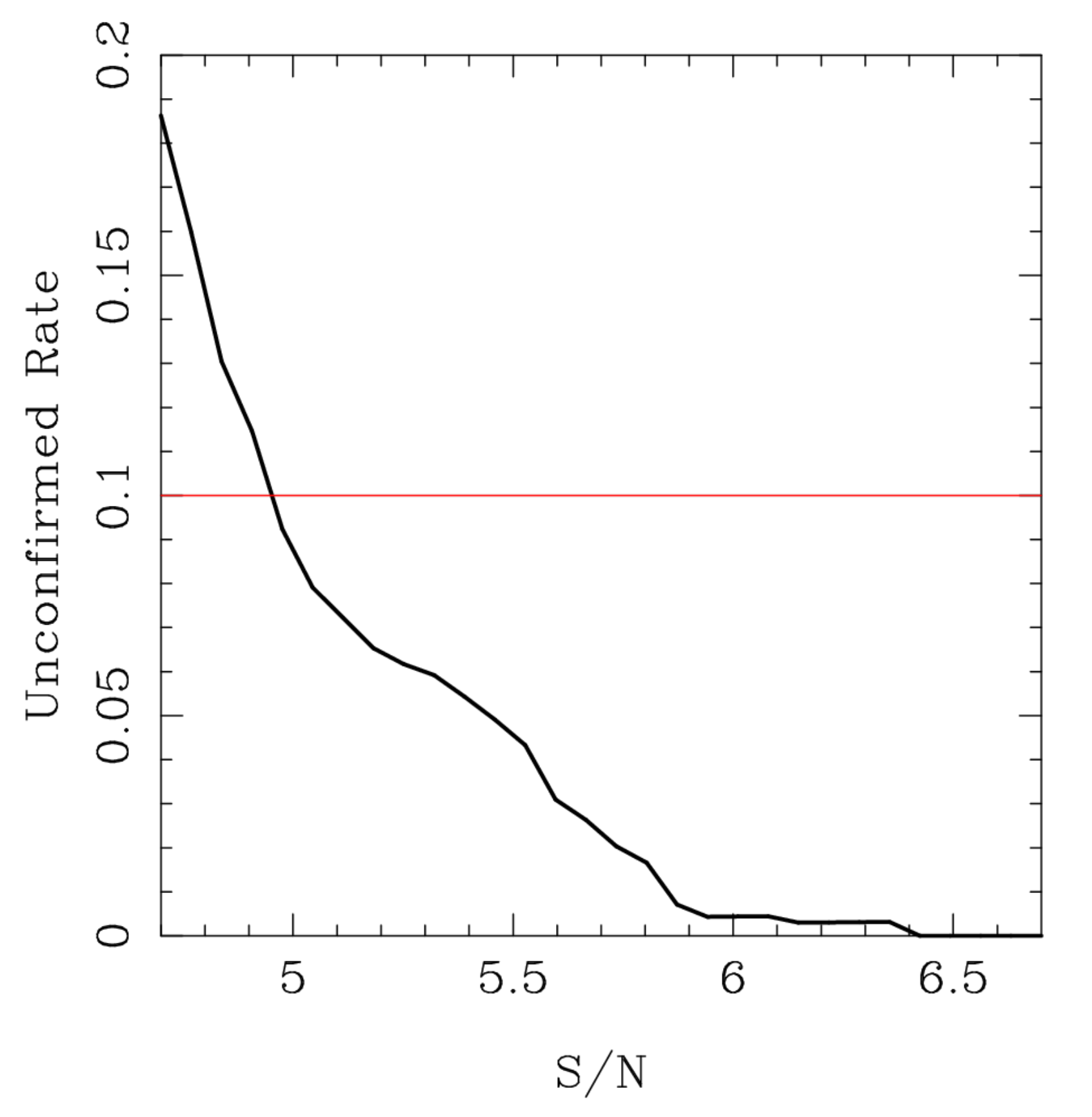}
\caption{Rate of unconfirmed detections, which is an upper limit to the false positive rate, from the study of overlapped pointings. Emission-line objects detected with a signal-to-noise above 5.5 are almost always real; it is only when the signal-to-noise drops below 5 that the rate of false positives becomes problematic.}
\label{fig:false}
\end{figure}

\begin{figure*}
\includegraphics[height=264pt]{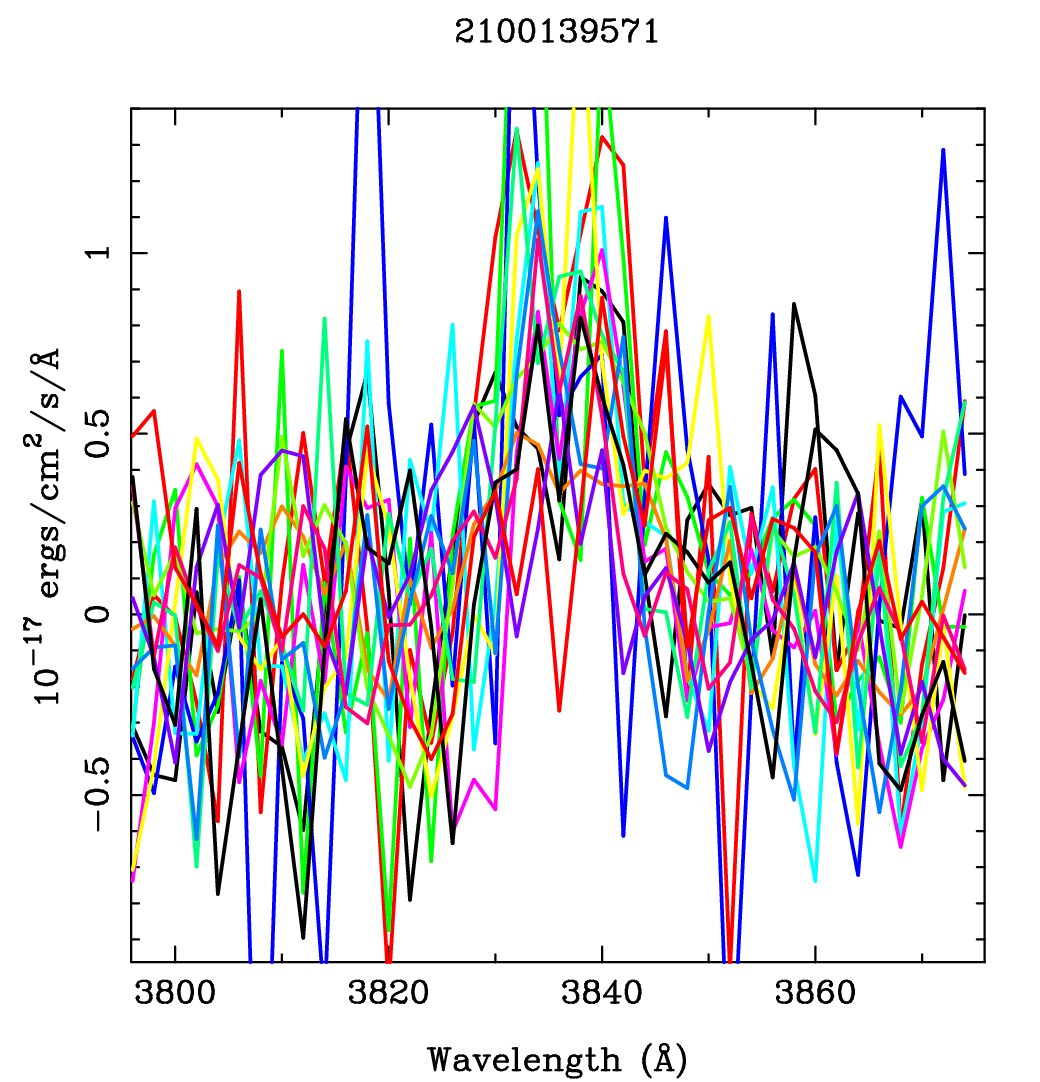}
\includegraphics[height=264pt]{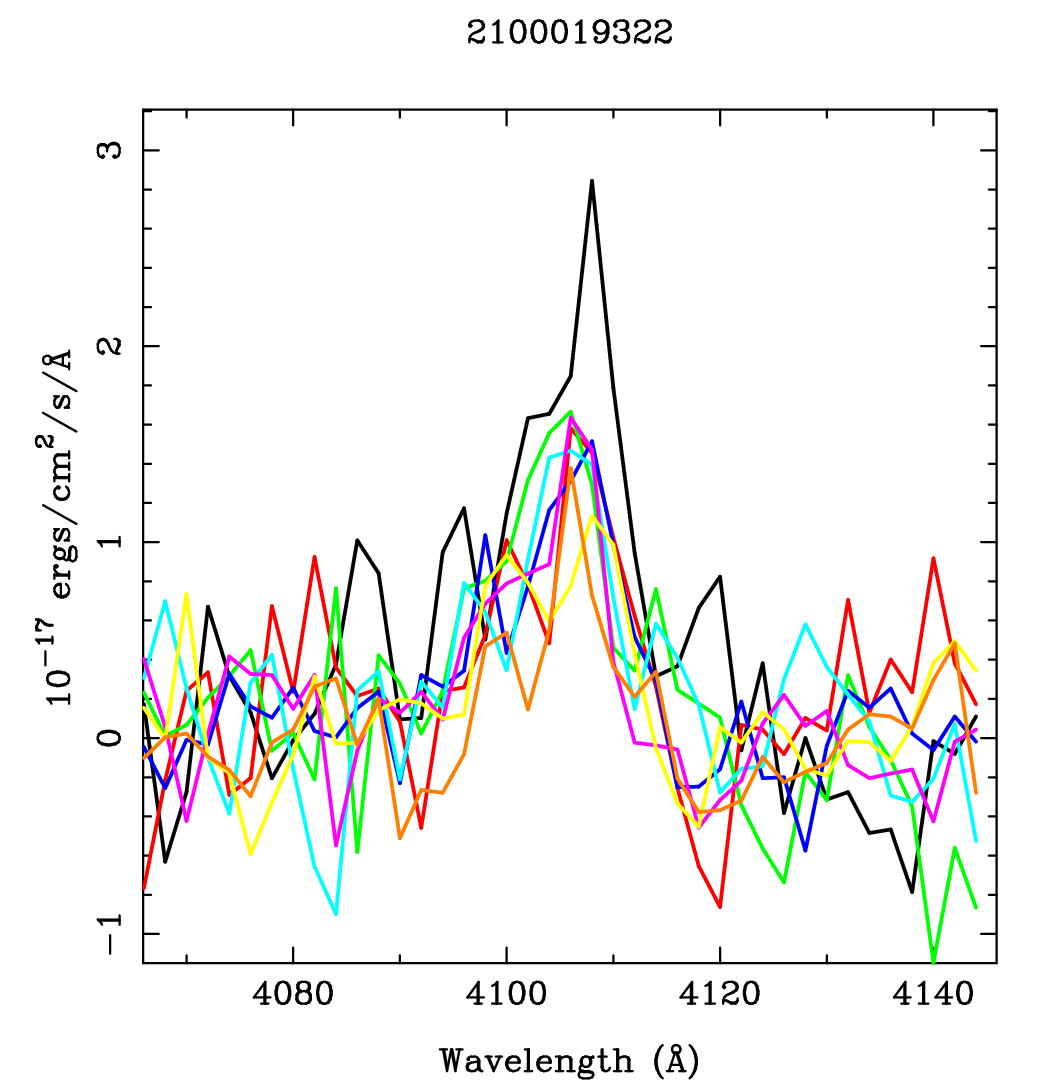}
\caption{Flux versus wavelength for example detections seen in repeat exposures for two different emission lines. Each color represents a different exposure set. All observing sets are plotted, but only a subset make it into the catalog.}
\label{fig:detection1}
\end{figure*}

At present, our false positive rate relies primarily on sources with repeat measurements. If a S/N=5 source in one field also has a significant detection in another observation in a different part of the array, it is clearly real. While most HETDEX fields are observed only once, there are three sets of fields where we intentionally repeated observations (up to 26 times!).  Moreover, there are also several regions of the sky which have overlapping fibers caused by our field placements when the array of IFUs was still incomplete. By cross-matching the one million sources in the HDR2 catalog with the 0.5 billion spectra taken to date, we can identify a significant set of overlapping observations.  

There are several fields dedicated to helping determine the false positive rate. These include deep fields and dedicated repeat fields.  In addition, as mentioned above, we also have a set of observations where some of the IFUs overlapped the positions of previous observations.  These serendipitous fields dominate in terms of numbers used for the false positive rate, but all data (i.e., dedicated repeats, deep fields, and serendipitous overlaps) are considered in the analysis. 

The deep fields used for the calculation of the false position rate are in GOODS-N\null. In these two fields, the exposure times four times longer than normal. One those fields was taken in moderately good conditions (image quality and transparency), and has flux limit is $1.8 \times$ deeper than most HETDEX data. The other field had poorer conditions, with a flux limit that is $\sim 1.4\times$ fainter than normal. The number of sources that overlap with these two deep fields is small, around 100, limiting its use for the false positive rate analysis.

In addition to the deep fields, we targeted several regions of the sky for repeat observations.  One of these fields is in SHELA \citep{SHELA} and has twelve $3 \times 6$~min dither sets at the same location (to within a few arcseconds).  In addition, four dedicated repeat fields are in COSMOS \citep{COSMOS}, each with five $3 \times 6$~min dither sets. Each of these visits consists of an observation with the standard HETDEX dither pattern, and was reduced independently of the other observations.  The result is a set of data that allows us to combine observations and compare a deep summed spectrum to its individual components. 

A unique feature of the HETDEX survey is its use of data with large fiber diameters, large gaps between the fibers, moderately-poor image quality, and the limited track time of an observation.  These effects cause the signal-to-noise of a detection to vary by a factor up to 1.5 depending on the image quality and the exact location relative to the dither pattern. For example, a source observed in good conditions and centered on a fiber will be easier to detect than an object whose center is located between three fibers. As a result, an individual deep exposure is not as valuable as a set of repeat observations where the source position moves around relative to the fiber dither pattern. We test this through input sources simulations and real data, and find sources can sometimes be detected using the nominal HETDEX exposure time but not on deeper exposure.  (Because of the way the HET tracks, it is almost impossible to repeat the exact same observation, with all the fibers at their exact same positions on two different nights.)  Thus, it is hard to generate a false positive rate from a single deep exposure. Thus we do not rely on the deep exposures for our false positive rates, but we do use these observations for confirmation when possible.

As stated, the areas of serendipitous overlap contribute most to the false positive study. The main reason is the large numbers of overlapping fibers, allowing us to significantly increase the number of sources in our study. Also, there are advantages associated with the observing conditions, as many of the dedicated repeat fields had strongly varying image quality and sky transparency.  In contrast, most of the fields with serendipitous overlap had observing conditions which were more typical for the HETDEX survey as as a whole.  

There are about 3000 sources that have been covered by more than 10 dither sets, and over 30,000 independent measures of these candidate galaxies. Still, even with this large database, there are difficulties, as a S/N=4.8 source might only appear in a single observation.  Thus, we use four methods to confirm whether a source is real.  The first method stacks all of the exposures to create a single, extremely deep (up to 9-hour) observation. Line fitting procedures can then be applied to the stacked frame to confirm the existence of an emission line. For about 70\% of the 3000 sources, we were able to use this technique to test the reality of a feature flagged by our detection software.

Unfortunately, stacking does not work for all sources due to variations in the observing conditions, the fact that many of the overlapping fibers are near the edges of an IFU, and the issue of luck with the source position (i.e., whether an object is centered on a fiber or at its edge).  Moreover, there are some cases where an emission line is very obvious in one subset of observations, but not in the overall stacked spectrum.  Thus, a second approach is to confirm the existence of an object by visually inspecting all the spectra of a given source.  This method implicitly takes observing conditions into account and uses a reviewer's experience with data to help with the classification. 

The third approach, which we performed for about 10\% of the sample, is to visually inspect not only the one-dimensional spectrum of each object, but the spectrum produced by each individual fiber in each individual observation.  This time-consuming procedure allowed us to better understand how the counts of an emission line are built up and to better judge the noise inherent in the observation.

The fourth approach to the false detection problem is to compare our HETDEX detections with the results of other surveys that offer higher signal-to-noise and/or extended wavelength coverage. A subset of our data can be investigated in this way, as a number of observations were taken in the COSMOS \citep{COSMOS} and GOODS-N \citep{GOODS} fields, where a substantial amount of archival data exists.  Unfortunately, most of the spectroscopy in this region targeted objects selected by their broadband magnitudes, so the utility of these data for confirming the emission-lines of continuum-faint sources is limited. We are in the process of using follow-up observations to help with the confirmation rate.

At the end of our four-tiered process, we created a list of real emission lines from the set of 3000 original detections and plotted the rate of confirmed detections as a function of the signal-to-noise of the detection. This curve is shown in Figure~\ref{fig:false} and provides our current best estimate for the false positive rate.  We note that these tests can only confirm whether a source is real; it is extremely difficult to prove that a source is false. Thus, the curve presented in the figure represents an upper limit to the rate of false positives; we call this measure the rate for unconfirmed sources. As pointed out in \S\ref{sec:requirements}, the HETDEX science requirements state that the fraction of false detections must be less than 10\%; as Figure~\ref{fig:false} shows, we currently reach this level at a signal-to-noise around 5.

\begin{figure*}
\includegraphics[width=500pt]{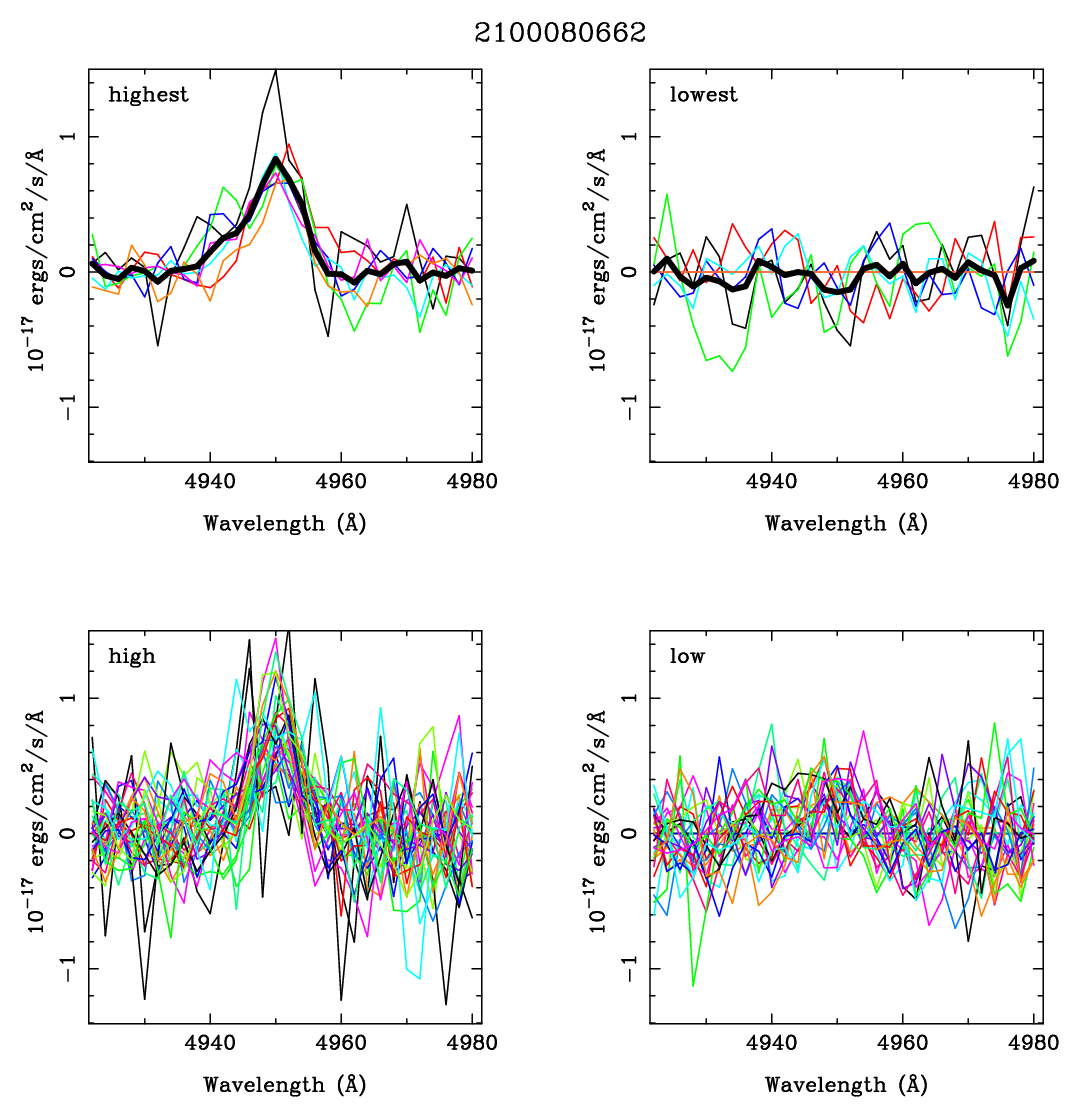}
\caption{Flux versus wavelength for the individual fiber spectra of an LAE with repeat HETDEX observations. Each line in all four panels is a different fiber spectrum. For this source there are 12 repeated frames, with a total of 78 individual fibers. We sort the fibers by summed flux in the central 4 bins, with the top-left panel containing the 7 fibers with the largest flux, the top-right panel displaying the 7 fibers with the lowest flux, and the bottom panels showing the intermediate cases, with the larger fluxes on the left and the lower fluxes on the right. There are some fibers that are flagged and set to zero, and these are ignored. The thick black lines in the top two panels represent the averaged spectra for those fibers.}
\label{fig:detection2}
\end{figure*}

Figure~\ref{fig:detection1} shows two examples of sources that have repeat observations. The lines with color represent the spectra from the individual observations. All overlapping observations are considered, but only a subset of these make it into the final catalog, since some have too low a signal-to-noise. These sources are flagged as LAEs.

We can further examine these data using the spectra of individual fibers. Figure~\ref{fig:detection2} shows an example of a source observed multiple times, where we plot the spectrum of every fiber within the extraction radius.  Many of the repeat sources have a distribution similar to that displayed.  We use the distribution of central fluxes over all fibers to help determine the statistical significance of the source being real. For this case, the line is obviously real.

There are two pieces of information we obtain from these multiple exposures. First, the summed spectrum provides the signal-to-noise necessary to differentiate against false detections. In the case of Figure~\ref{fig:detection2}, the source is considered real. Second, the detection probability, compared to that determined from the simulations, confirms the applicability of the simulations for computing our flux limits.   For example, for this source shown in the figure, the simulations give a probability of detection of 30\%, based on the measured counts in one dither set.  Thus we expect to detect this source on about one-third of the 12 dither sets.  Note that for the source shown in Figure~\ref{fig:detection2}, the emission line is resolved.

As the survey continues, there will be additional dedicated repeat observations and a significant number of additional fibers with overlapping spectra. We will continue to explore the false positive rate with these on-sky studies.

\section{Simulations, Completeness, and Flux Limits}
\label{sec:data-modeling}

An essential requirement for our cosmological measurements is to understand the completeness rate of a source of a given flux over each spectral and spatial resolution element in the survey. We measure completeness using input source simulations combined with an approximation of the noise properties for each element. 

Ideally, we would run a simulation for each spectral and spatial element in the survey, using a range of input source fluxes.  However, the amount of cpu time needed for this type of study is prohibitive and in the end not necessary.  Instead, our procedure is to select a set of observations that span a range of observing conditions, run an extensive suite of input source simulations on these data, and measure an approximate recovery rate using the appropriate noise models.  This approximation requires a trivial amount of cpu, and can be applied to all spectral and spatial elements within an observation.

As our focus is on high-redshift Ly$\alpha$ emitting galaxies, our source simulations use point sources; numerous studies have shown that, although Ly$\alpha$ emission from high-redshift sources can be extended, in the vast majority of objects, virtually all the emission fits within the 3\arcsec\  extraction aperture of the HETDEX observations \citep[e.g.,][]{guaita+15, momose+16, wisotzki+16, ouchi+20}.  Our choice of spectral line-widths is empirical:  we use the average of our detections.  Simulations with a range of line-widths show that the use of a fixed line-width for object detection does not bias our recovery fraction, as long as the true widths are less than $\sim 20$\,\AA\ FWHM.

We have identified 20 observations that span the range of observing conditions applicable to HETDEX\null. These observations form the basis for the simulations and subsequent calibration of the HETDEX flux limits. The image quality of these data ranges from 1.21\arcsec\ to 2.46\arcsec, and the throughput as measured at 4540~\AA\ varies between 0.105 and 0.184. These ranges span the majority of the data.


\begin{figure}
\includegraphics[width=242pt]{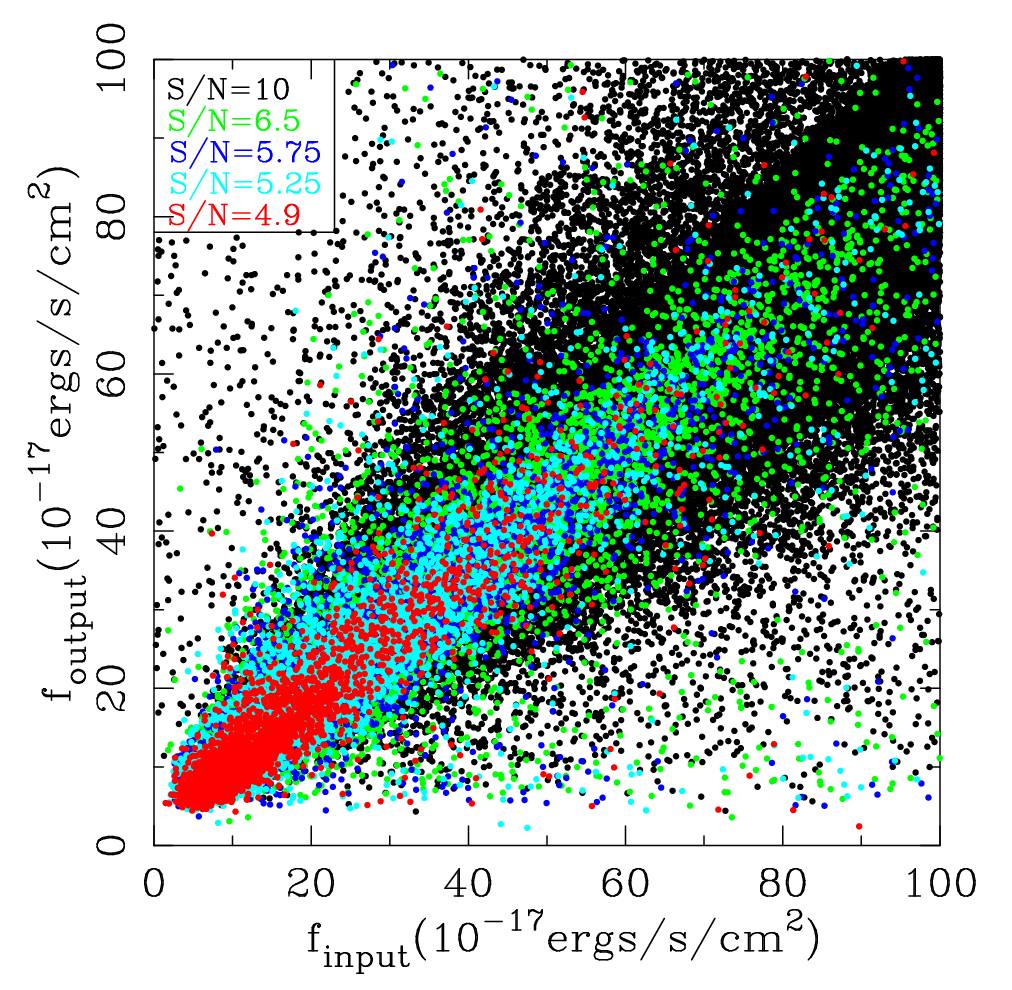}
\caption{Input versus output fluxes for the million simulated sources in one of the HETDEX science verification fields (20200428v020). The different colors represent the S/N of the detected source, with central S/N values of 4.9, 5.25, 5.75, 6.5, and 10 in red, light blue, dark blue, green, and black respectively.}
\label{fig:simul1}
\end{figure}

\begin{figure}
\includegraphics[width=240pt]{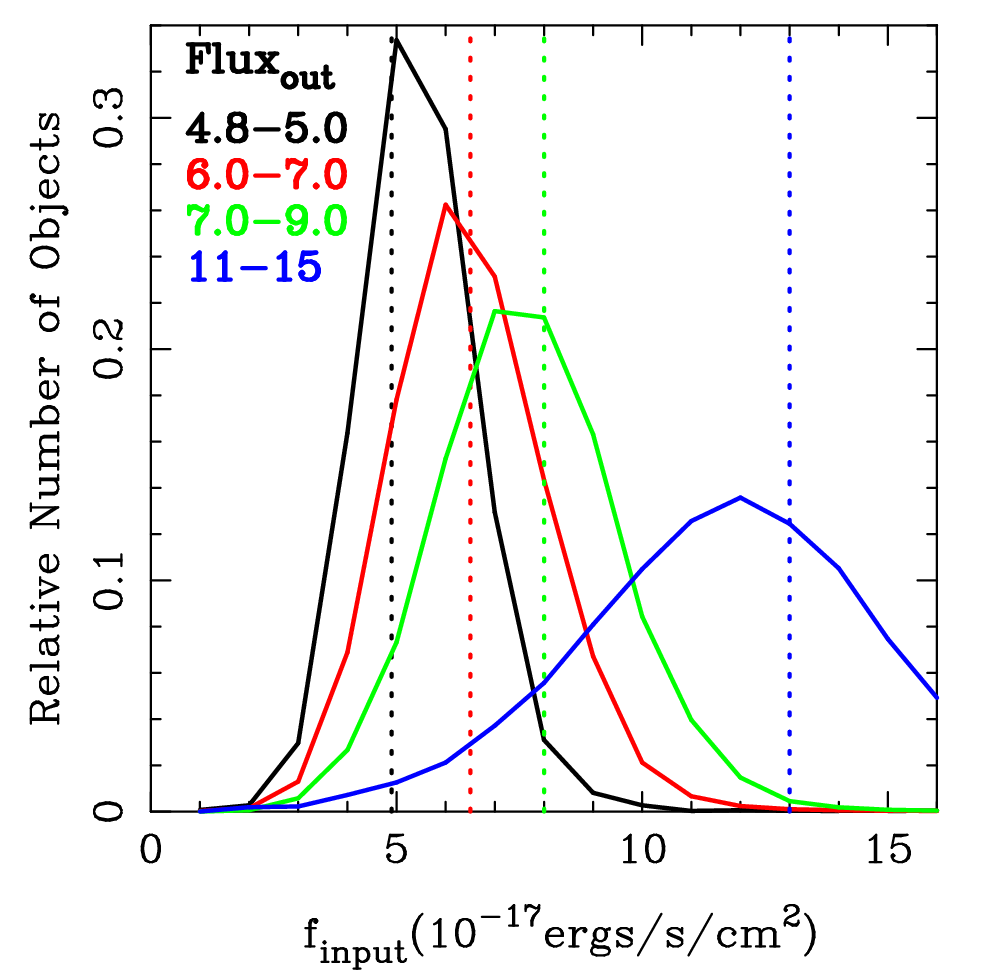}
\caption{Distribution of input fluxes (solid lines) for a given measured flux (vertical dotted lines). These distributions come from the 20 million simulated sources used in the science verification fields and are horizontal cuts through Fig.~\ref{fig:simul1}. The curves are all for the same S/N range of 5.5--6.0; the different colors represent the different measured fluxes, with the dotted lines representing the measured (output) flux. Black represents a measured line flux of 4.8--5.0, red is 6.0--7.0, green is 7.0--9.0, and blue is 11--15, in units of $10^{-17}$~ergs~s$^{-1}$~cm$^{-2}$.}
\label{fig:finfout}
\end{figure}

\begin{figure*}
\includegraphics[width=\textwidth, trim={0 10.3cm 0 50pt}, clip]{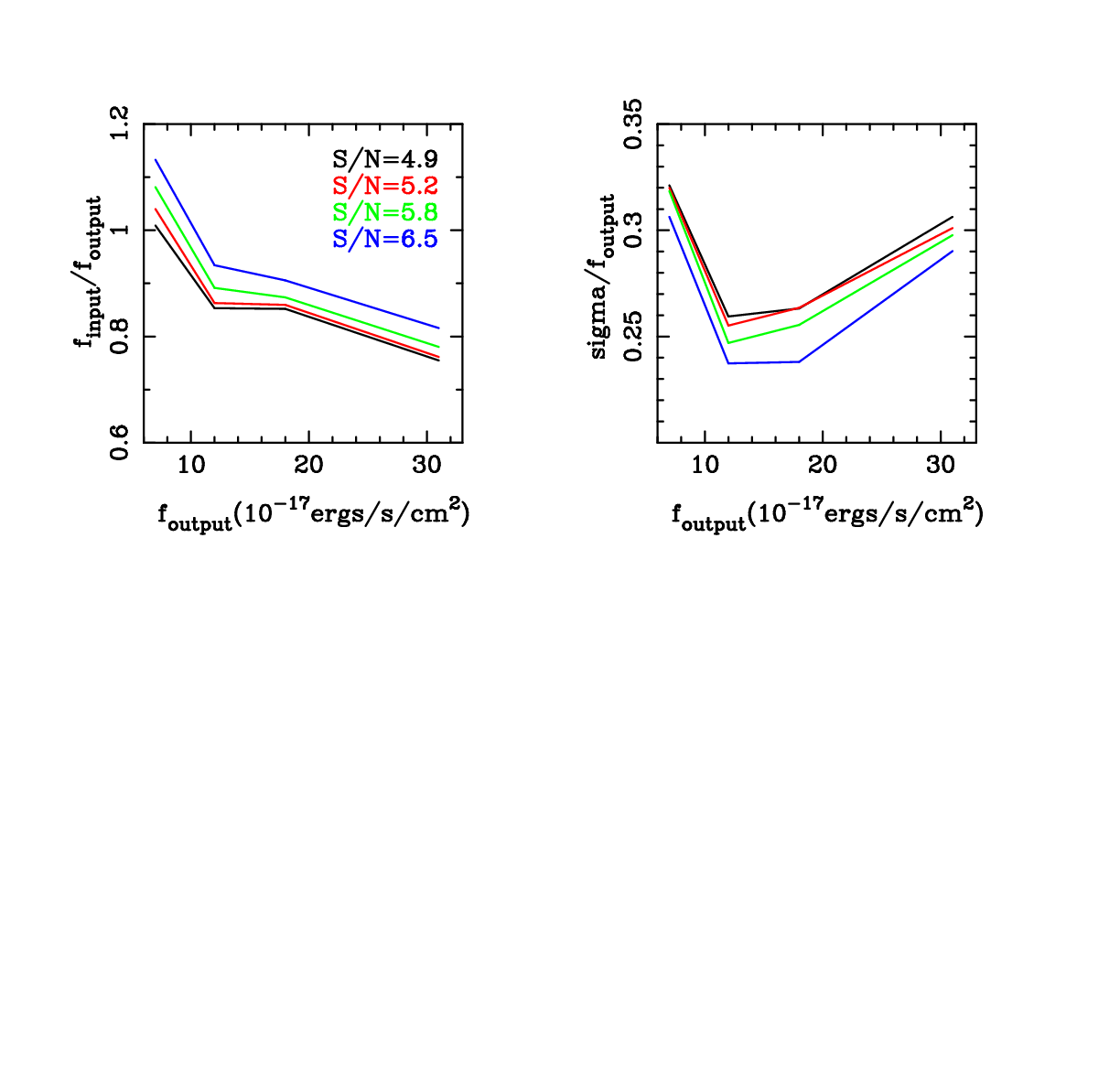}
\caption{Left: the mean ratio of input flux to the measured flux for our artificial emission-line simulations. The curves represent detections with a signal-to-noise of 4.9 (black), 5.2 (red), 5.8 (green), and 6.5 (blue). We use these curves to correct the output fluxes. Right: the relative scatter about the mean relations shown in the left panel with the same color-coding.  The dispersion is higher than the statistical uncertainty of the measurements caused (primarily) by uncertainties in the source positions.}
\label{fig:simul3}
\end{figure*}

\subsection{Simulations}
\label{subsec:simulations}

We simulate Ly$\alpha$ emitters by adding flux into the actual fiber-spectra database. We then use our HETDEX detection algorithms to generate a mock catalog of the sources. (Of course, since we are adding flux to the actual data, this catalog will also contain the list of real sources.)  Since there are three dithers, each with its own normalization, we take care to match both the PSF and spatial locations for the fibers. 

For each IFU, we generate a set of artificial emission lines that linearly span the flux range between 0 and $100 \times 10^{-17}$~erg~cm$^{-2}$~s$^{-1}$ and the wavelength range between 3500 and 5500\,\AA\null.  To avoid losing objects due to source crowding, we only add 200 objects at a time to a single frame\null. (For comparison, a frame will typically contain $\sim 5$ real sources.) We then run the simulation 66 times to produce a sample of 13,200 artificial emission-lines per frame.  Since HETDEX employs over 70 active IFUs, the simulation of one observation therefore contains close to one million artificial sources.

Once the simulated frames are created, we run the same line-detection algorithms that are used on the real HETDEX data.  Any source in the output catalog that is within 2\arcsec\ and 3\,\AA\ of an input position is considered to be a match. For each match, we track the difference in flux, wavelength, spatial location, and signal-to-noise;  we do not explore these trends on an IFU by IFU basis, but instead analyze the integrated results over all IFUs. 

The simulations determine the maximum precision attainable by our HETDEX observations, in terms of wavelength, flux, and completeness.  As expected, the wavelength accuracy is a function of signal-to-noise:  for emission lines measured with S/N bins of 4.8--5.2, 5.2--5.5, 5.5--6.5, and $>$6.5, the recovered wavelengths agree with the input values to within 0.63\,\AA, 0.56\,\AA, 0.49\,\AA, and 0.25\,\AA, respectively.  All are well within the required HETDEX specifications.

Figures~\ref{fig:simul1} and \ref{fig:finfout} show how our measured emission-line fluxes compare with the input values of the simulation.  Once again, we break the data down by the signal-to-noise measured by our detection algorithms. Both figures demonstrate that there is a bias and a scatter in the relation. Figure~\ref{fig:finfout} provides a horizontal cut in Figure~\ref{fig:simul1} at various output fluxes. The vertical dotted lines show the output flux, and the corresponding color shows the distribution of input fluxes that generate that output value. We use these distribution directly when estimating the expected input flux values.

These trends are better seen in Figure~\ref{fig:simul3}; the left-hand panel shows the systematic bias in the measured fluxes, while the right-hand figure shows the scatter about these lines, both as a function of measured flux and measured signal-to-noise.  The observed scatter is greater than the statistical error due primarily to uncertainties in the object centroids -- with prior knowledge of the centroid, the scatter decreases considerably. We typically measure an input flux to about 27\% accuracy.

\subsection{Completeness and Flux Limits}
\label{subsec:completeness}

The input sources that are not found in the output catalogs define the recovery rate.  We track these sources as a function of input source flux and as a function of wavelength.

\begin{figure*}
\includegraphics[width=\textwidth]{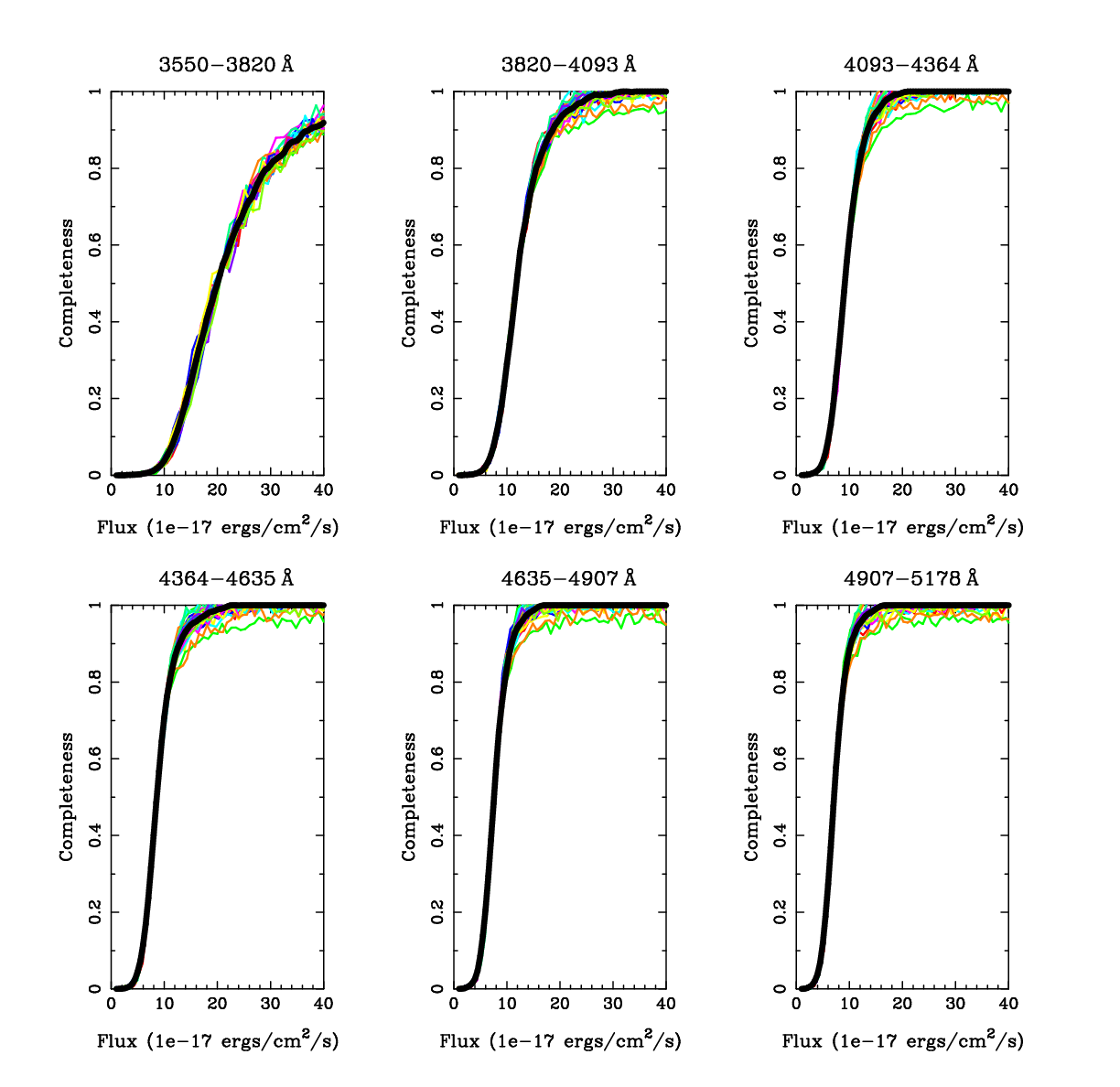}
\caption{Completeness curves from the simulations, where each panel is for a different wavelength region. The lines with color are from the individual calibration fields, and the thick black line is the average curve used to represent the completeness corrections. In the extreme blue (top left panel), we do not reach 100\% completeness until the flux is greater than $40 \times 10^{-17}$~ergs~cm$^2$~s$^{-1}$.}
\label{fig:simul2}
\end{figure*}

Figure~\ref{fig:simul2} shows the completeness corrections for all 20 of the science verification fields. Each curve has been scaled according to its 50\% completeness value. The recovery fractions are remarkably consistent:  while each field may have a different depth, and the shapes of the curves vary with wavelength, at a given wavelength, the curves for all 20 fields are virtually identical.   We can therefore average the data to provide the final shape for the completeness function.  This is given as the thick black line in the figure.   The completeness function at any location can be found from these curves, along with the region's 50\% completeness limit.

We estimate the 50\% recovery fractions from the noise properties of each spatial and spectral element.  It is not feasible to simulate this limit for every spatial and spectral element in our survey, as the observations contain over $10^{11}$ such elements. Instead, we generate an extracted spectrum based on an observation's PSF and sum the noise using the central four spectral pixels (i.e., a tophat function). This sum provides the $1\sigma$ noise model at a given spatial and spectral position. This approximation is then calibrated against the actual detection algorithm, where we use the source simulations to determine the scaling between the two methods. For example, if we want to determine the recovery curve for a sample using a S/N cut of 5.0, the 50\% completeness value is given by $4.6\times$ the $1\sigma$ noise. The value of 4.6 is determined by comparing the $1\sigma$ noise of the four summed spectral pixels to that found from the source simulation's 50\% completeness limit. We obtain a smaller value than a simple multiplication of 5.0 since emission lines are fitted to the instrumental spectral profile, as opposed to a tophat function of the $1\,\sigma$ noise model.

For each IFU, we define a grid from $-$26\arcsec\ to +26\arcsec\ on 2\arcsec\ spacing, and measure the flux limit as a function of wavelength and position, as described above. We call these models flux limit cubes. Edge effects are naturally included since the number of fibers decreases near the IFU edges and, hence, the noise there is larger. This technique basically makes a fuzzy IFU, which is primarily determined by the image quality of that observation.

We provide a grid of (RA,DEC) for each IFU, each with a flux limit curve defined in 2\,\AA\ wavelength bins. These grids, combined with the completeness curves, are inputs to the window function estimates used for the cosmological analysis. In addition to the flux limits derived from these grids, we can also measure flux limits at any specific spatial and spectral position. Depending on the accuracy and computing efficiency, we use one or the other.

\begin{figure*}
\hspace*{-0.9cm}\includegraphics[width=300pt, trim={0 0 0 8cm}]{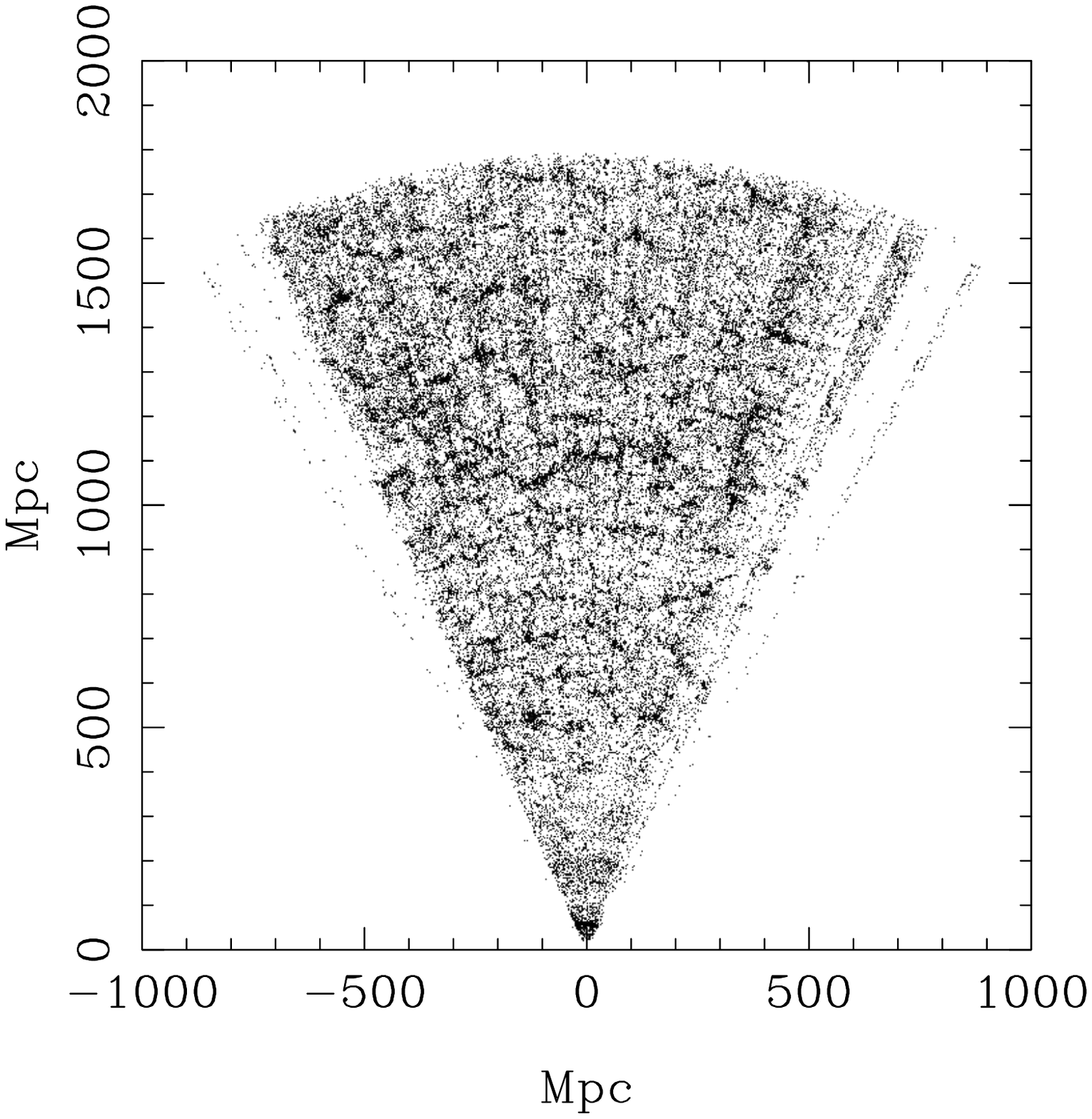}
\hspace*{-1.7cm}\includegraphics[width=300pt, trim={0 0 0 8cm}]{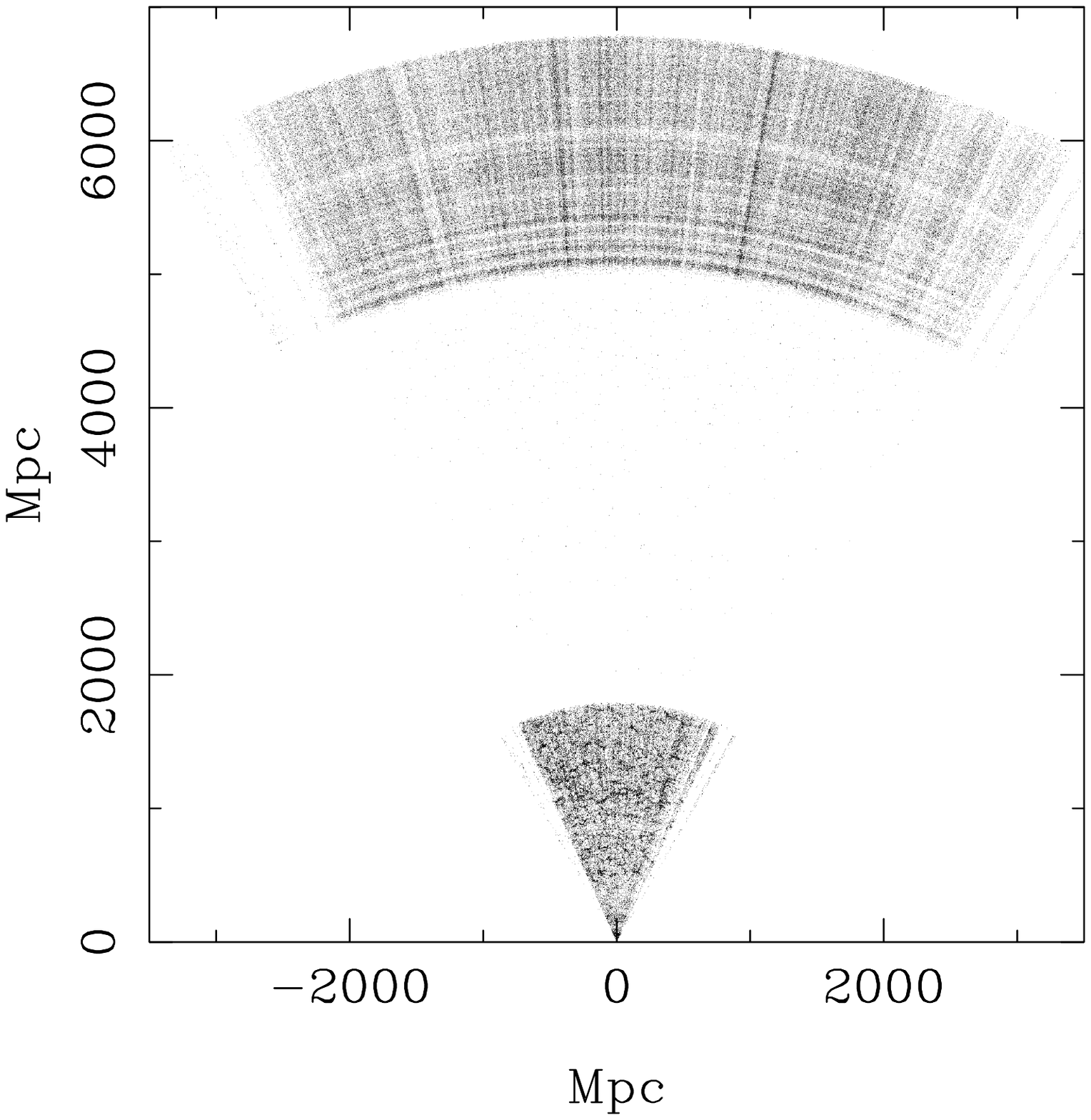}
\caption{Redshift wedge diagrams for galaxies recorded in HDR2\null.  The left panel  displays the distribution of $z < 0.5$ galaxies between $50^\circ < \delta < 52^\circ$, detected principally via their emission at \OII\ $\lambda 3727$; the right panel covers the declination range from $45^\circ < \delta < 55^\circ$ and plots both \OII\ emitters with $z < 0.5$ and the LAEs between $1.88 < z < 3.52$.  The (few) points in the redshift gap come from the identification of other emission lines. The range in right ascension runs from $155^\circ$ to $245^\circ$.  The radial features (over 
and under densities) are due to the incomplete spatial coverage of HDR2; the angular features (seen easily in the distant galaxies in the right panel) are due to variations in survey depth due to sky lines.  Both of these issues are understood and included in the flux limit cubes.}
\label{fig:slice}
\end{figure*}

\section{Line Identification}
\label{sec:line-identification}

This paper is focused on describing the pipeline used to generate the HETDEX emission line catalog. An important next step is to determine the line identification. \cite{leung+17} provide our initial basis for line discrimination using a Bayesian analysis with the equivalent width distributions of Ly$\alpha$ and \OII\ emitting galaxies. Davis et al.\ 2021 (in preparation) and Mentuch Cooper et al.\ 2021 (in preparation) describe the details of the line discrimination and various tests. A Line identification relies on multiple pieces of information, including defining an imaging counterpart, the presence (or absense) of additional features in the spectrum, the measurement of the continuum flux density in the HETDEX spectrum or ancillary imaging data, and the use of line expectations based on known distributions of equivalent widths \citep[e.g.,][]{gronwall+07, ciardullo+13}.

The primary method for classifying a HETDEX emission line as either Ly$\alpha$ or \OII\ $\lambda 3727$ is through the identification of an imaging counterpart and estimating the line's equivalent width. As discussed in \cite{leung+17} and \cite{ouchi+20}, a rest-frame equivalent width around 20\,\AA\ generally works well as a discriminant, though the optimum cut does change with redshift due to the evolution of both lines' equivalent width distributions \citep{ciardullo+12, ciardullo+13}.  Other lines that can cause confusion in our redshift determinations include \ion{Mg}{2} $\lambda 2799$, [\ion{O}{3}] $\lambda 5007$, \ion{C}{4} $\lambda 1550$, and \ion{C}{3}] $\lambda 1909$, and these features are considered and included in our line identification software.

Determination of the imaging counterpart also has uncertainties, due to the physical offset of an emission from the position of the host galaxy, astrometric uncertainties for the spatial position of the line, and most importantly, crowding. We study all imaging counterparts within 3\arcsec, and apply appropriate weighting given the object size, image quality of the observation, line flux, and wavelength. Fortunately the HETDEX continuum level for a 20-minute exposure is around $g=24$; while this does not allow us to detect the continuum emission from most LAEs, it does enable us to measure the continuum flux density of most nearby galaxies.  Thus, HETDEX continuum measurements serve as a strong check on the use of imaging catalogs for determining  the counterparts of emission lines.

Our required specification is to have no more than 2\% of objects classified as Ly$\alpha$ emitters to be \OII\ galaxies; though the other lines have similar criteria, they are not significant contaminants in our survey.  Our tests for line identification are presented in Davis et al.\ 2021 and Mentuch Cooper et al.\ 2021 (both in preparation).

The most secure method of confirming our line identifications is to compare the redshifts of other surveys to HETDEX redshifts.  We gather these redshifts from many published catalogs, and COSMOS \citep{COSMOS} is particularly useful. These comparisons are discussed in detail in Mentuch Cooper et al.\ (2021, in preparation). As a summary, we currently are using over 3800 sources with published redshifts. From this initial comparison, the contamination rate of LAEs by \OII\ galaxies is 4\%. Our target goal is 2\%, and this is easily achieved via slight modifications to our algorithm. However, before we apply any changes to our algorithm, we must vet the published spectroscopic catalogs, since those data have their own mis-classifications. This work is time consuming and will be publish separately (Mentuch Cooper et al.\ 2021, in preparation). For those sources in agreement (i.e., the 96\%) the redshift accuracy is excellent, and within our specification of under 180~km~s$^{-1}$.

\section{Summary}
\label{sec:summary}

A primary goal of HETDEX is to study the large scale clustering of LAEs at $1.88<z<3.52$ in order to measure the effect of dark energy. This paper describes the reduction and analysis procedures that will allow us to measure the Hubble expansion rate and angular diameter distance to better than 1\% accuracy. This level of precision is comparable with that expected from the ongoing measurements of the late-time expansion rate. Thus, the combination of all these experiments will provide a full picture of any potential evolution on dark energy, with HETDEX providing the high-redshift anchor.

Figure~\ref{fig:slice} shows redshift slices for the current sample described in this paper (and discussed in Mentuch Cooper et al.\  2021, in preparation). The scale is in Mpc, with the left panel displaying \OII\ emitting galaxies out to $z=0.5$, and the right panel showing the full radial extent of the survey using both the distant LAEs and the nearby \OII\ emitters.  The (few) galaxies in the redshift gap are objects identified via other emission lines.  There are a number of features seen in these redshift slices which fold back into our measurement of clustering.  For example, the radial features (the spokes) seen in both panels are due to non-uniform coverage of our observations to date; at the end of survey, we expect these spokes to be filled in. Similarly, the wedge diagrams also contain some regions which are missing data due the presence of large nearby galaxies (such as M101) and the scattered light from very bright stars, such as Mizar, Alcor, and three other 2nd magnitude stars of the Ursa Major constellation.  Obviously, these holes will remain, even when the survey is complete. Finally, the angular arc-like features seen in the wedge-diagram of the distant universe are due to variations in the survey depth caused by the presence of weak emission-lines in the sky. Since the line fluxes of LAEs are generally fainter than those of the \OII\ emitters, their detection is more sensitive to the existence of these features. These variations are tracked and calibrated out by the flux-limit analysis, but are not removed in this visualization.    

The large scale clustering of nearby ($z < 0.5$) sources is easily seen in the left panel of Fig.~\ref{fig:slice}.  The clustering at $z > 1.9$ is less obvious, due to the greater dependence on the observing conditions, the lower number density of LAEs, and the fact that high-$z$ galaxies are less clustered than local objects. Nevertheless, the example demonstrates that by using the flux limit analysis techniques presented in this paper, we can quantify the clustering of $1.88 < z < 3.52$ LAEs to high precision.

The science requirements to reach our goal are outlined in \S\ref{sec:requirements}. All design elements in the instrumentation and in the data analysis have been coordinated to reach these specifications, and this paper describes the reduction steps. The most difficult aspect is obtaining an adequate number of LAEs while keeping the number of false positives and line mis-identifications to a minimum. We show in Fig.~\ref{fig:ndets} that the current catalog of emission line sources reaches our specification. We obviously need to further demonstrate that these sources are indeed LAEs, and that we are finding them at a rate consistent with other studies. \cite{Zhang2021} provides our first LAE luminosity function as measured from HETDEX. While the focus of that paper is on AGNs, they include analysis of a large sample of LAEs, and demonstrate that the LAE luminosity function is consistent with previously published results. That sample contained 18,000 galaxies, a small fraction of the current catalog, and additional luminosity functions will be published.

For the false positive rate, the noise model is particularly important and must be well-calibrated.  We use a combination of repeat observations and simulations for this work, and while additional work is required, \S\ref{subsec:false-detections} shows that we are close to specification. The line mis-identification rate is best studied using known sources or follow-up with other spectrographs. As discussed in \S\ref{sec:line-identification}, we are already at 4\% contamination, and this number is likely to improve. Additionally \cite{grasshorn+19}, \cite{awan2020}, and \cite{farrow+21} outline techniques to measure and/or include the contamination from \OII\ from the galaxy power spectrum itself. Our target requirement remains a $3\,\sigma$ detection of the dark energy density at $z=2.4$, assuming a cosmological constant. There are physical and systematic unknowns in both the LAE properties (namely, the bias value and non-linear effects) and these could increase (or decrease) the uncertainties on the cosmological parameters. Our primary mitigation strategy against increased uncertainties is observing time. We designed the spring field to allow for a 30\% increase in LAE number density, if required. As we obtain more data and improve the analysis, and if a compelling theoretical model for dark energy emerges, we will determine whether additional time is warranted.

\acknowledgements

The dedication, innovation, and expertise of the staff of the Hobby-Eberly Telescope played a crucial role in the success of these observations. KG acknowledges support for this work from NSF-2008793.  SLF acknowledges support from NSF-1908817. M.K. acknowledges support by DFG grant KR 3338/4-1.

HETDEX is led by the University of Texas at Austin McDonald Observatory and Department of Astronomy with participation from the Ludwig-Maximilians-Universit\"at M\"unchen, Max-Planck-Institut f\"ur Extraterrestrische Physik (MPE), Leibniz-Institut f\"ur Astrophysik Potsdam (AIP), Texas A\&M University, The Pennsylvania State University, Institut f\"ur Astrophysik G\"ottingen, The University of Oxford, Max-Planck-Institut f\"ur Astrophysik (MPA), The University of Tokyo, and Missouri University of Science and Technology. In addition to Institutional support, HETDEX is funded by the National Science Foundation (grant AST-0926815), the State of Texas, the US Air Force (AFRL FA9451-04-2-0355), and generous support from private individuals and foundations.

The Hobby-Eberly Telescope (HET) is a joint project of the University of Texas at Austin, the Pennsylvania State University, Ludwig-Maximilians-Universit\"at M\"unchen, and Georg-August-Universit\"at G\"ottingen. The HET is named in honor of its principal benefactors, William P. Hobby and Robert E. Eberly.

The authors acknowledge the Texas Advanced Computing Center (TACC) at The University of Texas at Austin for providing high performance computing, visualization, and storage resources that have contributed to the research results reported within this paper. URL: http://www.tacc.utexas.edu

The Institute for Gravitation and the Cosmos is supported by the Eberly College of Science and the Office of the Senior Vice President for Research at the Pennsylvania State University. The Kavli IPMU is supported by World Premier International Research Center Initiative (WPI), MEXT, Japan. 

This work makes use of the Sloan Digital Sky Survey IV, with funding provided by the Alfred P. Sloan Foundation, the U.S. Department of Energy Office of Science, and the Participating Institutions. SDSS-IV acknowledges support and resources from the Center for High-Performance Computing at the University of Utah. The SDSS web site is www.sdss.org.

This work makes use of the Pan-STARRS1 Surveys (PS1) and the PS1 public science archive, which have been made possible through contributions by the Institute for Astronomy, the University of Hawaii, the Pan-STARRS Project Office, the Max-Planck Society and its participating institutes.

This work makes use of data from the European Space Agency (ESA) mission {\it Gaia} (\url{https://www.cosmos.esa.int/gaia}), processed by the {\it Gaia} Data Processing and Analysis Consortium (DPAC, \url{https://www.cosmos.esa.int/web/gaia/dpac/consortium}). Funding for the DPAC has been provided by national institutions, in particular the institutions participating in the {\it Gaia} Multilateral Agreement.

This work makes use of observations made with the NASA/ESA Hubble Space Telescope obtained from the Space Telescope Science Institute, which is operated by the Association of Universities for Research in Astronomy, Inc., under NASA contract NAS 5–26555. 


\clearpage
\bibliography{survey.bib}

\begin{thebibliography}{}
\expandafter\ifx\csname natexlab\endcsname\relax\def\natexlab#1{#1}\fi
\providecommand{\url}[1]{\href{#1}{#1}}

\bibitem[{{Abazajian} {et~al.}(2009){Abazajian}, {Adelman-McCarthy},
  {Ag{\"u}eros}, {Allam}, {Allende Prieto}, {An}, {Anderson}, {Anderson},
  {Annis}, {Bahcall}, {Bailer-Jones}, {Barentine}, {Bassett}, {Becker},
  {Beers}, {Bell}, {Belokurov}, {Berlind}, {Berman}, {Bernardi}, {Bickerton},
  {Bizyaev}, {Blakeslee}, {Blanton}, {Bochanski}, {Boroski}, {Brewington},
  {Brinchmann}, {Brinkmann}, {Brunner}, {Budav{\'a}ri}, {Carey}, {Carliles},
  {Carr}, {Castander}, {Cinabro}, {Connolly}, {Csabai}, {Cunha}, {Czarapata},
  {Davenport}, {de Haas}, {Dilday}, {Doi}, {Eisenstein}, {Evans}, {Evans},
  {Fan}, {Friedman}, {Frieman}, {Fukugita}, {G{\"a}nsicke}, {Gates},
  {Gillespie}, {Gilmore}, {Gonzalez}, {Gonzalez}, {Grebel}, {Gunn},
  {Gy{\"o}ry}, {Hall}, {Harding}, {Harris}, {Harvanek}, {Hawley}, {Hayes},
  {Heckman}, {Hendry}, {Hennessy}, {Hindsley}, {Hoblitt}, {Hogan}, {Hogg},
  {Holtzman}, {Hyde}, {Ichikawa}, {Ichikawa}, {Im}, {Ivezi{\'c}}, {Jester},
  {Jiang}, {Johnson}, {Jorgensen}, {Juri{\'c}}, {Kent}, {Kessler}, {Kleinman},
  {Knapp}, {Konishi}, {Kron}, {Krzesinski}, {Kuropatkin}, {Lampeitl},
  {Lebedeva}, {Lee}, {Lee}, {French Leger}, {L{\'e}pine}, {Li}, {Lima}, {Lin},
  {Long}, {Loomis}, {Loveday}, {Lupton}, {Magnier}, {Malanushenko},
  {Malanushenko}, {Mandelbaum}, {Margon}, {Marriner}, {Mart{\'\i}nez-Delgado},
  {Matsubara}, {McGehee}, {McKay}, {Meiksin}, {Morrison}, {Mullally}, {Munn},
  {Murphy}, {Nash}, {Nebot}, {Neilsen}, {Newberg}, {Newman}, {Nichol},
  {Nicinski}, {Nieto-Santisteban}, {Nitta}, {Okamura}, {Oravetz}, {Ostriker},
  {Owen}, {Padmanabhan}, {Pan}, {Park}, {Pauls}, {Peoples}, {Percival}, {Pier},
  {Pope}, {Pourbaix}, {Price}, {Purger}, {Quinn}, {Raddick}, {Re Fiorentin},
  {Richards}, {Richmond}, {Riess}, {Rix}, {Rockosi}, {Sako}, {Schlegel},
  {Schneider}, {Scholz}, {Schreiber}, {Schwope}, {Seljak}, {Sesar}, {Sheldon},
  {Shimasaku}, {Sibley}, {Simmons}, {Sivarani}, {Allyn Smith}, {Smith},
  {Smol{\v{c}}i{\'c}}, {Snedden}, {Stebbins}, {Steinmetz}, {Stoughton},
  {Strauss}, {SubbaRao}, {Suto}, {Szalay}, {Szapudi}, {Szkody}, {Tanaka},
  {Tegmark}, {Teodoro}, {Thakar}, {Tremonti}, {Tucker}, {Uomoto}, {Vanden
  Berk}, {Vandenberg}, {Vidrih}, {Vogeley}, {Voges}, {Vogt}, {Wadadekar},
  {Watters}, {Weinberg}, {West}, {White}, {Wilhite}, {Wonders}, {Yanny},
  {Yocum}, {York}, {Zehavi}, {Zibetti}, \& {Zucker}}]{SDSS-7}
{Abazajian}, K.~N., {Adelman-McCarthy}, J.~K., {Ag{\"u}eros}, M.~A., {et~al.}
  2009, \apjs, 182, 543

\bibitem[{{Adachi} {et~al.}(2020){Adachi}, {Aguilar Fa{\'u}ndez}, {Arnold},
  {Baccigalupi}, {Barron}, {Beck}, {Bianchini}, {Chapman}, {Cheung}, {Chinone},
  {Crowley}, {Dobbs}, {El Bouhargani}, {Elleflot}, {Errard}, {Fabbian}, {Feng},
  {Fujino}, {Galitzki}, {Goeckner-Wald}, {Groh}, {Hall}, {Hasegawa}, {Hazumi},
  {Hirose}, {Jaffe}, {Jeong}, {Kaneko}, {Katayama}, {Keating}, {Kikuchi},
  {Kisner}, {Kusaka}, {Lee}, {Leon}, {Linder}, {Lowry}, {Matsuda}, {Matsumura},
  {Minami}, {Navaroli}, {Nishino}, {Pham}, {Poletti}, {Reichardt}, {Segawa},
  {Siritanasak}, {Tajima}, {Takakura}, {Takatori}, {Tanabe}, {Teply}, {Tsai},
  {Verg{\`e}s}, {Westbrook}, {Zhou}, \& {Polarbear Collaboration}}]{adachi+20}
{Adachi}, S., {Aguilar Fa{\'u}ndez}, M.~A.~O., {Arnold}, K., {et~al.} 2020,
  \apj, 904, 65

\bibitem[{{Adams} {et~al.}(2011){Adams}, {Blanc}, {Hill}, {Gebhardt}, {Drory},
  {Hao}, {Bender}, {Byun}, {Ciardullo}, {Cornell}, {Finkelstein}, {Fry},
  {Gawiser}, {Gronwall}, {Hopp}, {Jeong}, {Kelz}, {Kelzenberg}, {Komatsu},
  {MacQueen}, {Murphy}, {Odoms}, {Roth}, {Schneider}, {Tufts}, \&
  {Wilkinson}}]{Adams2011}
{Adams}, J.~J., {Blanc}, G.~A., {Hill}, G.~J., {et~al.} 2011, \apjs, 192, 5

\bibitem[{{Addison} {et~al.}(2019){Addison}, {Bennett}, {Jeong}, {Komatsu}, \&
  {Weiland}}]{addison+19}
{Addison}, G.~E., {Bennett}, C.~L., {Jeong}, D., {Komatsu}, E., \& {Weiland},
  J.~L. 2019, \apj, 879, 15

\bibitem[{{Adelman-McCarthy} {et~al.}(2008){Adelman-McCarthy}, {Ag{\"u}eros},
  {Allam}, {Allende Prieto}, {Anderson}, {Anderson}, {Annis}, {Bahcall},
  {Bailer-Jones}, {Baldry}, {Barentine}, {Bassett}, {Becker}, {Beers}, {Bell},
  {Berlind}, {Bernardi}, {Blanton}, {Bochanski}, {Boroski}, {Brinchmann},
  {Brinkmann}, {Brunner}, {Budav{\'a}ri}, {Carliles}, {Carr}, {Castander},
  {Cinabro}, {Cool}, {Covey}, {Csabai}, {Cunha}, {Davenport}, {Dilday}, {Doi},
  {Eisenstein}, {Evans}, {Fan}, {Finkbeiner}, {Friedman}, {Frieman},
  {Fukugita}, {G{\"a}nsicke}, {Gates}, {Gillespie}, {Glazebrook}, {Gray},
  {Grebel}, {Gunn}, {Gurbani}, {Hall}, {Harding}, {Harvanek}, {Hawley},
  {Hayes}, {Heckman}, {Hendry}, {Hindsley}, {Hirata}, {Hogan}, {Hogg}, {Hyde},
  {Ichikawa}, {Ivezi{\'c}}, {Jester}, {Johnson}, {Jorgensen}, {Juri{\'c}},
  {Kent}, {Kessler}, {Kleinman}, {Knapp}, {Kron}, {Krzesinski}, {Kuropatkin},
  {Lamb}, {Lampeitl}, {Lebedeva}, {Lee}, {French Leger}, {L{\'e}pine}, {Lima},
  {Lin}, {Long}, {Loomis}, {Loveday}, {Lupton}, {Malanushenko}, {Malanushenko},
  {Mandelbaum}, {Margon}, {Marriner}, {Mart{\'\i}nez-Delgado}, {Matsubara},
  {McGehee}, {McKay}, {Meiksin}, {Morrison}, {Munn}, {Nakajima}, {Neilsen},
  {Newberg}, {Nichol}, {Nicinski}, {Nieto-Santisteban}, {Nitta}, {Okamura},
  {Owen}, {Oyaizu}, {Padmanabhan}, {Pan}, {Park}, {Peoples}, {Pier}, {Pope},
  {Purger}, {Raddick}, {Re Fiorentin}, {Richards}, {Richmond}, {Riess}, {Rix},
  {Rockosi}, {Sako}, {Schlegel}, {Schneider}, {Schreiber}, {Schwope}, {Seljak},
  {Sesar}, {Sheldon}, {Shimasaku}, {Sivarani}, {Allyn Smith}, {Snedden},
  {Steinmetz}, {Strauss}, {SubbaRao}, {Suto}, {Szalay}, {Szapudi}, {Szkody},
  {Tegmark}, {Thakar}, {Tremonti}, {Tucker}, {Uomoto}, {Vanden Berk},
  {Vandenberg}, {Vidrih}, {Vogeley}, {Voges}, {Vogt}, {Wadadekar}, {Weinberg},
  {West}, {White}, {Wilhite}, {Yanny}, {Yocum}, {York}, {Zehavi}, \&
  {Zucker}}]{SDSS-6}
{Adelman-McCarthy}, J.~K., {Ag{\"u}eros}, M.~A., {Allam}, S.~S., {et~al.} 2008,
  \apjs, 175, 297

\bibitem[{{Aiola} {et~al.}(2020){Aiola}, {Calabrese}, {Maurin}, {Naess},
  {Schmitt}, {Abitbol}, {Addison}, {Ade}, {Alonso}, {Amiri}, {Amodeo},
  {Angile}, {Austermann}, {Baildon}, {Battaglia}, {Beall}, {Bean}, {Becker},
  {Bond}, {Bruno}, {Calafut}, {Campusano}, {Carrero}, {Chesmore}, {Cho},
  {Choi}, {Clark}, {Cothard}, {Crichton}, {Crowley}, {Darwish}, {Datta},
  {Denison}, {Devlin}, {Duell}, {Duff}, {Duivenvoorden}, {Dunkley},
  {D{\"u}nner}, {Essinger-Hileman}, {Fankhanel}, {Ferraro}, {Fox}, {Fuzia},
  {Gallardo}, {Gluscevic}, {Golec}, {Grace}, {Gralla}, {Guan}, {Hall},
  {Halpern}, {Han}, {Hargrave}, {Hasselfield}, {Helton}, {Henderson},
  {Hensley}, {Hill}, {Hilton}, {Hilton}, {Hincks}, {Hlo{\v{z}}ek}, {Ho},
  {Hubmayr}, {Huffenberger}, {Hughes}, {Infante}, {Irwin}, {Jackson}, {Klein},
  {Knowles}, {Koopman}, {Kosowsky}, {Lakey}, {Li}, {Li}, {Li}, {Lokken},
  {Louis}, {Lungu}, {MacInnis}, {Madhavacheril}, {Maldonado}, {Mallaby-Kay},
  {Marsden}, {McMahon}, {Menanteau}, {Moodley}, {Morton}, {Namikawa}, {Nati},
  {Newburgh}, {Nibarger}, {Nicola}, {Niemack}, {Nolta}, {Orlowski-Sherer},
  {Page}, {Pappas}, {Partridge}, {Phakathi}, {Pisano}, {Prince}, {Puddu}, {Qu},
  {Rivera}, {Robertson}, {Rojas}, {Salatino}, {Schaan}, {Schillaci}, {Sehgal},
  {Sherwin}, {Sierra}, {Sievers}, {Sifon}, {Sikhosana}, {Simon}, {Spergel},
  {Staggs}, {Stevens}, {Storer}, {Sunder}, {Switzer}, {Thorne}, {Thornton},
  {Trac}, {Treu}, {Tucker}, {Vale}, {Van Engelen}, {Van Lanen}, {Vavagiakis},
  {Wagoner}, {Wang}, {Ward}, {Wollack}, {Xu}, {Zago}, \& {Zhu}}]{aiola+20}
{Aiola}, S., {Calabrese}, E., {Maurin}, L., {et~al.} 2020, \jcap, 2020, 047

\bibitem[{{Alam} {et~al.}(2021){Alam}, {Aubert}, {Avila}, {Balland},
  {Bautista}, {Bershady}, {Bizyaev}, {Blanton}, {Bolton}, {Bovy}, {Brinkmann},
  {Brownstein}, {Burtin}, {Chabanier}, {Chapman}, {Choi}, {Chuang}, {Comparat},
  {Cousinou}, {Cuceu}, {Dawson}, {de la Torre}, {de Mattia}, {Agathe}, {des
  Bourboux}, {Escoffier}, {Etourneau}, {Farr}, {Font-Ribera}, {Frinchaboy},
  {Fromenteau}, {Gil-Mar{\'\i}n}, {Le Goff}, {Gonzalez-Morales},
  {Gonzalez-Perez}, {Grabowski}, {Guy}, {Hawken}, {Hou}, {Kong}, {Parker},
  {Klaene}, {Kneib}, {Lin}, {Long}, {Lyke}, {de la Macorra}, {Martini},
  {Masters}, {Mohammad}, {Moon}, {Mueller}, {Mu{\~n}oz-Guti{\'e}rrez}, {Myers},
  {Nadathur}, {Neveux}, {Newman}, {Noterdaeme}, {Oravetz}, {Oravetz},
  {Palanque-Delabrouille}, {Pan}, {Paviot}, {Percival}, {P{\'e}rez-R{\`a}fols},
  {Petitjean}, {Pieri}, {Prakash}, {Raichoor}, {Ravoux}, {Rezaie}, {Rich},
  {Ross}, {Rossi}, {Ruggeri}, {Ruhlmann-Kleider}, {S{\'a}nchez}, {S{\'a}nchez},
  {S{\'a}nchez-Gallego}, {Sayres}, {Schneider}, {Seo}, {Shafieloo}, {Slosar},
  {Smith}, {Stermer}, {Tamone}, {Tinker}, {Tojeiro}, {Vargas-Maga{\~n}a},
  {Variu}, {Wang}, {Weaver}, {Weijmans}, {Y{\`e}che}, {Zarrouk}, {Zhao},
  {Zhao}, \& {Zheng}}]{alam2021}
{Alam}, S., {Aubert}, M., {Avila}, S., {et~al.} 2021, \prd, 103, 083533

\bibitem[{{Allende Prieto} {et~al.}(2009){Allende Prieto}, {Hubeny}, \&
  {Smith}}]{allende-prieto+09}
{Allende Prieto}, C., {Hubeny}, I., \& {Smith}, J.~A. 2009, \mnras, 396, 759

\bibitem[{{Awan} \& {Gawiser}(2020)}]{awan2020}
{Awan}, H., \& {Gawiser}, E. 2020, \apj, 890, 78

\bibitem[{Bautista {et~al.}(2020)Bautista, Paviot, Vargas Magaña,
  de la Torre, Fromenteau, Gil-Marín, Ross, Burtin, Dawson, Hou, Kneib,
  de Mattia, Percival, Rossi, Tojeiro, Zhao, Zhao, Alam, Brownstein, Chapman,
  Choi, Chuang, Escoffier, de la Macorra, du Mas des Bourboux, Mohammad,
  Moon, Müller, Nadathur, Newman, Schneider, Seo, \& Wang}]{bautista2020}
Bautista, J.~E., Paviot, R., Vargas Magaña, M., {et~al.} 2020, Monthly
  Notices of the Royal Astronomical Society, 500, 736.
\newblock \url{https://doi.org/10.1093/mnras/staa2800}

\bibitem[{{Beers} {et~al.}(1990){Beers}, {Flynn}, \& {Gebhardt}}]{beers90}
{Beers}, T.~C., {Flynn}, K., \& {Gebhardt}, K. 1990, \aj, 100, 32

\bibitem[{{Bennett} {et~al.}(2013){Bennett}, {Larson}, {Weiland}, {Jarosik},
  {Hinshaw}, {Odegard}, {Smith}, {Hill}, {Gold}, {Halpern}, {Komatsu}, {Nolta},
  {Page}, {Spergel}, {Wollack}, {Dunkley}, {Kogut}, {Limon}, {Meyer}, {Tucker},
  \& {Wright}}]{Bennett+13}
{Bennett}, C.~L., {Larson}, D., {Weiland}, J.~L., {et~al.} 2013, \apjs, 208, 20

\bibitem[{{Betoule} {et~al.}(2014){Betoule}, {Kessler}, {Guy}, {Mosher},
  {Hardin}, {Biswas}, {Astier}, {El-Hage}, {Konig}, {Kuhlmann}, {Marriner},
  {Pain}, {Regnault}, {Balland}, {Bassett}, {Brown}, {Campbell}, {Carlberg},
  {Cellier-Holzem}, {Cinabro}, {Conley}, {D'Andrea}, {DePoy}, {Doi}, {Ellis},
  {Fabbro}, {Filippenko}, {Foley}, {Frieman}, {Fouchez}, {Galbany}, {Goobar},
  {Gupta}, {Hill}, {Hlozek}, {Hogan}, {Hook}, {Howell}, {Jha}, {Le Guillou},
  {Leloudas}, {Lidman}, {Marshall}, {M{\"o}ller}, {Mour{\~a}o}, {Neveu},
  {Nichol}, {Olmstead}, {Palanque-Delabrouille}, {Perlmutter}, {Prieto},
  {Pritchet}, {Richmond}, {Riess}, {Ruhlmann-Kleider}, {Sako}, {Schahmaneche},
  {Schneider}, {Smith}, {Sollerman}, {Sullivan}, {Walton}, \&
  {Wheeler}}]{betoule+14}
{Betoule}, M., {Kessler}, R., {Guy}, J., {et~al.} 2014, \aap, 568, A22

\bibitem[{{Bhatti} {et~al.}(2010){Bhatti}, {Richmond}, {Ford}, \&
  {Petro}}]{bhatti+10}
{Bhatti}, W.~A., {Richmond}, M.~W., {Ford}, H.~C., \& {Petro}, L.~D. 2010,
  \apjs, 186, 233

\bibitem[{{Bianchi} {et~al.}(2017){Bianchi}, {Shiao}, \& {Thilker}}]{galex2}
{Bianchi}, L., {Shiao}, B., \& {Thilker}, D. 2017, \apjs, 230, 24

\bibitem[{{Bohlin} {et~al.}(2014){Bohlin}, {Gordon}, \& {Tremblay}}]{bohlin+14}
{Bohlin}, R.~C., {Gordon}, K.~D., \& {Tremblay}, P.~E. 2014, \pasp, 126, 711

\bibitem[{{Bond} {et~al.}(2009){Bond}, {Gawiser}, {Gronwall}, {Ciardullo},
  {Altmann}, \& {Schawinski}}]{bond2009}
{Bond}, N.~A., {Gawiser}, E., {Gronwall}, C., {et~al.} 2009, \apj, 705, 639

\bibitem[{{Bressan} {et~al.}(2012){Bressan}, {Marigo}, {Girardi}, {Salasnich},
  {Dal Cero}, {Rubele}, \& {Nanni}}]{bressan+12}
{Bressan}, A., {Marigo}, P., {Girardi}, L., {et~al.} 2012, \mnras, 427, 127

\bibitem[{{Byrohl} {et~al.}(2019){Byrohl}, {Saito}, \& {Behrens}}]{byrohl2019}
{Byrohl}, C., {Saito}, S., \& {Behrens}, C. 2019, \mnras, 489, 3472

\bibitem[{{Capitanio} {et~al.}(2017){Capitanio}, {Lallement}, {Vergely},
  {Elyajouri}, \& {Monreal-Ibero}}]{capitanio+17}
{Capitanio}, L., {Lallement}, R., {Vergely}, J.~L., {Elyajouri}, M., \&
  {Monreal-Ibero}, A. 2017, \aap, 606, A65

\bibitem[{{Castelli} \& {Kurucz}(2003)}]{castelli+04}
{Castelli}, F., \& {Kurucz}, R.~L. 2003, in Modelling of Stellar Atmospheres,
  ed. N.~{Piskunov}, W.~W. {Weiss}, \& D.~F. {Gray}, Vol. 210, A20

\bibitem[{{Cenarro} {et~al.}(2007){Cenarro}, {Peletier},
  {S{\'a}nchez-Bl{\'a}zquez}, {Selam}, {Toloba}, {Cardiel},
  {Falc{\'o}n-Barroso}, {Gorgas}, {Jim{\'e}nez-Vicente}, \&
  {Vazdekis}}]{cenarro+07}
{Cenarro}, A.~J., {Peletier}, R.~F., {S{\'a}nchez-Bl{\'a}zquez}, P., {et~al.}
  2007, \mnras, 374, 664

\bibitem[{{Chambers} {et~al.}(2016){Chambers}, {Magnier}, {Metcalfe},
  {Flewelling}, {Huber}, {Waters}, {Denneau}, {Draper}, {Farrow}, {Finkbeiner},
  {Holmberg}, {Koppenhoefer}, {Price}, {Rest}, {Saglia}, {Schlafly}, {Smartt},
  {Sweeney}, {Wainscoat}, {Burgett}, {Chastel}, {Grav}, {Heasley}, {Hodapp},
  {Jedicke}, {Kaiser}, {Kudritzki}, {Luppino}, {Lupton}, {Monet}, {Morgan},
  {Onaka}, {Shiao}, {Stubbs}, {Tonry}, {White}, {Ba{\~n}ados}, {Bell},
  {Bender}, {Bernard}, {Boegner}, {Boffi}, {Botticella}, {Calamida},
  {Casertano}, {Chen}, {Chen}, {Cole}, {Deacon}, {Frenk}, {Fitzsimmons},
  {Gezari}, {Gibbs}, {Goessl}, {Goggia}, {Gourgue}, {Goldman}, {Grant},
  {Grebel}, {Hambly}, {Hasinger}, {Heavens}, {Heckman}, {Henderson}, {Henning},
  {Holman}, {Hopp}, {Ip}, {Isani}, {Jackson}, {Keyes}, {Koekemoer}, {Kotak},
  {Le}, {Liska}, {Long}, {Lucey}, {Liu}, {Martin}, {Masci}, {McLean}, {Mindel},
  {Misra}, {Morganson}, {Murphy}, {Obaika}, {Narayan}, {Nieto-Santisteban},
  {Norberg}, {Peacock}, {Pier}, {Postman}, {Primak}, {Rae}, {Rai}, {Riess},
  {Riffeser}, {Rix}, {R{\"o}ser}, {Russel}, {Rutz}, {Schilbach}, {Schultz},
  {Scolnic}, {Strolger}, {Szalay}, {Seitz}, {Small}, {Smith}, {Soderblom},
  {Taylor}, {Thomson}, {Taylor}, {Thakar}, {Thiel}, {Thilker}, {Unger},
  {Urata}, {Valenti}, {Wagner}, {Walder}, {Walter}, {Watters}, {Werner},
  {Wood-Vasey}, \& {Wyse}}]{ps1}
{Chambers}, K.~C., {Magnier}, E.~A., {Metcalfe}, N., {et~al.} 2016, arXiv
  e-prints, arXiv:1612.05560

\bibitem[{{Chiang} {et~al.}(2013){Chiang}, {Wullstein}, {Jeong}, {Komatsu},
  {Blanc}, {Ciardullo}, {Drory}, {Fabricius}, {Finkelstein}, {Gebhardt},
  {Gronwall}, {Hagen}, {Hill}, {Jee}, {Jogee}, {Landriau}, {Mentuch Cooper},
  {Schneider}, \& {Tuttle}}]{chiang+13}
{Chiang}, C.-T., {Wullstein}, P., {Jeong}, D., {et~al.} 2013, \jcap, 2013, 030

\bibitem[{{Ciardullo} {et~al.}(2012){Ciardullo}, {Gronwall}, {Wolf},
  {McCathran}, {Bond}, {Gawiser}, {Guaita}, {Feldmeier}, {Treister}, {Padilla},
  {Francke}, {Matkovi{\'c}}, {Altmann}, \& {Herrera}}]{ciardullo+12}
{Ciardullo}, R., {Gronwall}, C., {Wolf}, C., {et~al.} 2012, \apj, 744, 110

\bibitem[{{Ciardullo} {et~al.}(2013){Ciardullo}, {Gronwall}, {Adams}, {Blanc},
  {Gebhardt}, {Finkelstein}, {Jogee}, {Hill}, {Drory}, {Hopp}, {Schneider},
  {Zeimann}, \& {Dalton}}]{ciardullo+13}
{Ciardullo}, R., {Gronwall}, C., {Adams}, J.~J., {et~al.} 2013, \apj, 769, 83

\bibitem[{{Coupon} {et~al.}(2009){Coupon}, {Ilbert}, {Kilbinger}, {McCracken},
  {Mellier}, {Arnouts}, {Bertin}, {Hudelot}, {Schultheis}, {Le F{\`e}vre}, {Le
  Brun}, {Guzzo}, {Bardelli}, {Zucca}, {Bolzonella}, {Garilli}, {Zamorani},
  {Zanichelli}, {Tresse}, \& {Aussel}}]{CFHTLS}
{Coupon}, J., {Ilbert}, O., {Kilbinger}, M., {et~al.} 2009, \aap, 500, 981

\bibitem[{{Cowie} \& {Hu}(1998)}]{cowie+98}
{Cowie}, L.~L., \& {Hu}, E.~M. 1998, \aj, 115, 1319

\bibitem[{{Dawson} {et~al.}(2013){Dawson}, {Schlegel}, {Ahn}, {Anderson},
  {Aubourg}, {Bailey}, {Barkhouser}, {Bautista}, {Beifiori}, {Berlind},
  {Bhardwaj}, {Bizyaev}, {Blake}, {Blanton}, {Blomqvist}, {Bolton}, {Borde},
  {Bovy}, {Brandt}, {Brewington}, {Brinkmann}, {Brown}, {Brownstein}, {Bundy},
  {Busca}, {Carithers}, {Carnero}, {Carr}, {Chen}, {Comparat}, {Connolly},
  {Cope}, {Croft}, {Cuesta}, {da Costa}, {Davenport}, {Delubac}, {de Putter},
  {Dhital}, {Ealet}, {Ebelke}, {Eisenstein}, {Escoffier}, {Fan}, {Filiz Ak},
  {Finley}, {Font-Ribera}, {G{\'e}nova-Santos}, {Gunn}, {Guo}, {Haggard},
  {Hall}, {Hamilton}, {Harris}, {Harris}, {Ho}, {Hogg}, {Holder}, {Honscheid},
  {Huehnerhoff}, {Jordan}, {Jordan}, {Kauffmann}, {Kazin}, {Kirkby}, {Klaene},
  {Kneib}, {Le Goff}, {Lee}, {Long}, {Loomis}, {Lundgren}, {Lupton}, {Maia},
  {Makler}, {Malanushenko}, {Malanushenko}, {Mandelbaum}, {Manera}, {Maraston},
  {Margala}, {Masters}, {McBride}, {McDonald}, {McGreer}, {McMahon}, {Mena},
  {Miralda-Escud{\'e}}, {Montero-Dorta}, {Montesano}, {Muna}, {Myers},
  {Naugle}, {Nichol}, {Noterdaeme}, {Nuza}, {Olmstead}, {Oravetz}, {Oravetz},
  {Owen}, {Padmanabhan}, {Palanque-Delabrouille}, {Pan}, {Parejko},
  {P{\^a}ris}, {Percival}, {P{\'e}rez-Fournon}, {P{\'e}rez-R{\`a}fols},
  {Petitjean}, {Pfaffenberger}, {Pforr}, {Pieri}, {Prada}, {Price-Whelan},
  {Raddick}, {Rebolo}, {Rich}, {Richards}, {Rockosi}, {Roe}, {Ross}, {Ross},
  {Rossi}, {Rubi{\~n}o-Martin}, {Samushia}, {S{\'a}nchez}, {Sayres}, {Schmidt},
  {Schneider}, {Sc{\'o}ccola}, {Seo}, {Shelden}, {Sheldon}, {Shen}, {Shu},
  {Slosar}, {Smee}, {Snedden}, {Stauffer}, {Steele}, {Strauss}, {Streblyanska},
  {Suzuki}, {Swanson}, {Tal}, {Tanaka}, {Thomas}, {Tinker}, {Tojeiro},
  {Tremonti}, {Vargas Maga{\~n}a}, {Verde}, {Viel}, {Wake}, {Watson}, {Weaver},
  {Weinberg}, {Weiner}, {West}, {White}, {Wood-Vasey}, {Yeche}, {Zehavi},
  {Zhao}, \& {Zheng}}]{BOSS}
{Dawson}, K.~S., {Schlegel}, D.~J., {Ahn}, C.~P., {et~al.} 2013, \aj, 145, 10

\bibitem[{{Dawson} {et~al.}(2016){Dawson}, {Kneib}, {Percival}, {Alam},
  {Albareti}, {Anderson}, {Armengaud}, {Aubourg}, {Bailey}, {Bautista},
  {Berlind}, {Bershady}, {Beutler}, {Bizyaev}, {Blanton}, {Blomqvist},
  {Bolton}, {Bovy}, {Brandt}, {Brinkmann}, {Brownstein}, {Burtin}, {Busca},
  {Cai}, {Chuang}, {Clerc}, {Comparat}, {Cope}, {Croft}, {Cruz-Gonzalez}, {da
  Costa}, {Cousinou}, {Darling}, {de la Macorra}, {de la Torre}, {Delubac}, {du
  Mas des Bourboux}, {Dwelly}, {Ealet}, {Eisenstein}, {Eracleous}, {Escoffier},
  {Fan}, {Finoguenov}, {Font-Ribera}, {Frinchaboy}, {Gaulme}, {Georgakakis},
  {Green}, {Guo}, {Guy}, {Ho}, {Holder}, {Huehnerhoff}, {Hutchinson}, {Jing},
  {Jullo}, {Kamble}, {Kinemuchi}, {Kirkby}, {Kitaura}, {Klaene}, {Laher},
  {Lang}, {Laurent}, {Le Goff}, {Li}, {Liang}, {Lima}, {Lin}, {Lin}, {Lin},
  {Long}, {Lundgren}, {MacDonald}, {Geimba Maia}, {Malanushenko},
  {Malanushenko}, {Mariappan}, {McBride}, {McGreer}, {M{\'e}nard}, {Merloni},
  {Meza}, {Montero-Dorta}, {Muna}, {Myers}, {Nandra}, {Naugle}, {Newman},
  {Noterdaeme}, {Nugent}, {Ogando}, {Olmstead}, {Oravetz}, {Oravetz},
  {Padmanabhan}, {Palanque-Delabrouille}, {Pan}, {Parejko}, {P{\^a}ris},
  {Peacock}, {Petitjean}, {Pieri}, {Pisani}, {Prada}, {Prakash}, {Raichoor},
  {Reid}, {Rich}, {Ridl}, {Rodriguez-Torres}, {Carnero Rosell}, {Ross},
  {Rossi}, {Ruan}, {Salvato}, {Sayres}, {Schneider}, {Schlegel}, {Seljak},
  {Seo}, {Sesar}, {Shandera}, {Shu}, {Slosar}, {Sobreira}, {Streblyanska},
  {Suzuki}, {Taylor}, {Tao}, {Tinker}, {Tojeiro}, {Vargas-Maga{\~n}a}, {Wang},
  {Weaver}, {Weinberg}, {White}, {Wood-Vasey}, {Yeche}, {Zhai}, {Zhao}, {Zhao},
  {Zheng}, {Ben Zhu}, \& {Zou}}]{eBOSS}
{Dawson}, K.~S., {Kneib}, J.-P., {Percival}, W.~J., {et~al.} 2016, \aj, 151, 44

\bibitem[{{DES Collaboration} {et~al.}(2021){DES Collaboration}, {Abbott},
  {Aguena}, {Alarcon}, {Allam}, {Alves}, {Amon}, {Andrade-Oliveira}, {Annis},
  {Avila}, {Bacon}, {Baxter}, {Bechtol}, {Becker}, {Bernstein}, {Bhargava},
  {Birrer}, {Blazek}, {Brandao-Souza}, {Bridle}, {Brooks}, {Buckley-Geer},
  {Burke}, {Camacho}, {Campos}, {Carnero Rosell}, {Carrasco Kind}, {Carretero},
  {Castander}, {Cawthon}, {Chang}, {Chen}, {Chen}, {Choi}, {Conselice},
  {Cordero}, {Costanzi}, {Crocce}, {da Costa}, {da Silva Pereira}, {Davis},
  {Davis}, {De Vicente}, {DeRose}, {Desai}, {Di Valentino}, {Diehl},
  {Dietrich}, {Dodelson}, {Doel}, {Doux}, {Drlica-Wagner}, {Eckert}, {Eifler},
  {Elsner}, {Elvin-Poole}, {Everett}, {Evrard}, {Fang}, {Farahi}, {Fernandez},
  {Ferrero}, {Fert{\'e}}, {Fosalba}, {Friedrich}, {Frieman},
  {Garc{\'\i}a-Bellido}, {Gatti}, {Gaztanaga}, {Gerdes}, {Giannantonio},
  {Giannini}, {Gruen}, {Gruendl}, {Gschwend}, {Gutierrez}, {Harrison},
  {Hartley}, {Herner}, {Hinton}, {Hollowood}, {Honscheid}, {Hoyle}, {Huff},
  {Huterer}, {Jain}, {James}, {Jarvis}, {Jeffrey}, {Jeltema}, {Kovacs},
  {Krause}, {Kron}, {Kuehn}, {Kuropatkin}, {Lahav}, {Leget}, {Lemos}, {Liddle},
  {Lidman}, {Lima}, {Lin}, {MacCrann}, {Maia}, {Marshall}, {Martini},
  {McCullough}, {Melchior}, {Mena-Fern{\'a}ndez}, {Menanteau}, {Miquel},
  {Mohr}, {Morgan}, {Muir}, {Myles}, {Nadathur}, {Navarro-Alsina}, {Nichol},
  {Ogando}, {Omori}, {Palmese}, {Pandey}, {Park}, {Paz-Chinch{\'o}n},
  {Petravick}, {Pieres}, {Plazas Malag{\'o}n}, {Porredon}, {Prat}, {Raveri},
  {Rodriguez-Monroy}, {Rollins}, {Romer}, {Roodman}, {Rosenfeld}, {Ross},
  {Rykoff}, {Samuroff}, {S{\'a}nchez}, {Sanchez}, {Sanchez}, {Sanchez Cid},
  {Scarpine}, {Schubnell}, {Scolnic}, {Secco}, {Serrano}, {Sevilla-Noarbe},
  {Sheldon}, {Shin}, {Smith}, {Soares-Santos}, {Suchyta}, {Swanson}, {Tabbutt},
  {Tarle}, {Thomas}, {To}, {Troja}, {Troxel}, {Tucker}, {Tutusaus}, {Varga},
  {Walker}, {Weaverdyck}, {Weller}, {Yanny}, {Yin}, {Zhang}, \&
  {Zuntz}}]{des2021a}
{DES Collaboration}, {Abbott}, T.~M.~C., {Aguena}, M., {et~al.} 2021, arXiv
  e-prints, arXiv:2105.13549

\bibitem[{{DESI Collaboration} {et~al.}(2016){DESI Collaboration}, {Aghamousa},
  {Aguilar}, {Ahlen}, {Alam}, {Allen}, {Allende Prieto}, {Annis}, {Bailey},
  {Balland}, {Ballester}, {Baltay}, {Beaufore}, {Bebek}, {Beers}, {Bell},
  {Bernal}, {Besuner}, {Beutler}, {Blake}, {Bleuler}, {Blomqvist}, {Blum},
  {Bolton}, {Briceno}, {Brooks}, {Brownstein}, {Buckley-Geer}, {Burden},
  {Burtin}, {Busca}, {Cahn}, {Cai}, {Cardiel-Sas}, {Carlberg}, {Carton},
  {Casas}, {Castander}, {Cervantes-Cota}, {Claybaugh}, {Close}, {Coker},
  {Cole}, {Comparat}, {Cooper}, {Cousinou}, {Crocce}, {Cuby}, {Cunningham},
  {Davis}, {Dawson}, {de la Macorra}, {De Vicente}, {Delubac}, {Derwent},
  {Dey}, {Dhungana}, {Ding}, {Doel}, {Duan}, {Ealet}, {Edelstein},
  {Eftekharzadeh}, {Eisenstein}, {Elliott}, {Escoffier}, {Evatt}, {Fagrelius},
  {Fan}, {Fanning}, {Farahi}, {Farihi}, {Favole}, {Feng}, {Fernandez},
  {Findlay}, {Finkbeiner}, {Fitzpatrick}, {Flaugher}, {Flender}, {Font-Ribera},
  {Forero-Romero}, {Fosalba}, {Frenk}, {Fumagalli}, {Gaensicke}, {Gallo},
  {Garcia-Bellido}, {Gaztanaga}, {Pietro Gentile Fusillo}, {Gerard},
  {Gershkovich}, {Giannantonio}, {Gillet}, {Gonzalez-de-Rivera},
  {Gonzalez-Perez}, {Gott}, {Graur}, {Gutierrez}, {Guy}, {Habib}, {Heetderks},
  {Heetderks}, {Heitmann}, {Hellwing}, {Herrera}, {Ho}, {Holland}, {Honscheid},
  {Huff}, {Hutchinson}, {Huterer}, {Hwang}, {Illa Laguna}, {Ishikawa},
  {Jacobs}, {Jeffrey}, {Jelinsky}, {Jennings}, {Jiang}, {Jimenez}, {Johnson},
  {Joyce}, {Jullo}, {Juneau}, {Kama}, {Karcher}, {Karkar}, {Kehoe}, {Kennamer},
  {Kent}, {Kilbinger}, {Kim}, {Kirkby}, {Kisner}, {Kitanidis}, {Kneib},
  {Koposov}, {Kovacs}, {Koyama}, {Kremin}, {Kron}, {Kronig}, {Kueter-Young},
  {Lacey}, {Lafever}, {Lahav}, {Lambert}, {Lampton}, {Landriau}, {Lang},
  {Lauer}, {Le Goff}, {Le Guillou}, {Le Van Suu}, {Lee}, {Lee}, {Leitner},
  {Lesser}, {Levi}, {L'Huillier}, {Li}, {Liang}, {Lin}, {Linder}, {Loebman},
  {Luki{\'c}}, {Ma}, {MacCrann}, {Magneville}, {Makarem}, {Manera}, {Manser},
  {Marshall}, {Martini}, {Massey}, {Matheson}, {McCauley}, {McDonald},
  {McGreer}, {Meisner}, {Metcalfe}, {Miller}, {Miquel}, {Moustakas}, {Myers},
  {Naik}, {Newman}, {Nichol}, {Nicola}, {Nicolati da Costa}, {Nie}, {Niz},
  {Norberg}, {Nord}, {Norman}, {Nugent}, {O'Brien}, {Oh}, {Olsen}, {Padilla},
  {Padmanabhan}, {Padmanabhan}, {Palanque-Delabrouille}, {Palmese},
  {Pappalardo}, {P{\^a}ris}, {Park}, {Patej}, {Peacock}, {Peiris}, {Peng},
  {Percival}, {Perruchot}, {Pieri}, {Pogge}, {Pollack}, {Poppett}, {Prada},
  {Prakash}, {Probst}, {Rabinowitz}, {Raichoor}, {Ree}, {Refregier}, {Regal},
  {Reid}, {Reil}, {Rezaie}, {Rockosi}, {Roe}, {Ronayette}, {Roodman}, {Ross},
  {Ross}, {Rossi}, {Rozo}, {Ruhlmann-Kleider}, {Rykoff}, {Sabiu}, {Samushia},
  {Sanchez}, {Sanchez}, {Schlegel}, {Schneider}, {Schubnell}, {Secroun},
  {Seljak}, {Seo}, {Serrano}, {Shafieloo}, {Shan}, {Sharples}, {Sholl},
  {Shourt}, {Silber}, {Silva}, {Sirk}, {Slosar}, {Smith}, {Smoot}, {Som},
  {Song}, {Sprayberry}, {Staten}, {Stefanik}, {Tarle}, {Sien Tie}, {Tinker},
  {Tojeiro}, {Valdes}, {Valenzuela}, {Valluri}, {Vargas-Magana}, {Verde},
  {Walker}, {Wang}, {Wang}, {Weaver}, {Weaverdyck}, {Wechsler}, {Weinberg},
  {White}, {Yang}, {Yeche}, {Zhang}, {Zhao}, {Zheng}, {Zhou}, {Zhou}, {Zhu},
  {Zou}, \& {Zu}}]{DESI}
{DESI Collaboration}, {Aghamousa}, A., {Aguilar}, J., {et~al.} 2016, arXiv
  e-prints, arXiv:1611.00036

\bibitem[{{Dey} {et~al.}(2019){Dey}, {Schlegel}, {Lang}, {Blum}, {Burleigh},
  {Fan}, {Findlay}, {Finkbeiner}, {Herrera}, {Juneau}, {Landriau}, {Levi},
  {McGreer}, {Meisner}, {Myers}, {Moustakas}, {Nugent}, {Patej}, {Schlafly},
  {Walker}, {Valdes}, {Weaver}, {Y{\`e}che}, {Zou}, {Zhou}, {Abareshi},
  {Abbott}, {Abolfathi}, {Aguilera}, {Alam}, {Allen}, {Alvarez}, {Annis},
  {Ansarinejad}, {Aubert}, {Beechert}, {Bell}, {BenZvi}, {Beutler}, {Bielby},
  {Bolton}, {Brice{\~n}o}, {Buckley-Geer}, {Butler}, {Calamida}, {Carlberg},
  {Carter}, {Casas}, {Castander}, {Choi}, {Comparat}, {Cukanovaite}, {Delubac},
  {DeVries}, {Dey}, {Dhungana}, {Dickinson}, {Ding}, {Donaldson}, {Duan},
  {Duckworth}, {Eftekharzadeh}, {Eisenstein}, {Etourneau}, {Fagrelius},
  {Farihi}, {Fitzpatrick}, {Font-Ribera}, {Fulmer}, {G{\"a}nsicke},
  {Gaztanaga}, {George}, {Gerdes}, {Gontcho}, {Gorgoni}, {Green}, {Guy},
  {Harmer}, {Hernandez}, {Honscheid}, {Huang}, {James}, {Jannuzi}, {Jiang},
  {Joyce}, {Karcher}, {Karkar}, {Kehoe}, {Kneib}, {Kueter-Young}, {Lan},
  {Lauer}, {Le Guillou}, {Le Van Suu}, {Lee}, {Lesser}, {Perreault Levasseur},
  {Li}, {Mann}, {Marshall}, {Mart{\'\i}nez-V{\'a}zquez}, {Martini}, {du Mas des
  Bourboux}, {McManus}, {Meier}, {M{\'e}nard}, {Metcalfe},
  {Mu{\~n}oz-Guti{\'e}rrez}, {Najita}, {Napier}, {Narayan}, {Newman}, {Nie},
  {Nord}, {Norman}, {Olsen}, {Paat}, {Palanque-Delabrouille}, {Peng},
  {Poppett}, {Poremba}, {Prakash}, {Rabinowitz}, {Raichoor}, {Rezaie},
  {Robertson}, {Roe}, {Ross}, {Ross}, {Rudnick}, {Safonova}, {Saha},
  {S{\'a}nchez}, {Savary}, {Schweiker}, {Scott}, {Seo}, {Shan}, {Silva},
  {Slepian}, {Soto}, {Sprayberry}, {Staten}, {Stillman}, {Stupak}, {Summers},
  {Sien Tie}, {Tirado}, {Vargas-Maga{\~n}a}, {Vivas}, {Wechsler}, {Williams},
  {Yang}, {Yang}, {Yapici}, {Zaritsky}, {Zenteno}, {Zhang}, {Zhang}, {Zhou}, \&
  {Zhou}}]{DECaLS}
{Dey}, A., {Schlegel}, D.~J., {Lang}, D., {et~al.} 2019, \aj, 157, 168

\bibitem[{de Mattia {et~al.}(2020)de Mattia, Ruhlmann-Kleider, Raichoor,
  Ross, Tamone, Zhao, Alam, Avila, Burtin, Bautista, Beutler, Brinkmann,
  Brownstein, Chapman, Chuang, Comparat, du Mas des Bourboux, Dawson,
  de la Macorra, Gil-Marín, Gonzalez-Perez, Gorgoni, Hou, Kong, Lin,
  Nadathur, Newman, Mueller, Percival, Rezaie, Rossi, Schneider, Tiwari, Vivek,
  Wang, \& Zhao}]{demattia2021}
de Mattia, A., Ruhlmann-Kleider, V., Raichoor, A., {et~al.} 2020, Monthly
  Notices of the Royal Astronomical Society, 501, 5616.
\newblock \url{https://doi.org/10.1093/mnras/staa3891}

\bibitem[{{Di Valentino}(2021)}]{divalentino21}
{Di Valentino}, E. 2021, \mnras, 502, 2065

\bibitem[{{Di Valentino} {et~al.}(2021){Di Valentino}, {Mena}, {Pan},
  {Visinelli}, {Yang}, {Melchiorri}, {Mota}, {Riess}, \&
  {Silk}}]{divalentino+21}
{Di Valentino}, E., {Mena}, O., {Pan}, S., {et~al.} 2021, arXiv e-prints,
  arXiv:2103.01183

\bibitem[{{Doi} {et~al.}(2010){Doi}, {Tanaka}, {Fukugita}, {Gunn}, {Yasuda},
  {Ivezi{\'c}}, {Brinkmann}, {de Haars}, {Kleinman}, {Krzesinski}, \& {French
  Leger}}]{doi+10}
{Doi}, M., {Tanaka}, M., {Fukugita}, M., {et~al.} 2010, \aj, 139, 1628

\bibitem[{du~Mas~des Bourboux {et~al.}(2020)du~Mas~des Bourboux, Rich,
  Font-Ribera, de~Sainte~Agathe, Farr, Etourneau, Goff, Cuceu, Balland,
  Bautista, Blomqvist, Brinkmann, Brownstein, Chabanier, Chaussidon, Dawson,
  Gonz{\'{a}}lez-Morales, Guy, Lyke, de~la Macorra, Mueller, Myers, Nitschelm,
  Guti{\'{e}}rrez, Palanque-Delabrouille, Parker, Percival,
  P{\'{e}}rez-R{\`{a}}fols, Petitjean, Pieri, Ravoux, Rossi, Schneider, Seo,
  Slosar, Stermer, Vivek, Y{\`{e}}che, \& Youles}]{dumas2020}
du~Mas~des Bourboux, H., Rich, J., Font-Ribera, A., {et~al.} 2020, The
  Astrophysical Journal, 901, 153.
\newblock \url{https://doi.org/10.3847/1538-4357/abb085}

\bibitem[{{Dutcher} {et~al.}(2021){Dutcher}, {Balkenhol}, {Ade}, {Ahmed},
  {Anderes}, {Anderson}, {Archipley}, {Avva}, {Aylor}, {Barry}, {Basu Thakur},
  {Benabed}, {Bender}, {Benson}, {Bianchini}, {Bleem}, {Bouchet}, {Bryant},
  {Byrum}, {Carlstrom}, {Carter}, {Cecil}, {Chang}, {Chaubal}, {Chen}, {Cho},
  {Chou}, {Cliche}, {Crawford}, {Cukierman}, {Daley}, {de Haan}, {Denison},
  {Dibert}, {Ding}, {Dobbs}, {Everett}, {Feng}, {Ferguson}, {Foster}, {Fu},
  {Galli}, {Gambrel}, {Gardner}, {Goeckner-Wald}, {Gualtieri}, {Guns}, {Gupta},
  {Guyser}, {Halverson}, {Harke-Hosemann}, {Harrington}, {Henning}, {Hilton},
  {Hivon}, {Holder}, {Holzapfel}, {Hood}, {Howe}, {Huang}, {Irwin}, {Jeong},
  {Jonas}, {Jones}, {Khaire}, {Knox}, {Kofman}, {Korman}, {Kubik}, {Kuhlmann},
  {Kuo}, {Lee}, {Leitch}, {Lowitz}, {Lu}, {Meyer}, {Michalik}, {Millea},
  {Montgomery}, {Nadolski}, {Natoli}, {Nguyen}, {Noble}, {Novosad}, {Omori},
  {Padin}, {Pan}, {Paschos}, {Pearson}, {Posada}, {Prabhu}, {Quan},
  {Raghunathan}, {Rahlin}, {Reichardt}, {Riebel}, {Riedel}, {Rouble}, {Ruhl},
  {Sayre}, {Schiappucci}, {Shirokoff}, {Smecher}, {Sobrin}, {Stark}, {Stephen},
  {Story}, {Suzuki}, {Thompson}, {Thorne}, {Tucker}, {Umilta}, {Vale},
  {Vanderlinde}, {Vieira}, {Wang}, {Whitehorn}, {Wu}, {Yefremenko}, {Yoon},
  {Young}, \& {SPT-3G Collaboration}}]{dutcher+21}
{Dutcher}, D., {Balkenhol}, L., {Ade}, P.~A.~R., {et~al.} 2021, \prd, 104,
  022003

\bibitem[{{Falc{\'o}n-Barroso} {et~al.}(2011){Falc{\'o}n-Barroso},
  {S{\'a}nchez-Bl{\'a}zquez}, {Vazdekis}, {Ricciardelli}, {Cardiel}, {Cenarro},
  {Gorgas}, \& {Peletier}}]{falcon-barroso+11}
{Falc{\'o}n-Barroso}, J., {S{\'a}nchez-Bl{\'a}zquez}, P., {Vazdekis}, A.,
  {et~al.} 2011, \aap, 532, A95

\bibitem[{{Farrow} {et~al.}(2021){Farrow}, {S{\'a}nchez}, {Ciardullo}, {Mentuch
  Cooper}, {Davis}, {Fabricius}, {Gawiser}, {Grasshorn Gebhardt}, {Gebhardt},
  {Hill}, {Jeong}, {Komatsu}, {Landriau}, {Liu}, {Saito}, {Snigula}, \&
  {Wold}}]{farrow+21}
{Farrow}, D.~J., {S{\'a}nchez}, A.~G., {Ciardullo}, R., {et~al.} 2021, arXiv
  e-prints, arXiv:2104.04613

\bibitem[{{Filippenko}(1982)}]{filippenko82}
{Filippenko}, A.~V. 1982, \pasp, 94, 715

\bibitem[{{Flewelling} {et~al.}(2020){Flewelling}, {Magnier}, {Chambers},
  {Heasley}, {Holmberg}, {Huber}, {Sweeney}, {Waters}, {Calamida}, {Casertano},
  {Chen}, {Farrow}, {Hasinger}, {Henderson}, {Long}, {Metcalfe}, {Narayan},
  {Nieto-Santisteban}, {Norberg}, {Rest}, {Saglia}, {Szalay}, {Thakar},
  {Tonry}, {Valenti}, {Werner}, {White}, {Denneau}, {Draper}, {Hodapp},
  {Jedicke}, {Kaiser}, {Kudritzki}, {Price}, {Wainscoat}, {Chastel}, {McLean},
  {Postman}, \& {Shiao}}]{flewelling+20}
{Flewelling}, H.~A., {Magnier}, E.~A., {Chambers}, K.~C., {et~al.} 2020, \apjs,
  251, 7

\bibitem[{{Freedman} {et~al.}(2019){Freedman}, {Madore}, {Hatt}, {Hoyt},
  {Jang}, {Beaton}, {Burns}, {Lee}, {Monson}, {Neeley}, {Phillips}, {Rich}, \&
  {Seibert}}]{freedman+19}
{Freedman}, W.~L., {Madore}, B.~F., {Hatt}, D., {et~al.} 2019, \apj, 882, 34

\bibitem[{{Frieman} {et~al.}(2008){Frieman}, {Turner}, \&
  {Huterer}}]{Turner+08}
{Frieman}, J.~A., {Turner}, M.~S., \& {Huterer}, D. 2008, \araa, 46, 385

\bibitem[{{Gaia Collaboration} {et~al.}(2018){Gaia Collaboration}, {Brown},
  {Vallenari}, {Prusti}, {de Bruijne}, {Babusiaux}, {Bailer-Jones}, {Biermann},
  {Evans}, {Eyer}, {Jansen}, {Jordi}, {Klioner}, {Lammers}, {Lindegren},
  {Luri}, {Mignard}, {Panem}, {Pourbaix}, {Randich}, {Sartoretti}, {Siddiqui},
  {Soubiran}, {van Leeuwen}, {Walton}, {Arenou}, {Bastian}, {Cropper},
  {Drimmel}, {Katz}, {Lattanzi}, {Bakker}, {Cacciari}, {Casta{\~n}eda},
  {Chaoul}, {Cheek}, {De Angeli}, {Fabricius}, {Guerra}, {Holl}, {Masana},
  {Messineo}, {Mowlavi}, {Nienartowicz}, {Panuzzo}, {Portell}, {Riello},
  {Seabroke}, {Tanga}, {Th{\'e}venin}, {Gracia-Abril}, {Comoretto},
  {Garcia-Reinaldos}, {Teyssier}, {Altmann}, {Andrae}, {Audard},
  {Bellas-Velidis}, {Benson}, {Berthier}, {Blomme}, {Burgess}, {Busso},
  {Carry}, {Cellino}, {Clementini}, {Clotet}, {Creevey}, {Davidson}, {De
  Ridder}, {Delchambre}, {Dell'Oro}, {Ducourant},
  {Fern{\'a}ndez-Hern{\'a}ndez}, {Fouesneau}, {Fr{\'e}mat}, {Galluccio},
  {Garc{\'\i}a-Torres}, {Gonz{\'a}lez-N{\'u}{\~n}ez}, {Gonz{\'a}lez-Vidal},
  {Gosset}, {Guy}, {Halbwachs}, {Hambly}, {Harrison}, {Hern{\'a}ndez},
  {Hestroffer}, {Hodgkin}, {Hutton}, {Jasniewicz}, {Jean-Antoine-Piccolo},
  {Jordan}, {Korn}, {Krone-Martins}, {Lanzafame}, {Lebzelter}, {L{\"o}ffler},
  {Manteiga}, {Marrese}, {Mart{\'\i}n-Fleitas}, {Moitinho}, {Mora}, {Muinonen},
  {Osinde}, {Pancino}, {Pauwels}, {Petit}, {Recio-Blanco}, {Richards},
  {Rimoldini}, {Robin}, {Sarro}, {Siopis}, {Smith}, {Sozzetti}, {S{\"u}veges},
  {Torra}, {van Reeven}, {Abbas}, {Abreu Aramburu}, {Accart}, {Aerts},
  {Altavilla}, {{\'A}lvarez}, {Alvarez}, {Alves}, {Anderson}, {Andrei},
  {Anglada Varela}, {Antiche}, {Antoja}, {Arcay}, {Astraatmadja}, {Bach},
  {Baker}, {Balaguer-N{\'u}{\~n}ez}, {Balm}, {Barache}, {Barata}, {Barbato},
  {Barblan}, {Barklem}, {Barrado}, {Barros}, {Barstow}, {Bartholom{\'e}
  Mu{\~n}oz}, {Bassilana}, {Becciani}, {Bellazzini}, {Berihuete}, {Bertone},
  {Bianchi}, {Bienaym{\'e}}, {Blanco-Cuaresma}, {Boch}, {Boeche}, {Bombrun},
  {Borrachero}, {Bossini}, {Bouquillon}, {Bourda}, {Bragaglia}, {Bramante},
  {Breddels}, {Bressan}, {Brouillet}, {Br{\"u}semeister}, {Brugaletta},
  {Bucciarelli}, {Burlacu}, {Busonero}, {Butkevich}, {Buzzi}, {Caffau},
  {Cancelliere}, {Cannizzaro}, {Cantat-Gaudin}, {Carballo}, {Carlucci},
  {Carrasco}, {Casamiquela}, {Castellani}, {Castro-Ginard}, {Charlot},
  {Chemin}, {Chiavassa}, {Cocozza}, {Costigan}, {Cowell}, {Crifo}, {Crosta},
  {Crowley}, {Cuypers}, {Dafonte}, {Damerdji}, {Dapergolas}, {David}, {David},
  {de Laverny}, {De Luise}, {De March}, {de Martino}, {de Souza}, {de Torres},
  {Debosscher}, {del Pozo}, {Delbo}, {Delgado}, {Delgado}, {Di Matteo},
  {Diakite}, {Diener}, {Distefano}, {Dolding}, {Drazinos}, {Dur{\'a}n},
  {Edvardsson}, {Enke}, {Eriksson}, {Esquej}, {Eynard Bontemps}, {Fabre},
  {Fabrizio}, {Faigler}, {Falc{\~a}o}, {Farr{\`a}s Casas}, {Federici},
  {Fedorets}, {Fernique}, {Figueras}, {Filippi}, {Findeisen}, {Fonti},
  {Fraile}, {Fraser}, {Fr{\'e}zouls}, {Gai}, {Galleti}, {Garabato},
  {Garc{\'\i}a-Sedano}, {Garofalo}, {Garralda}, {Gavel}, {Gavras}, {Gerssen},
  {Geyer}, {Giacobbe}, {Gilmore}, {Girona}, {Giuffrida}, {Glass}, {Gomes},
  {Granvik}, {Gueguen}, {Guerrier}, {Guiraud}, {Guti{\'e}rrez-S{\'a}nchez},
  {Haigron}, {Hatzidimitriou}, {Hauser}, {Haywood}, {Heiter}, {Helmi}, {Heu},
  {Hilger}, {Hobbs}, {Hofmann}, {Holland}, {Huckle}, {Hypki}, {Icardi},
  {Jan{\ss}en}, {Jevardat de Fombelle}, {Jonker}, {Juh{\'a}sz}, {Julbe},
  {Karampelas}, {Kewley}, {Klar}, {Kochoska}, {Kohley}, {Kolenberg},
  {Kontizas}, {Kontizas}, {Koposov}, {Kordopatis}, {Kostrzewa-Rutkowska},
  {Koubsky}, {Lambert}, {Lanza}, {Lasne}, {Lavigne}, {Le Fustec}, {Le
  Poncin-Lafitte}, {Lebreton}, {Leccia}, {Leclerc}, {Lecoeur-Taibi},
  {Lenhardt}, {Leroux}, {Liao}, {Licata}, {Lindstr{\o}m}, {Lister}, {Livanou},
  {Lobel}, {L{\'o}pez}, {Managau}, {Mann}, {Mantelet}, {Marchal}, {Marchant},
  {Marconi}, {Marinoni}, {Marschalk{\'o}}, {Marshall}, {Martino}, {Marton},
  {Mary}, {Massari}, {Matijevi{\v{c}}}, {Mazeh}, {McMillan}, {Messina},
  {Michalik}, {Millar}, {Molina}, {Molinaro}, {Moln{\'a}r}, {Montegriffo},
  {Mor}, {Morbidelli}, {Morel}, {Morris}, {Mulone}, {Muraveva}, {Musella},
  {Nelemans}, {Nicastro}, {Noval}, {O'Mullane}, {Ord{\'e}novic},
  {Ord{\'o}{\~n}ez-Blanco}, {Osborne}, {Pagani}, {Pagano}, {Pailler},
  {Palacin}, {Palaversa}, {Panahi}, {Pawlak}, {Piersimoni}, {Pineau}, {Plachy},
  {Plum}, {Poggio}, {Poujoulet}, {Pr{\v{s}}a}, {Pulone}, {Racero}, {Ragaini},
  {Rambaux}, {Ramos-Lerate}, {Regibo}, {Reyl{\'e}}, {Riclet}, {Ripepi}, {Riva},
  {Rivard}, {Rixon}, {Roegiers}, {Roelens}, {Romero-G{\'o}mez}, {Rowell},
  {Royer}, {Ruiz-Dern}, {Sadowski}, {Sagrist{\`a} Sell{\'e}s}, {Sahlmann},
  {Salgado}, {Salguero}, {Sanna}, {Santana-Ros}, {Sarasso}, {Savietto},
  {Schultheis}, {Sciacca}, {Segol}, {Segovia}, {S{\'e}gransan}, {Shih},
  {Siltala}, {Silva}, {Smart}, {Smith}, {Solano}, {Solitro}, {Sordo}, {Soria
  Nieto}, {Souchay}, {Spagna}, {Spoto}, {Stampa}, {Steele},
  {Steidelm{\"u}ller}, {Stephenson}, {Stoev}, {Suess}, {Surdej}, {Szabados},
  {Szegedi-Elek}, {Tapiador}, {Taris}, {Tauran}, {Taylor}, {Teixeira},
  {Terrett}, {Teyssandier}, {Thuillot}, {Titarenko}, {Torra Clotet}, {Turon},
  {Ulla}, {Utrilla}, {Uzzi}, {Vaillant}, {Valentini}, {Valette}, {van Elteren},
  {Van Hemelryck}, {van Leeuwen}, {Vaschetto}, {Vecchiato}, {Veljanoski},
  {Viala}, {Vicente}, {Vogt}, {von Essen}, {Voss}, {Votruba}, {Voutsinas},
  {Walmsley}, {Weiler}, {Wertz}, {Wevers}, {Wyrzykowski}, {Yoldas},
  {{\v{Z}}erjal}, {Ziaeepour}, {Zorec}, {Zschocke}, {Zucker}, {Zurbach}, \&
  {Zwitter}}]{Gaia2018}
{Gaia Collaboration}, {Brown}, A.~G.~A., {Vallenari}, A., {et~al.} 2018, \aap,
  616, A1

\bibitem[{{Gawiser} {et~al.}(2007){Gawiser}, {Francke}, {Lai}, {Schawinski},
  {Gronwall}, {Ciardullo}, {Quadri}, {Orsi}, {Barrientos}, {Blanc}, {Fazio},
  {Feldmeier}, {Huang}, {Infante}, {Lira}, {Padilla}, {Taylor}, {Treister},
  {Urry}, {van Dokkum}, \& {Virani}}]{gawiser+07}
{Gawiser}, E., {Francke}, H., {Lai}, K., {et~al.} 2007, \apj, 671, 278

\bibitem[{{Giavalisco} {et~al.}(2004){Giavalisco}, {Ferguson}, {Koekemoer},
  {Dickinson}, {Alexander}, {Bauer}, {Bergeron}, {Biagetti}, {Brandt},
  {Casertano}, {Cesarsky}, {Chatzichristou}, {Conselice}, {Cristiani}, {Da
  Costa}, {Dahlen}, {de Mello}, {Eisenhardt}, {Erben}, {Fall}, {Fassnacht},
  {Fosbury}, {Fruchter}, {Gardner}, {Grogin}, {Hook}, {Hornschemeier}, {Idzi},
  {Jogee}, {Kretchmer}, {Laidler}, {Lee}, {Livio}, {Lucas}, {Madau},
  {Mobasher}, {Moustakas}, {Nonino}, {Padovani}, {Papovich}, {Park},
  {Ravindranath}, {Renzini}, {Richardson}, {Riess}, {Rosati}, {Schirmer},
  {Schreier}, {Somerville}, {Spinrad}, {Stern}, {Stiavelli}, {Strolger},
  {Urry}, {Vandame}, {Williams}, \& {Wolf}}]{GOODS}
{Giavalisco}, M., {Ferguson}, H.~C., {Koekemoer}, A.~M., {et~al.} 2004, \apjl,
  600, L93

\bibitem[{Gil-Marín {et~al.}(2020)Gil-Marín, Bautista, Paviot,
  Vargas-Magaña, de la Torre, Fromenteau, Alam, Ávila, Burtin, Chuang,
  Dawson, Hou, de Mattia, Mohammad, Müller, Nadathur, Neveux, Percival,
  Raichoor, Rezaie, Ross, Rossi, Ruhlmann-Kleider, Smith, Tamone, Tinker,
  Tojeiro, Wang, Zhao, Zhao, Brinkmann, Brownstein, Choi, Escoffier,
  de la Macorra, Moon, Newman, Schneider, Seo, \& Vivek}]{gilmarin2020}
Gil-Marín, H., Bautista, J.~E., Paviot, R., {et~al.} 2020, Monthly Notices of
  the Royal Astronomical Society, 498, 2492.
\newblock \url{https://doi.org/10.1093/mnras/staa2455}

\bibitem[{{Grasshorn Gebhardt} {et~al.}(2019){Grasshorn Gebhardt}, {Jeong},
  {Awan}, {Bridge}, {Ciardullo}, {Farrow}, {Gebhardt}, {Hill}, {Komatsu},
  {Molina}, {Paulino-Afonso}, {Saito}, {Schneider}, \&
  {Zeimann}}]{grasshorn+19}
{Grasshorn Gebhardt}, H.~S., {Jeong}, D., {Awan}, H., {et~al.} 2019, \apj, 876,
  32

\bibitem[{{Green} {et~al.}(2014){Green}, {Schlafly}, {Finkbeiner}, {Juri{\'c}},
  {Rix}, {Burgett}, {Chambers}, {Draper}, {Flewelling}, {Kudritzki}, {Magnier},
  {Martin}, {Metcalfe}, {Tonry}, {Wainscoat}, \& {Waters}}]{green+14}
{Green}, G.~M., {Schlafly}, E.~F., {Finkbeiner}, D.~P., {et~al.} 2014, \apj,
  783, 114

\bibitem[{{Gronwall} {et~al.}(2007){Gronwall}, {Ciardullo}, {Hickey},
  {Gawiser}, {Feldmeier}, {van Dokkum}, {Urry}, {Herrera}, {Lehmer}, {Infante},
  {Orsi}, {Marchesini}, {Blanc}, {Francke}, {Lira}, \&
  {Treister}}]{gronwall+07}
{Gronwall}, C., {Ciardullo}, R., {Hickey}, T., {et~al.} 2007, \apj, 667, 79

\bibitem[{{Guaita} {et~al.}(2010){Guaita}, {Gawiser}, {Padilla}, {Francke},
  {Bond}, {Gronwall}, {Ciardullo}, {Feldmeier}, {Sinawa}, {Blanc}, \&
  {Virani}}]{guaita+10}
{Guaita}, L., {Gawiser}, E., {Padilla}, N., {et~al.} 2010, \apj, 714, 255

\bibitem[{{Guaita} {et~al.}(2015){Guaita}, {Melinder}, {Hayes}, {{\"O}stlin},
  {Gonzalez}, {Micheva}, {Adamo}, {Mas-Hesse}, {Sandberg}, {Ot{\'\i}-Floranes},
  {Schaerer}, {Verhamme}, {Freeland}, {Orlitov{\'a}}, {Laursen}, {Cannon},
  {Duval}, {Rivera-Thorsen}, {Herenz}, {Kunth}, {Atek}, {Puschnig}, {Gruyters},
  \& {Pardy}}]{guaita+15}
{Guaita}, L., {Melinder}, J., {Hayes}, M., {et~al.} 2015, \aap, 576, A51

\bibitem[{{Gurung-L{\'o}pez} {et~al.}(2021){Gurung-L{\'o}pez}, {Saito},
  {Baugh}, {Bonoli}, {Lacey}, \& {Orsi}}]{gurung2021}
{Gurung-L{\'o}pez}, S., {Saito}, S., {Baugh}, C.~M., {et~al.} 2021, \mnras,
  500, 603

\bibitem[{{Hawkins} {et~al.}(2021){Hawkins}, {Zeimann}, {Sneden}, {Cooper},
  {Gebhardt}, {Bond}, {Carrillo}, {Casey}, {Castanheira}, {Ciardullo}, {Davis},
  {Farrow}, {Finkelstein}, {Hill}, {Kelz}, {Liu}, {Shetrone}, {Schneider},
  {Starkenburg}, {Steinmetz}, {Wheeler}, \& {Hetdex
  Collaboration}}]{hawkins+21}
{Hawkins}, K., {Zeimann}, G., {Sneden}, C., {et~al.} 2021, \apj, 911, 108

\bibitem[{{Hayes} \& {Latham}(1975)}]{hayes-latham75}
{Hayes}, D.~S., \& {Latham}, D.~W. 1975, \apj, 197, 593

\bibitem[{{Herenz} \& {Wisotzki}(2017)}]{herenz2017}
{Herenz}, E.~C., \& {Wisotzki}, L. 2017, \aap, 602, A111

\bibitem[{{Heymans} {et~al.}(2021){Heymans}, {Tr{\"o}ster}, {Asgari}, {Blake},
  {Hildebrandt}, {Joachimi}, {Kuijken}, {Lin}, {S{\'a}nchez}, {van den Busch},
  {Wright}, {Amon}, {Bilicki}, {de Jong}, {Crocce}, {Dvornik}, {Erben},
  {Fortuna}, {Getman}, {Giblin}, {Glazebrook}, {Hoekstra}, {Joudaki},
  {Kannawadi}, {K{\"o}hlinger}, {Lidman}, {Miller}, {Napolitano}, {Parkinson},
  {Schneider}, {Shan}, {Valentijn}, {Verdoes Kleijn}, \& {Wolf}}]{heymans+21}
{Heymans}, C., {Tr{\"o}ster}, T., {Asgari}, M., {et~al.} 2021, \aap, 646, A140

\bibitem[{{Hikage} {et~al.}(2019){Hikage}, {Oguri}, {Hamana}, {More},
  {Mandelbaum}, {Takada}, {K{\"o}hlinger}, {Miyatake}, {Nishizawa}, {Aihara},
  {Armstrong}, {Bosch}, {Coupon}, {Ducout}, {Ho}, {Hsieh}, {Komiyama},
  {Lanusse}, {Leauthaud}, {Lupton}, {Medezinski}, {Mineo}, {Miyama},
  {Miyazaki}, {Murata}, {Murayama}, {Shirasaki}, {Sif{\'o}n}, {Simet},
  {Speagle}, {Spergel}, {Strauss}, {Sugiyama}, {Tanaka}, {Utsumi}, {Wang}, \&
  {Yamada}}]{hikage+19}
{Hikage}, C., {Oguri}, M., {Hamana}, T., {et~al.} 2019, \pasj, 71, 43

\bibitem[{{Hill} \& {HETDEX Consortium}(2016)}]{hill2016}
{Hill}, G.~J., \& {HETDEX Consortium}. 2016, in Astronomical Society of the
  Pacific Conference Series, Vol. 507, Multi-Object Spectroscopy in the Next
  Decade: Big Questions, Large Surveys, and Wide Fields, ed. I.~{Skillen},
  M.~{Balcells}, \& S.~{Trager}, 393

\bibitem[{{Horne}(1986)}]{horne86}
{Horne}, K. 1986, \pasp, 98, 609

\bibitem[{{Indahl} {et~al.}(2016){Indahl}, {Hill}, {Drory}, {Gebhardt},
  {Tuttle}, {Ramsey}, {Ziemann}, {Chonis}, {Peterson}, {Peterson}, {Vattiat},
  {Li}, \& {Hao}}]{indahl2016}
{Indahl}, B.~L., {Hill}, G.~J., {Drory}, N., {et~al.} 2016, in Society of
  Photo-Optical Instrumentation Engineers (SPIE) Conference Series, Vol. 9908,
  Ground-based and Airborne Instrumentation for Astronomy VI, ed. C.~J.
  {Evans}, L.~{Simard}, \& H.~{Takami}, 990880

\bibitem[{{Jeong} \& {Komatsu}(2006)}]{jeong2006}
{Jeong}, D., \& {Komatsu}, E. 2006, \apj, 651, 619

\bibitem[{{Jeong} \& {Komatsu}(2009)}]{jeong2009}
---. 2009, \apj, 691, 569

\bibitem[{{Joudaki} {et~al.}(2020){Joudaki}, {Hildebrandt}, {Traykova},
  {Chisari}, {Heymans}, {Kannawadi}, {Kuijken}, {Wright}, {Asgari}, {Erben},
  {Hoekstra}, {Joachimi}, {Miller}, {Tr{\"o}ster}, \& {van den
  Busch}}]{joudaki+20}
{Joudaki}, S., {Hildebrandt}, H., {Traykova}, D., {et~al.} 2020, \aap, 638, L1

\bibitem[{{Kelz} {et~al.}(2014){Kelz}, {Jahn}, {Haynes}, {Hill}, {Lee},
  {Murphy}, {Neumann}, {Nicklas}, {Rutowska}, {Sandin}, {Streicher}, {Tuttle},
  {Fabricius}, {Bauer}, {Vattiat}, {Anwand}, \& {Savage}}]{kelz2014}
{Kelz}, A., {Jahn}, T., {Haynes}, D., {et~al.} 2014, in Society of
  Photo-Optical Instrumentation Engineers (SPIE) Conference Series, Vol. 9147,
  Ground-based and Airborne Instrumentation for Astronomy V, ed. S.~K.
  {Ramsay}, I.~S. {McLean}, \& H.~{Takami}, 914775

\bibitem[{{Khostovan} {et~al.}(2019){Khostovan}, {Sobral}, {Mobasher},
  {Matthee}, {Cochrane}, {Chartab}, {Jafariyazani}, {Paulino-Afonso}, {Santos},
  \& {Calhau}}]{khostovan19}
{Khostovan}, A.~A., {Sobral}, D., {Mobasher}, B., {et~al.} 2019, \mnras, 489,
  555

\bibitem[{{Koekemoer} {et~al.}(2011){Koekemoer}, {Faber}, {Ferguson}, {Grogin},
  {Kocevski}, {Koo}, {Lai}, {Lotz}, {Lucas}, {McGrath}, {Ogaz}, {Rajan},
  {Riess}, {Rodney}, {Strolger}, {Casertano}, {Castellano}, {Dahlen},
  {Dickinson}, {Dolch}, {Fontana}, {Giavalisco}, {Grazian}, {Guo}, {Hathi},
  {Huang}, {van der Wel}, {Yan}, {Acquaviva}, {Alexander}, {Almaini}, {Ashby},
  {Barden}, {Bell}, {Bournaud}, {Brown}, {Caputi}, {Cassata}, {Challis},
  {Chary}, {Cheung}, {Cirasuolo}, {Conselice}, {Roshan Cooray}, {Croton},
  {Daddi}, {Dav{\'e}}, {de Mello}, {de Ravel}, {Dekel}, {Donley}, {Dunlop},
  {Dutton}, {Elbaz}, {Fazio}, {Filippenko}, {Finkelstein}, {Frazer}, {Gardner},
  {Garnavich}, {Gawiser}, {Gruetzbauch}, {Hartley}, {H{\"a}ussler},
  {Herrington}, {Hopkins}, {Huang}, {Jha}, {Johnson}, {Kartaltepe},
  {Khostovan}, {Kirshner}, {Lani}, {Lee}, {Li}, {Madau}, {McCarthy},
  {McIntosh}, {McLure}, {McPartland}, {Mobasher}, {Moreira}, {Mortlock},
  {Moustakas}, {Mozena}, {Nandra}, {Newman}, {Nielsen}, {Niemi}, {Noeske},
  {Papovich}, {Pentericci}, {Pope}, {Primack}, {Ravindranath}, {Reddy},
  {Renzini}, {Rix}, {Robaina}, {Rosario}, {Rosati}, {Salimbeni}, {Scarlata},
  {Siana}, {Simard}, {Smidt}, {Snyder}, {Somerville}, {Spinrad}, {Straughn},
  {Telford}, {Teplitz}, {Trump}, {Vargas}, {Villforth}, {Wagner}, {Wandro},
  {Wechsler}, {Weiner}, {Wiklind}, {Wild}, {Wilson}, {Wuyts}, \&
  {Yun}}]{koekemoer+11}
{Koekemoer}, A.~M., {Faber}, S.~M., {Ferguson}, H.~C., {et~al.} 2011, \apjs,
  197, 36

\bibitem[{{Komatsu} {et~al.}(2014){Komatsu}, {Bennett}, {Barnes}, {Bean},
  {Bennett}, {Dor{\'e}}, {Dunkley}, {Gold}, {Greason}, {Halpern}, {Hill},
  {Hinshaw}, {Jarosik}, {Kogut}, {Komatsu}, {Larson}, {Limon}, {Meyer},
  {Nolta}, {Odegard}, {Page}, {Peiris}, {Smith}, {Spergel}, {Tucker}, {Verde},
  {Weiland}, {Wollack}, \& {Wright}}]{komatsu+14}
{Komatsu}, E., {Bennett}, C.~L., {Barnes}, C., {et~al.} 2014, Progress of
  Theoretical and Experimental Physics, 2014, 06B102

\bibitem[{{Kurucz}(1970)}]{kurucz70}
{Kurucz}, R.~L. 1970, SAO Special Report, 309

\bibitem[{{Kurucz}(1979)}]{kurucz79}
---. 1979, \apjs, 40, 1

\bibitem[{{Kurucz} {et~al.}(1984){Kurucz}, {Furenlid}, {Brault}, \&
  {Testerman}}]{kurucz+84}
{Kurucz}, R.~L., {Furenlid}, I., {Brault}, J., \& {Testerman}, L. 1984, {Solar
  flux atlas from 296 to 1300 nm}

\bibitem[{{Kusakabe} {et~al.}(2018){Kusakabe}, {Shimasaku}, {Ouchi},
  {Nakajima}, {Goto}, {Hashimoto}, {Konno}, {Harikane}, {Silverman}, \&
  {Capak}}]{kusakabe+18}
{Kusakabe}, H., {Shimasaku}, K., {Ouchi}, M., {et~al.} 2018, \pasj, 70, 4

\bibitem[{{Laureijs} {et~al.}(2011){Laureijs}, {Amiaux}, {Arduini},
  {Augu{\`e}res}, {Brinchmann}, {Cole}, {Cropper}, {Dabin}, {Duvet}, {Ealet},
  {Garilli}, {Gondoin}, {Guzzo}, {Hoar}, {Hoekstra}, {Holmes}, {Kitching},
  {Maciaszek}, {Mellier}, {Pasian}, {Percival}, {Rhodes}, {Saavedra Criado},
  {Sauvage}, {Scaramella}, {Valenziano}, {Warren}, {Bender}, {Castander},
  {Cimatti}, {Le F{\`e}vre}, {Kurki-Suonio}, {Levi}, {Lilje}, {Meylan},
  {Nichol}, {Pedersen}, {Popa}, {Rebolo Lopez}, {Rix}, {Rottgering},
  {Zeilinger}, {Grupp}, {Hudelot}, {Massey}, {Meneghetti}, {Miller}, {Paltani},
  {Paulin-Henriksson}, {Pires}, {Saxton}, {Schrabback}, {Seidel}, {Walsh},
  {Aghanim}, {Amendola}, {Bartlett}, {Baccigalupi}, {Beaulieu}, {Benabed},
  {Cuby}, {Elbaz}, {Fosalba}, {Gavazzi}, {Helmi}, {Hook}, {Irwin}, {Kneib},
  {Kunz}, {Mannucci}, {Moscardini}, {Tao}, {Teyssier}, {Weller}, {Zamorani},
  {Zapatero Osorio}, {Boulade}, {Foumond}, {Di Giorgio}, {Guttridge}, {James},
  {Kemp}, {Martignac}, {Spencer}, {Walton}, {Bl{\"u}mchen}, {Bonoli},
  {Bortoletto}, {Cerna}, {Corcione}, {Fabron}, {Jahnke}, {Ligori}, {Madrid},
  {Martin}, {Morgante}, {Pamplona}, {Prieto}, {Riva}, {Toledo}, {Trifoglio},
  {Zerbi}, {Abdalla}, {Douspis}, {Grenet}, {Borgani}, {Bouwens}, {Courbin},
  {Delouis}, {Dubath}, {Fontana}, {Frailis}, {Grazian}, {Koppenh{\"o}fer},
  {Mansutti}, {Melchior}, {Mignoli}, {Mohr}, {Neissner}, {Noddle}, {Poncet},
  {Scodeggio}, {Serrano}, {Shane}, {Starck}, {Surace}, {Taylor},
  {Verdoes-Kleijn}, {Vuerli}, {Williams}, {Zacchei}, {Altieri}, {Escudero
  Sanz}, {Kohley}, {Oosterbroek}, {Astier}, {Bacon}, {Bardelli}, {Baugh},
  {Bellagamba}, {Benoist}, {Bianchi}, {Biviano}, {Branchini}, {Carbone},
  {Cardone}, {Clements}, {Colombi}, {Conselice}, {Cresci}, {Deacon}, {Dunlop},
  {Fedeli}, {Fontanot}, {Franzetti}, {Giocoli}, {Garcia-Bellido}, {Gow},
  {Heavens}, {Hewett}, {Heymans}, {Holland}, {Huang}, {Ilbert}, {Joachimi},
  {Jennins}, {Kerins}, {Kiessling}, {Kirk}, {Kotak}, {Krause}, {Lahav}, {van
  Leeuwen}, {Lesgourgues}, {Lombardi}, {Magliocchetti}, {Maguire}, {Majerotto},
  {Maoli}, {Marulli}, {Maurogordato}, {McCracken}, {McLure}, {Melchiorri},
  {Merson}, {Moresco}, {Nonino}, {Norberg}, {Peacock}, {Pello}, {Penny},
  {Pettorino}, {Di Porto}, {Pozzetti}, {Quercellini}, {Radovich}, {Rassat},
  {Roche}, {Ronayette}, {Rossetti}, {Sartoris}, {Schneider}, {Semboloni},
  {Serjeant}, {Simpson}, {Skordis}, {Smadja}, {Smartt}, {Spano}, {Spiro},
  {Sullivan}, {Tilquin}, {Trotta}, {Verde}, {Wang}, {Williger}, {Zhao},
  {Zoubian}, \& {Zucca}}]{EUCLID}
{Laureijs}, R., {Amiaux}, J., {Arduini}, S., {et~al.} 2011, arXiv e-prints,
  arXiv:1110.3193

\bibitem[{{Lemaux} {et~al.}(2021){Lemaux}, {Fuller}, {Brada{\v{c}}},
  {Pentericci}, {Hoag}, {Strait}, {Treu}, {Alvarez}, {Bolan}, {Gandhi},
  {Huang}, {Jones}, {Mason}, {Pelliccia}, {Ribeiro}, {Ryan}, {Schmidt},
  {Vanzella}, {Khusanova}, {Le F{\`e}vre}, {Guaita}, {Hathi}, {Koekemoer}, \&
  {Pforr}}]{lemaux2021}
{Lemaux}, B.~C., {Fuller}, S., {Brada{\v{c}}}, M., {et~al.} 2021, \mnras, 504,
  3662

\bibitem[{{Leung} {et~al.}(2017){Leung}, {Acquaviva}, {Gawiser}, {Ciardullo},
  {Komatsu}, {Malz}, {Zeimann}, {Bridge}, {Drory}, {Feldmeier}, {Finkelstein},
  {Gebhardt}, {Gronwall}, {Hagen}, {Hill}, \& {Schneider}}]{leung+17}
{Leung}, A.~S., {Acquaviva}, V., {Gawiser}, E., {et~al.} 2017, \apj, 843, 130

\bibitem[{{Luo} {et~al.}(2015){Luo}, {Zhao}, {Zhao}, {Deng}, {Liu}, {Jing},
  {Wang}, {Zhang}, {Shi}, {Cui}, {Chu}, {Li}, {Bai}, {Wu}, {Cai}, {Cao}, {Cao},
  {Carlin}, {Chen}, {Chen}, {Chen}, {Chen}, {Chen}, {Chen}, {Chen},
  {Christlieb}, {Chu}, {Cui}, {Dong}, {Du}, {Fan}, {Feng}, {Fu}, {Gao}, {Gong},
  {Gu}, {Guo}, {Han}, {He}, {Hou}, {Hou}, {Hou}, {Hu}, {Hu}, {Hu}, {Huo},
  {Jia}, {Jiang}, {Jiang}, {Jiang}, {Jin}, {Kong}, {Kong}, {Lei}, {Li}, {Li},
  {Li}, {Li}, {Li}, {Li}, {Li}, {Li}, {Li}, {Li}, {Li}, {Li}, {Liang}, {Lin},
  {Liu}, {Liu}, {Liu}, {Liu}, {Lu}, {Luo}, {Mao}, {Newberg}, {Ni}, {Qi}, {Qi},
  {Shen}, {Shi}, {Song}, {Song}, {Su}, {Su}, {Tang}, {Tao}, {Tian}, {Wang},
  {Wang}, {Wang}, {Wang}, {Wang}, {Wang}, {Wang}, {Wang}, {Wang}, {Wang},
  {Wang}, {Wang}, {Wang}, {Wang}, {Wang}, {Wang}, {Wang}, {Wang}, {Wang},
  {Wang}, {Wei}, {Wei}, {Wu}, {Wu}, {Wu}, {Wu}, {Xing}, {Xu}, {Xu}, {Xu},
  {Yan}, {Yang}, {Yang}, {Yang}, {Yang}, {Yao}, {Yu}, {Yuan}, {Yuan}, {Yuan},
  {Yuan}, {Zhai}, {Zhang}, {Zhang}, {Zhang}, {Zhang}, {Zhang}, {Zhang},
  {Zhang}, {Zhang}, {Zhao}, {Zhou}, {Zhou}, {Zhu}, {Zhu}, {Zou}, \&
  {Zuo}}]{luo+15}
{Luo}, A.~L., {Zhao}, Y.-H., {Zhao}, G., {et~al.} 2015, Research in Astronomy
  and Astrophysics, 15, 1095

\bibitem[{{Marigo} {et~al.}(2017){Marigo}, {Girardi}, {Bressan}, {Rosenfield},
  {Aringer}, {Chen}, {Dussin}, {Nanni}, {Pastorelli}, {Rodrigues}, {Trabucchi},
  {Bladh}, {Dalcanton}, {Groenewegen}, {Montalb{\'a}n}, \& {Wood}}]{marigo+17}
{Marigo}, P., {Girardi}, L., {Bressan}, A., {et~al.} 2017, \apj, 835, 77

\bibitem[{{Massey} {et~al.}(1988){Massey}, {Strobel}, {Barnes}, \&
  {Anderson}}]{massey+88}
{Massey}, P., {Strobel}, K., {Barnes}, J.~V., \& {Anderson}, E. 1988, \apj,
  328, 315

\bibitem[{{McCullagh} {et~al.}(2016){McCullagh}, {Jeong}, \&
  {Szalay}}]{mccullagh2016}
{McCullagh}, N., {Jeong}, D., \& {Szalay}, A.~S. 2016, \mnras, 455, 2945

\bibitem[{{Momose} {et~al.}(2016){Momose}, {Ouchi}, {Nakajima}, {Ono},
  {Shibuya}, {Shimasaku}, {Yuma}, {Mori}, \& {Umemura}}]{momose+16}
{Momose}, R., {Ouchi}, M., {Nakajima}, K., {et~al.} 2016, \mnras, 457, 2318

\bibitem[{{Morrissey} {et~al.}(2007){Morrissey}, {Conrow}, {Barlow}, {Small},
  {Seibert}, {Wyder}, {Budav{\'a}ri}, {Arnouts}, {Friedman}, {Forster},
  {Martin}, {Neff}, {Schiminovich}, {Bianchi}, {Donas}, {Heckman}, {Lee},
  {Madore}, {Milliard}, {Rich}, {Szalay}, {Welsh}, \& {Yi}}]{galex1}
{Morrissey}, P., {Conrow}, T., {Barlow}, T.~A., {et~al.} 2007, \apjs, 173, 682

\bibitem[{{Muzahid} {et~al.}(2020){Muzahid}, {Schaye}, {Marino}, {Cantalupo},
  {Brinchmann}, {Contini}, {Wendt}, {Wisotzki}, {Zabl}, {Bouch{\'e}},
  {Akhlaghi}, {Chen}, {Claeyssens}, {Johnson}, {Leclercq}, {Maseda}, {Matthee},
  {Richard}, {Urrutia}, \& {Verhamme}}]{muzahid+20}
{Muzahid}, S., {Schaye}, J., {Marino}, R.~A., {et~al.} 2020, \mnras, 496, 1013

\bibitem[{{Narayan} {et~al.}(2019){Narayan}, {Matheson}, {Saha}, {Axelrod},
  {Calamida}, {Olszewski}, {Claver}, {Mandel}, {Bohlin}, {Holberg}, {Deustua},
  {Rest}, {Stubbs}, {Shanahan}, {Vaz}, {Zenteno}, {Strampelli}, {Hubeny},
  {Points}, {Sabbi}, \& {Mackenty}}]{narayan+19}
{Narayan}, G., {Matheson}, T., {Saha}, A., {et~al.} 2019, \apjs, 241, 20

\bibitem[{{Oke}(1990)}]{oke90}
{Oke}, J.~B. 1990, \aj, 99, 1621

\bibitem[{{Ouchi} {et~al.}(2020){Ouchi}, {Ono}, \& {Shibuya}}]{ouchi+20}
{Ouchi}, M., {Ono}, Y., \& {Shibuya}, T. 2020, \araa, 58, 617

\bibitem[{{Ouchi} {et~al.}(2008){Ouchi}, {Shimasaku}, {Akiyama}, {Simpson},
  {Saito}, {Ueda}, {Furusawa}, {Sekiguchi}, {Yamada}, {Kodama}, {Kashikawa},
  {Okamura}, {Iye}, {Takata}, {Yoshida}, \& {Yoshida}}]{ouchi+08}
{Ouchi}, M., {Shimasaku}, K., {Akiyama}, M., {et~al.} 2008, \apjs, 176, 301

\bibitem[{{Padmanabhan} {et~al.}(2008){Padmanabhan}, {Schlegel}, {Finkbeiner},
  {Barentine}, {Blanton}, {Brewington}, {Gunn}, {Harvanek}, {Hogg},
  {Ivezi{\'c}}, {Johnston}, {Kent}, {Kleinman}, {Knapp}, {Krzesinski}, {Long},
  {Neilsen}, {Nitta}, {Loomis}, {Lupton}, {Roweis}, {Snedden}, {Strauss}, \&
  {Tucker}}]{padmanabhan+08}
{Padmanabhan}, N., {Schlegel}, D.~J., {Finkbeiner}, D.~P., {et~al.} 2008, \apj,
  674, 1217

\bibitem[{{Papovich} {et~al.}(2016){Papovich}, {Shipley}, {Mehrtens}, {Lanham},
  {Lacy}, {Ciardullo}, {Finkelstein}, {Bassett}, {Behroozi}, {Blanc}, {de
  Jong}, {DePoy}, {Drory}, {Gawiser}, {Gebhardt}, {Gronwall}, {Hill}, {Hopp},
  {Jogee}, {Kawinwanichakij}, {Marshall}, {McLinden}, {Mentuch Cooper},
  {Somerville}, {Steinmetz}, {Tran}, {Tuttle}, {Viero}, {Wechsler}, \&
  {Zeimann}}]{SHELA}
{Papovich}, C., {Shipley}, H.~V., {Mehrtens}, N., {et~al.} 2016, \apjs, 224, 28

\bibitem[{{Perlmutter} {et~al.}(1999){Perlmutter}, {Aldering}, {Goldhaber},
  {Knop}, {Nugent}, {Castro}, {Deustua}, {Fabbro}, {Goobar}, {Groom}, {Hook},
  {Kim}, {Kim}, {Lee}, {Nunes}, {Pain}, {Pennypacker}, {Quimby}, {Lidman},
  {Ellis}, {Irwin}, {McMahon}, {Ruiz-Lapuente}, {Walton}, {Schaefer}, {Boyle},
  {Filippenko}, {Matheson}, {Fruchter}, {Panagia}, {Newberg}, {Couch}, \&
  {Project}}]{perlmutter+99}
{Perlmutter}, S., {Aldering}, G., {Goldhaber}, G., {et~al.} 1999, \apj, 517,
  565

\bibitem[{{Planck Collaboration VI}(2020)}]{Planck2020}
{Planck Collaboration VI}. 2020, \aap, 641, A6

\bibitem[{{Pullen} {et~al.}(2016){Pullen}, {Hirata}, {Dor{\'e}}, \&
  {Raccanelli}}]{pullen+16}
{Pullen}, A.~R., {Hirata}, C.~M., {Dor{\'e}}, O., \& {Raccanelli}, A. 2016,
  \pasj, 68, 12

\bibitem[{{Rhoads} {et~al.}(2000){Rhoads}, {Malhotra}, {Dey}, {Stern},
  {Spinrad}, \& {Jannuzi}}]{rhoads+00}
{Rhoads}, J.~E., {Malhotra}, S., {Dey}, A., {et~al.} 2000, \apjl, 545, L85

\bibitem[{{Riess} {et~al.}(2021){Riess}, {Casertano}, {Yuan}, {Bowers},
  {Macri}, {Zinn}, \& {Scolnic}}]{riess+21}
{Riess}, A.~G., {Casertano}, S., {Yuan}, W., {et~al.} 2021, \apjl, 908, L6

\bibitem[{{Riess} {et~al.}(1998){Riess}, {Filippenko}, {Challis},
  {Clocchiatti}, {Diercks}, {Garnavich}, {Gilliland}, {Hogan}, {Jha},
  {Kirshner}, {Leibundgut}, {Phillips}, {Reiss}, {Schmidt}, {Schommer},
  {Smith}, {Spyromilio}, {Stubbs}, {Suntzeff}, \& {Tonry}}]{riess+98}
{Riess}, A.~G., {Filippenko}, A.~V., {Challis}, P., {et~al.} 1998, \aj, 116,
  1009

\bibitem[{{Roddier}(1981)}]{roddier81}
{Roddier}, F. 1981, Progess in Optics, 19, 281

\bibitem[{{Roeser} {et~al.}(2010){Roeser}, {Demleitner}, \& {Schilbach}}]{USNO}
{Roeser}, S., {Demleitner}, M., \& {Schilbach}, E. 2010, \aj, 139, 2440

\bibitem[{{Runnholm} {et~al.}(2021){Runnholm}, {Gronke}, \&
  {Hayes}}]{runnholm+21}
{Runnholm}, A., {Gronke}, M., \& {Hayes}, M. 2021, \pasp, 133, 034507

\bibitem[{{Schlafly} \& {Finkbeiner}(2011)}]{schlafly+11}
{Schlafly}, E.~F., \& {Finkbeiner}, D.~P. 2011, \apj, 737, 103

\bibitem[{{Schlegel} {et~al.}(1998){Schlegel}, {Finkbeiner}, \&
  {Davis}}]{schlegel+98}
{Schlegel}, D.~J., {Finkbeiner}, D.~P., \& {Davis}, M. 1998, \apj, 500, 525

\bibitem[{{Scolnic} {et~al.}(2018){Scolnic}, {Jones}, {Rest}, {Pan},
  {Chornock}, {Foley}, {Huber}, {Kessler}, {Narayan}, {Riess}, {Rodney},
  {Berger}, {Brout}, {Challis}, {Drout}, {Finkbeiner}, {Lunnan}, {Kirshner},
  {Sanders}, {Schlafly}, {Smartt}, {Stubbs}, {Tonry}, {Wood-Vasey}, {Foley},
  {Hand}, {Johnson}, {Burgett}, {Chambers}, {Draper}, {Hodapp}, {Kaiser},
  {Kudritzki}, {Magnier}, {Metcalfe}, {Bresolin}, {Gall}, {Kotak}, {McCrum}, \&
  {Smith}}]{scolnic+18}
{Scolnic}, D.~M., {Jones}, D.~O., {Rest}, A., {et~al.} 2018, \apj, 859, 101

\bibitem[{{Scoville} {et~al.}(2007){Scoville}, {Aussel}, {Brusa}, {Capak},
  {Carollo}, {Elvis}, {Giavalisco}, {Guzzo}, {Hasinger}, {Impey}, {Kneib},
  {LeFevre}, {Lilly}, {Mobasher}, {Renzini}, {Rich}, {Sanders}, {Schinnerer},
  {Schminovich}, {Shopbell}, {Taniguchi}, \& {Tyson}}]{COSMOS}
{Scoville}, N., {Aussel}, H., {Brusa}, M., {et~al.} 2007, \apjs, 172, 1

\bibitem[{{Shapley} {et~al.}(2003){Shapley}, {Steidel}, {Pettini}, \&
  {Adelberger}}]{shapley+03}
{Shapley}, A.~E., {Steidel}, C.~C., {Pettini}, M., \& {Adelberger}, K.~L. 2003,
  \apj, 588, 65

\bibitem[{{Shibuya} {et~al.}(2014){Shibuya}, {Ouchi}, {Nakajima}, {Hashimoto},
  {Ono}, {Rauch}, {Gauthier}, {Shimasaku}, {Goto}, {Mori}, \&
  {Umemura.}}]{shibuya2014}
{Shibuya}, T., {Ouchi}, M., {Nakajima}, K., {et~al.} 2014, \apj, 788, 74

\bibitem[{{Shoji} {et~al.}(2009){Shoji}, {Jeong}, \& {Komatsu}}]{shoji+09}
{Shoji}, M., {Jeong}, D., \& {Komatsu}, E. 2009, \apj, 693, 1404

\bibitem[{{Skrutskie} {et~al.}(2006){Skrutskie}, {Cutri}, {Stiening},
  {Weinberg}, {Schneider}, {Carpenter}, {Beichman}, {Capps}, {Chester},
  {Elias}, {Huchra}, {Liebert}, {Lonsdale}, {Monet}, {Price}, {Seitzer},
  {Jarrett}, {Kirkpatrick}, {Gizis}, {Howard}, {Evans}, {Fowler}, {Fullmer},
  {Hurt}, {Light}, {Kopan}, {Marsh}, {McCallon}, {Tam}, {Van Dyk}, \&
  {Wheelock}}]{2MASS}
{Skrutskie}, M.~F., {Cutri}, R.~M., {Stiening}, R., {et~al.} 2006, \aj, 131,
  1163

\bibitem[{{Spergel} {et~al.}(2015){Spergel}, {Gehrels}, {Baltay}, {Bennett},
  {Breckinridge}, {Donahue}, {Dressler}, {Gaudi}, {Greene}, {Guyon}, {Hirata},
  {Kalirai}, {Kasdin}, {Macintosh}, {Moos}, {Perlmutter}, {Postman},
  {Rauscher}, {Rhodes}, {Wang}, {Weinberg}, {Benford}, {Hudson}, {Jeong},
  {Mellier}, {Traub}, {Yamada}, {Capak}, {Colbert}, {Masters}, {Penny},
  {Savransky}, {Stern}, {Zimmerman}, {Barry}, {Bartusek}, {Carpenter}, {Cheng},
  {Content}, {Dekens}, {Demers}, {Grady}, {Jackson}, {Kuan}, {Kruk}, {Melton},
  {Nemati}, {Parvin}, {Poberezhskiy}, {Peddie}, {Ruffa}, {Wallace}, {Whipple},
  {Wollack}, \& {Zhao}}]{RST}
{Spergel}, D., {Gehrels}, N., {Baltay}, C., {et~al.} 2015, arXiv e-prints,
  arXiv:1503.03757

\bibitem[{{Stetson}(1987)}]{stetson87}
{Stetson}, P.~B. 1987, \pasp, 99, 191

\bibitem[{{Stetson}(1990)}]{stetson90}
---. 1990, \pasp, 102, 932

\bibitem[{{Takada} {et~al.}(2014){Takada}, {Ellis}, {Chiba}, {Greene},
  {Aihara}, {Arimoto}, {Bundy}, {Cohen}, {Dor{\'e}}, {Graves}, {Gunn},
  {Heckman}, {Hirata}, {Ho}, {Kneib}, {Le F{\`e}vre}, {Lin}, {More},
  {Murayama}, {Nagao}, {Ouchi}, {Seiffert}, {Silverman}, {Sodr{\'e}},
  {Spergel}, {Strauss}, {Sugai}, {Suto}, {Takami}, \& {Wyse}}]{PFS}
{Takada}, M., {Ellis}, R.~S., {Chiba}, M., {et~al.} 2014, \pasj, 66, R1

\bibitem[{{Tegmark}(1996)}]{tegmark96}
{Tegmark}, M. 1996, \apjl, 470, L81

\bibitem[{{Trainor} {et~al.}(2015){Trainor}, {Steidel}, {Strom}, \&
  {Rudie}}]{trainor+15}
{Trainor}, R.~F., {Steidel}, C.~C., {Strom}, A.~L., \& {Rudie}, G.~C. 2015,
  \apj, 809, 89

\bibitem[{{Vargas-Magana} {et~al.}(2019){Vargas-Magana}, {Brooks}, {Levi}, \&
  {Tarle}}]{vargas-magana+19}
{Vargas-Magana}, M., {Brooks}, D.~D., {Levi}, M.~M., \& {Tarle}, G.~G. 2019,
  arXiv e-prints, arXiv:1901.01581

\bibitem[{{Wahba}(1990)}]{wahba+90}
{Wahba}, G. 1990, in CBMS-NSF Regional Conference Series in Applied Mathematics

\bibitem[{{Wisotzki} {et~al.}(2016){Wisotzki}, {Bacon}, {Blaizot},
  {Brinchmann}, {Herenz}, {Schaye}, {Bouch{\'e}}, {Cantalupo}, {Contini},
  {Carollo}, {Caruana}, {Courbot}, {Emsellem}, {Kamann}, {Kerutt}, {Leclercq},
  {Lilly}, {Patr{\'\i}cio}, {Sandin}, {Steinmetz}, {Straka}, {Urrutia},
  {Verhamme}, {Weilbacher}, \& {Wendt}}]{wisotzki+16}
{Wisotzki}, L., {Bacon}, R., {Blaizot}, J., {et~al.} 2016, \aap, 587, A98

\bibitem[{{Wong} {et~al.}(2020){Wong}, {Suyu}, {Chen}, {Rusu}, {Millon},
  {Sluse}, {Bonvin}, {Fassnacht}, {Taubenberger}, {Auger}, {Birrer}, {Chan},
  {Courbin}, {Hilbert}, {Tihhonova}, {Treu}, {Agnello}, {Ding}, {Jee},
  {Komatsu}, {Shajib}, {Sonnenfeld}, {Blandford}, {Koopmans}, {Marshall}, \&
  {Meylan}}]{wong+20}
{Wong}, K.~C., {Suyu}, S.~H., {Chen}, G. C.~F., {et~al.} 2020, \mnras, 498,
  1420

\bibitem[{{Wright} {et~al.}(2010){Wright}, {Eisenhardt}, {Mainzer}, {Ressler},
  {Cutri}, {Jarrett}, {Kirkpatrick}, {Padgett}, {McMillan}, {Skrutskie},
  {Stanford}, {Cohen}, {Walker}, {Mather}, {Leisawitz}, {Gautier}, {McLean},
  {Benford}, {Lonsdale}, {Blain}, {Mendez}, {Irace}, {Duval}, {Liu}, {Royer},
  {Heinrichsen}, {Howard}, {Shannon}, {Kendall}, {Walsh}, {Larsen}, {Cardon},
  {Schick}, {Schwalm}, {Abid}, {Fabinsky}, {Naes}, \& {Tsai}}]{WISE}
{Wright}, E.~L., {Eisenhardt}, P. R.~M., {Mainzer}, A.~K., {et~al.} 2010, \aj,
  140, 1868

\bibitem[{{Xiang} {et~al.}(2017){Xiang}, {Liu}, {Yuan}, {Huo}, {Huang}, {Wang},
  {Chen}, {Ren}, {Zhang}, {Tian}, {Yang}, {Shi}, {Zhao}, {Li}, {Zhao}, {Cui},
  {Li}, {Hou}, {Zhang}, {Zhang}, {Wang}, {Wu}, {Cao}, {Yan}, {Yan}, {Luo},
  {Zhang}, {Bai}, {Yuan}, {Dong}, {Lei}, \& {Li}}]{xiang+17}
{Xiang}, M.~S., {Liu}, X.~W., {Yuan}, H.~B., {et~al.} 2017, \mnras, 467, 1890

\bibitem[{{Yanny} {et~al.}(2009){Yanny}, {Rockosi}, {Newberg}, {Knapp},
  {Adelman-McCarthy}, {Alcorn}, {Allam}, {Allende Prieto}, {An}, {Anderson},
  {Anderson}, {Bailer-Jones}, {Bastian}, {Beers}, {Bell}, {Belokurov},
  {Bizyaev}, {Blythe}, {Bochanski}, {Boroski}, {Brinchmann}, {Brinkmann},
  {Brewington}, {Carey}, {Cudworth}, {Evans}, {Evans}, {Gates}, {G{\"a}nsicke},
  {Gillespie}, {Gilmore}, {Nebot Gomez-Moran}, {Grebel}, {Greenwell}, {Gunn},
  {Jordan}, {Jordan}, {Harding}, {Harris}, {Hendry}, {Holder}, {Ivans},
  {Ivezi{\v{c}}}, {Jester}, {Johnson}, {Kent}, {Kleinman}, {Kniazev},
  {Krzesinski}, {Kron}, {Kuropatkin}, {Lebedeva}, {Lee}, {French Leger},
  {L{\'e}pine}, {Levine}, {Lin}, {Long}, {Loomis}, {Lupton}, {Malanushenko},
  {Malanushenko}, {Margon}, {Martinez-Delgado}, {McGehee}, {Monet}, {Morrison},
  {Munn}, {Neilsen}, {Nitta}, {Norris}, {Oravetz}, {Owen}, {Padmanabhan},
  {Pan}, {Peterson}, {Pier}, {Platson}, {Re Fiorentin}, {Richards}, {Rix},
  {Schlegel}, {Schneider}, {Schreiber}, {Schwope}, {Sibley}, {Simmons},
  {Snedden}, {Allyn Smith}, {Stark}, {Stauffer}, {Steinmetz}, {Stoughton},
  {SubbaRao}, {Szalay}, {Szkody}, {Thakar}, {Sivarani}, {Tucker}, {Uomoto},
  {Vanden Berk}, {Vidrih}, {Wadadekar}, {Watters}, {Wilhelm}, {Wyse}, {Yarger},
  \& {Zucker}}]{yanny+09}
{Yanny}, B., {Rockosi}, C., {Newberg}, H.~J., {et~al.} 2009, \aj, 137, 4377

\bibitem[{{York} {et~al.}(2000){York}, {Adelman}, {Anderson}, {Anderson},
  {Annis}, {Bahcall}, {Bakken}, {Barkhouser}, {Bastian}, {Berman}, {Boroski},
  {Bracker}, {Briegel}, {Briggs}, {Brinkmann}, {Brunner}, {Burles}, {Carey},
  {Carr}, {Castander}, {Chen}, {Colestock}, {Connolly}, {Crocker}, {Csabai},
  {Czarapata}, {Davis}, {Doi}, {Dombeck}, {Eisenstein}, {Ellman}, {Elms},
  {Evans}, {Fan}, {Federwitz}, {Fiscelli}, {Friedman}, {Frieman}, {Fukugita},
  {Gillespie}, {Gunn}, {Gurbani}, {de Haas}, {Haldeman}, {Harris}, {Hayes},
  {Heckman}, {Hennessy}, {Hindsley}, {Holm}, {Holmgren}, {Huang}, {Hull},
  {Husby}, {Ichikawa}, {Ichikawa}, {Ivezi{\'c}}, {Kent}, {Kim}, {Kinney},
  {Klaene}, {Kleinman}, {Kleinman}, {Knapp}, {Korienek}, {Kron}, {Kunszt},
  {Lamb}, {Lee}, {Leger}, {Limmongkol}, {Lindenmeyer}, {Long}, {Loomis},
  {Loveday}, {Lucinio}, {Lupton}, {MacKinnon}, {Mannery}, {Mantsch}, {Margon},
  {McGehee}, {McKay}, {Meiksin}, {Merelli}, {Monet}, {Munn}, {Narayanan},
  {Nash}, {Neilsen}, {Neswold}, {Newberg}, {Nichol}, {Nicinski}, {Nonino},
  {Okada}, {Okamura}, {Ostriker}, {Owen}, {Pauls}, {Peoples}, {Peterson},
  {Petravick}, {Pier}, {Pope}, {Pordes}, {Prosapio}, {Rechenmacher}, {Quinn},
  {Richards}, {Richmond}, {Rivetta}, {Rockosi}, {Ruthmansdorfer}, {Sandford},
  {Schlegel}, {Schneider}, {Sekiguchi}, {Sergey}, {Shimasaku}, {Siegmund},
  {Smee}, {Smith}, {Snedden}, {Stone}, {Stoughton}, {Strauss}, {Stubbs},
  {SubbaRao}, {Szalay}, {Szapudi}, {Szokoly}, {Thakar}, {Tremonti}, {Tucker},
  {Uomoto}, {Vanden Berk}, {Vogeley}, {Waddell}, {Wang}, {Watanabe},
  {Weinberg}, {Yanny}, {Yasuda}, \& {SDSS Collaboration}}]{york+00}
{York}, D.~G., {Adelman}, J., {Anderson}, John~E., J., {et~al.} 2000, \aj, 120,
  1579

\bibitem[{{Zhang} {et~al.}(2021){Zhang}, {Ouchi}, {Gebhardt}, {Mentuch Cooper},
  {Liu}, {Davis}, {Jeong}, {Farrow}, {Finkelstein}, {Gawiser}, {Hill},
  {Harikane}, {Kakuma}, {Acquaviva}, {Casey}, {Fabricius}, {Hopp}, {Jarvis},
  {Landriau}, {Mawatari}, {Mukae}, {Ono}, {Sakai}, \& {Schneider}}]{Zhang2021}
{Zhang}, Y., {Ouchi}, M., {Gebhardt}, K., {et~al.} 2021, arXiv e-prints,
  arXiv:2105.11497

\end{thebibliography}


\end{document}